\pdfoutput=1

\documentclass[11pt,twoside,a4paper,cmspaper,final,collab]{cms-tdr}
\def\svnVersion{282731}\def\cmsCernNo{2013-037}\def\cmsCernDate{\today}\def\cmsMessage{Published in Physical Review D as \href{http://dx.doi.org/10.1103/PhysRevD.91.052018}{\doi{10.1103/PhysRevD.91.052018.}}}

\begin{document}\cmsNoteHeader{SUS-13-004}

\hyphenation{had-ron-i-za-tion}
\hyphenation{cal-or-i-me-ter}
\hyphenation{de-vices}
\RCS$Revision: 282593 $
\RCS$HeadURL: svn+ssh://svn.cern.ch/reps/tdr2/papers/SUS-13-004/trunk/SUS-13-004.tex $
\RCS$Id: SUS-13-004.tex 282593 2015-03-27 21:03:04Z woodson $
\newlength\cmsFigWidth
\ifthenelse{\boolean{cms@external}}{\setlength\cmsFigWidth{0.85\columnwidth}}{\setlength\cmsFigWidth{0.65\textwidth}}
\ifthenelse{\boolean{cms@external}}{\providecommand{\cmsLeft}{top}}{\providecommand{\cmsLeft}{left}}
\ifthenelse{\boolean{cms@external}}{\providecommand{\cmsRight}{bottom}}{\providecommand{\cmsRight}{right}}
\ifthenelse{\boolean{cms@external}}{\providecommand{\CL}{C.L.\xspace}}{\providecommand{\CL}{CL\xspace}}
\ifthenelse{\boolean{cms@external}}{\providecommand{\cmsUpperLeft}{top}}{\providecommand{\cmsUpperLeft}{upper left}}
\ifthenelse{\boolean{cms@external}}{\providecommand{\cmsUpperRight}{middle}}{\providecommand{\cmsUpperRight}{upper right}}
\newcommand{\CLs}{\ensuremath{\mathrm{CL}_\mathrm{s}}\xspace}
\newcommand{\MR}{\ensuremath{M_\mathrm{R}}\xspace}
\newcommand{\MRz}{\ensuremath{M_\mathrm{R}^0}\xspace}
\newcommand{\Rtwo}{\ensuremath{\mathrm{R}^2}\xspace}
\newcommand{\Rtwoz}{\ensuremath{\mathrm{R}^2_0}\xspace}
\newcommand{\R}{\ensuremath{\mathrm{R}}\xspace}
\newcommand{\MRT}{\ensuremath{M_\mathrm{T}^\mathrm{R}}\xspace}
\providecommand{\cPV}{\ensuremath{\cmsSymbolFace{V}}\xspace}
\cmsNoteHeader{SUS-13-004}
\title{Search for supersymmetry using razor variables in events with \texorpdfstring{\PQb-tagged jets in $\Pp\Pp$ collisions at $\sqrt{s}=8$\TeV}{b-tagged jets in pp collisions at sqrt(s) = 8 TeV}}

\date{\today}

\abstract{An inclusive search for supersymmetry in events with at
  least one b-tagged jet is performed using proton-proton collision
  data collected by the CMS experiment in 2012 at a center-of-mass
  energy of 8\TeV. The data set size corresponds to an integrated
  luminosity of 19.3\fbinv. The two-dimensional distribution of the
  razor variables \Rtwo and \MR is studied in events with and without
  leptons. The data are found to be consistent with the expected
  background, which is modeled with an empirical function. Exclusion
  limits on supersymmetric particle masses at a 95\% confidence level
  are derived in several simplified supersymmetric scenarios for
  several choices of the branching fractions. By combining the
  likelihoods of a search in events without leptons and a search that
  requires a single lepton (electron or muon), an improved bound on
  the top-squark mass is obtained. Assuming the lightest
  supersymmetric particle to be stable, weakly interacting, and to
  have a mass of 100\GeV, the branching-fraction-dependent
  (-independent) production of gluinos is excluded for gluino masses
  up to 1310 (1175)\GeV. The corresponding limit for top-squark pair
  production is 730 (645)\GeV.}

\hypersetup{%
 pdfauthor={CMS Collaboration},%
 pdftitle={Search for supersymmetry using razor variables in events with
 b-tagged jets in pp collisions at sqrt(s)=8 TeV},%
 pdfsubject={CMS},%
 pdfkeywords={CMS, physics, razor, SUSY, b-tag}}

\maketitle

\section{Introduction}
Supersymmetry (SUSY) is a proposed symmetry of nature that introduces
a bosonic (fermionic) partner for every standard model (SM) fermion
(boson)~\cite{Wess,Golfand,Volkov,Chamseddine,Kane,Fayet,Barbieri,Hall,Ramond}. Supersymmetric
extensions of the SM that include a stable new particle at the
electroweak scale are well motivated because they may explain the
origin of dark matter. The discovery of the Higgs
boson~\cite{ATLAShiggs, CMShiggs, CMShiggslong} at the CERN LHC has
renewed interest in ``natural'' SUSY models, which minimize the
fine-tuning associated with the observed value of the Higgs boson mass
due to its radiative corrections. In the typical spectrum of these
models, the lightest neutralino and chargino are the lightest (LSP)
and next-to-lightest (NLSP) SUSY particles,
respectively~\cite{deCarlos:1993yy,Anderson:1994dz,Kitano:2005wc,Asano:2010ut,Strumia:2011dv,naturalSUSY}.
Charginos and neutralinos are fermions, corresponding to a quantum
mixture of the SUSY partners of the electroweak and Higgs bosons.  The
bottom and top squarks are the lightest squarks. The gluino is heavier
than these particles but potentially accessible at the LHC. Events are
thus characterized by an abundance of jets originating from the
hadronization of bottom quarks, a feature that we exploit in this
study. Previous searches for natural SUSY by the
CMS~\cite{1LepMVA,SUS12024,Chatrchyan:2014lfa,Chatrchyan:2013iqa,Chatrchyan:2013fea}
and ATLAS
Collaborations~\cite{Aad:2013wta,Aad:2014lra,Aad:2014pda,Aad:2014bva,Aad:2014qaa}
at the LHC have probed gluino masses up to 1300\GeV and top squark
masses up to 700\GeV under the assumptions of specific decay modes for
the SUSY particles.

We present an inclusive search for gluinos and top squarks in the
context of natural SUSY. Natural SUSY spectra include a gluino, the
third-generation squarks, a chargino, and a neutralino, representing
the minimum particle content needed in SUSY theories to
stabilize the Higgs boson mass. Within the context of natural SUSY,
several simplified
models~\cite{ArkaniHamed:2007fw,Alwall:2008ag,Alwall:2008va,Alves:2011sq,Alves:2011wf,Graesser:2012qy}
are considered (Section~\ref{sec:sms}), defined by a specific
production mechanism of SUSY particle pairs, with at most two decay
channels for each production mode.

The search is performed using events with two or more jets, at least
one of which is identified as originating from a bottom quark (jet b
tagging). The study is based on the data collected by the CMS
Collaboration in proton-proton collisions at $\sqrt{s}=8\TeV$ in 2012,
corresponding to an integrated luminosity of 19.3\fbinv. We
distinguish the signal from the SM background through their different
shapes in the razor variables \MR and $\Rtwo$~\cite{rogan,razor2010}.
This search extends the results we presented at
7\TeV~\cite{razorPRL,razorPRD} using the same analysis procedure. The
razor variables have also been used by the ATLAS Collaboration to
perform a multi-channel search for SUSY at 7\TeV~\cite{Aad:2012naa}.

The razor variables \MR and  \Rtwo are motivated by the generic process of the pair production of two
heavy particles (e.g., squarks or gluinos), each decaying to an
undetected particle (the stable, weakly interacting LSP $\chiz_1$)
plus visible particles. The LSP is assumed to escape without
detection, leading to an imbalance $\ptvecmiss$ in the momentum
perpendicular to the beam axis. Each event is treated as a dijet-like event
and the four-momenta of the two jets are used to compute \MR and $\MRT$, defined as
\begin{align}
 \label{eq:MRstar}
 \MR &\equiv
 \sqrt{
(\abs{\vec{p}^{j_{1}}}+\abs{\vec{p}^{j_{2}}})^2 -({p}^{j_1}_z+{p}^{j_2}_z)^2},\\
\MRT &\equiv \sqrt{ \frac{\ETm(\pt^{j_1}+\pt^{j_2}) -
\ptvecmiss \cdot
 (\ptvec^{\,j_1}+\ptvec^{\,j_2}) }{2}},
\end{align}
where $\vec{p}_{j_i}$, $\ptvec^{\,j_i}$, and
$p^{j_i}_z$ are the momentum of the $i$th jet, its
transverse component with respect to the beam axis, and its
longitudinal component, respectively, with $\ETm$ the magnitude of $\ptvecmiss$. While
$\MRT$ quantifies the transverse momentum imbalance,
$\MR$ estimates the mass scale of new-physics particle
production in the event. The razor dimensionless ratio is defined as
\begin{equation}
\R \equiv \frac{\MRT}{\MR}.
\end{equation}

In this search, each event is reduced to a two-jet topology by
clustering the selected objects (jets and leptons) into two megajets~\cite{razor2010,razorPRL,razorPRD}.
All possible assignments of objects to the megajets are considered,
with the requirement that a megajet consist of at least one
object. The sum of the four-momenta of the objects assigned to a
megajet defines the megajet four-momentum.  When more than two objects
are reconstructed, more than one megajet assignment is possible. We
select the assignment that minimizes the sum of the
invariant masses of the two megajets.

The analysis is performed on several exclusive data sets, referred to
as razor boxes, differing in the lepton and jet
multiplicity. Each box with fewer than two identified leptons
(electrons or muons) is analyzed in exclusive \PQb-tagged jet multiplicity bins in order to
maximize the sensitivity to both direct and cascade production of
third-generation squarks.  For a given box and \PQb-tagged jet
multiplicity, the shape of the SM background distribution is evaluated
in two rectangular
regions of the ($\MR$, $\Rtwo$) plane (sidebands),
selected so that potential bias due to contributions from signal
events is negligible. The background shape is then extrapolated to the
signal-sensitive region of the ($\MR$, $\Rtwo$)
plane. The results are interpreted in the context of several SUSY simplified
models by performing a hypothesis test. The test compares the
background-only and signal-plus-background possibilities
through simultaneous examination of the data in the two sidebands and the signal-sensitive
region~\cite{LHCCLs}.
In addition, we combine the results from the razor boxes
with those from our previous search~\cite{1LepMVA} for top-squark
production in the single-lepton (electron or muon) channel to obtain
an improved bound on top-squark pair production with respect to
previous CMS studies. For this combination, only the razor boxes
without an identified lepton (hadronic boxes) are used, so that the
event samples from the two studies are mutually exclusive.

This paper is organized as follows. Section~\ref{sec:sms} presents
the spectra of the simplified natural SUSY models
examined in this analysis. The CMS detector is briefly described in
Section~\ref{sec:cms}. The event selection and razor variables are
defined in Sections~\ref{sec:selection} and~\ref{sec:razVar},
respectively. The statistical model used to describe the SM
backgrounds as well as the comparisons between the predicted and observed
event yields in the search regions are shown in Section~\ref{sec:bmodel},
followed by a summary of the limit-setting procedure in
Section~\ref{sec:limit}. The interpretation of the results and a summary are presented
in Sections~\ref{sec:interp} and~\ref{sec:conclusion}, respectively.

\section{Simplified natural SUSY models}
\label{sec:sms}

\begin{figure}[htb!]
\centering
\includegraphics[width=\cmsFigWidth]{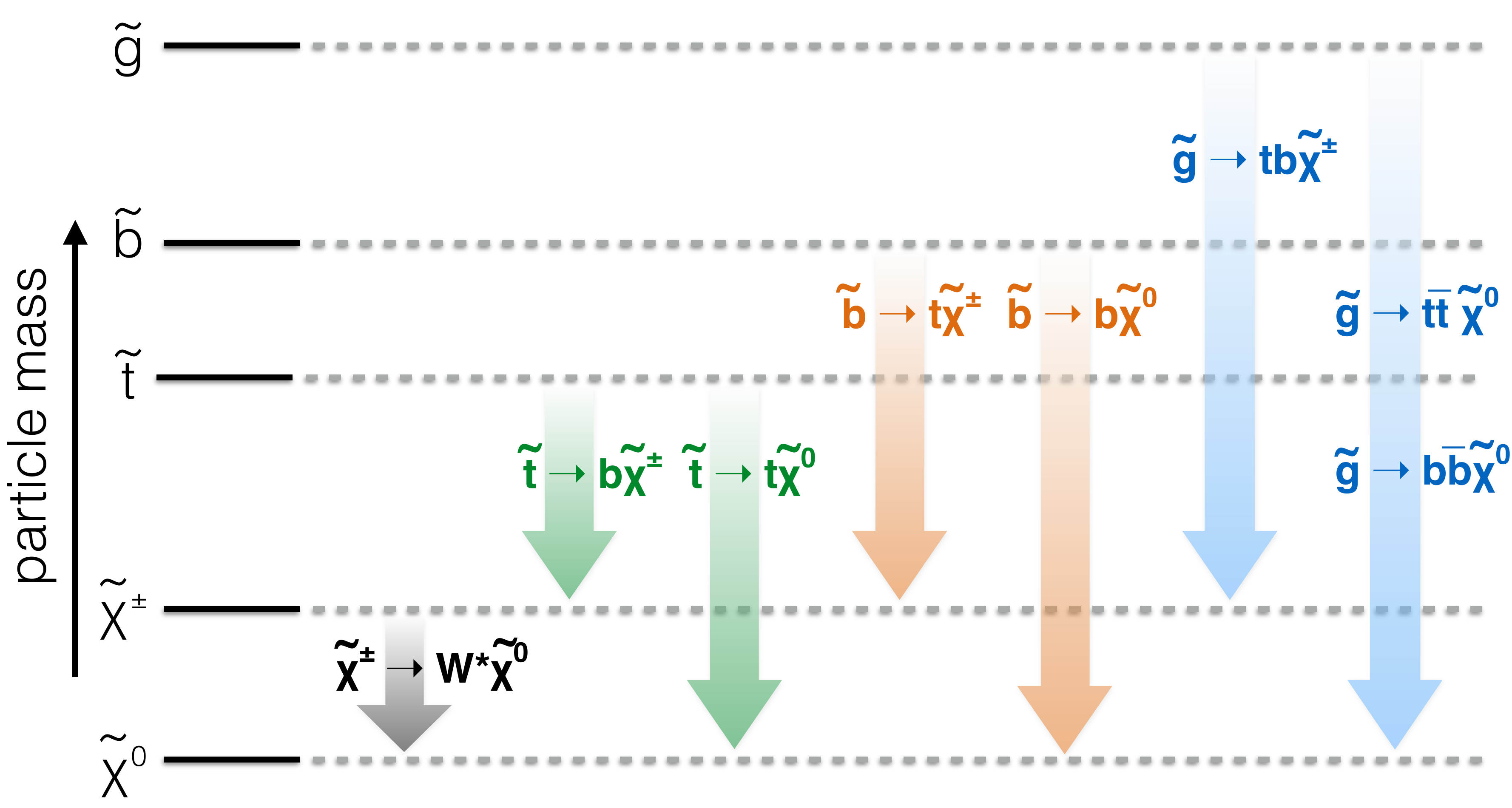}
\caption{\label{fig:spectrum} The simplified natural SUSY spectrum
  considered in this paper, along with the assumed decay modes.}
\end{figure}

In this paper, natural simplified SUSY scenarios are used to interpret
results. The LSP is the lightest neutralino $\chiz_1$ while the NLSP
is the lightest chargino $\chipm_1$.  They are both higgsinos and
their mass splitting is taken to be 5\GeV. The NLSP decays to the LSP
and a virtual $\PW$ boson ($\chipm_1 \to \PW^{\ast} \chiz_1$). The
other SUSY particles accessible at the LHC are the gluino and the
lightest top and bottom squarks. All other SUSY particles are
assumed to be too heavy to participate in the interactions. The SUSY
particles and their possible decay modes within this natural SUSY
spectrum are summarized in Fig.~\ref{fig:spectrum}.

\begin{figure*}[thb!]
\centering
\includegraphics[width=0.32\textwidth]{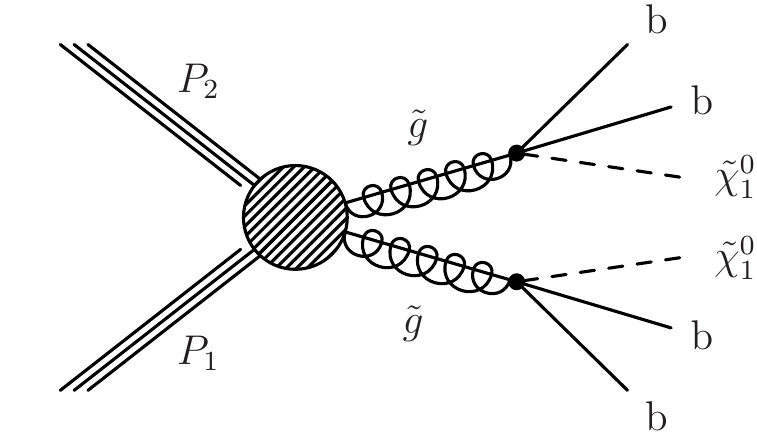}
\includegraphics[width=0.32\textwidth]{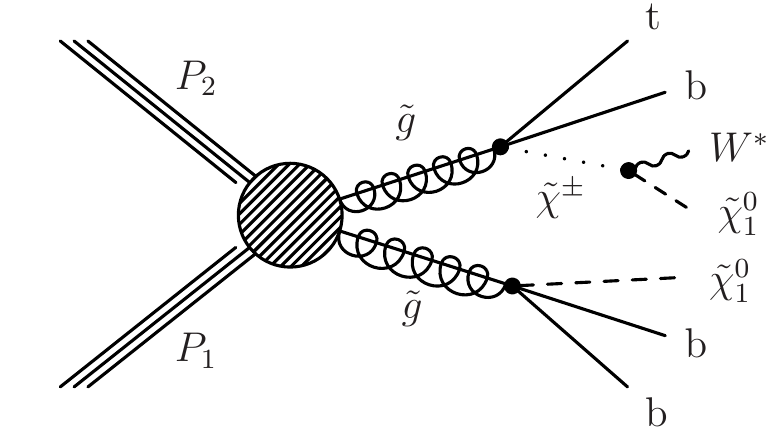}
\includegraphics[width=0.32\textwidth]{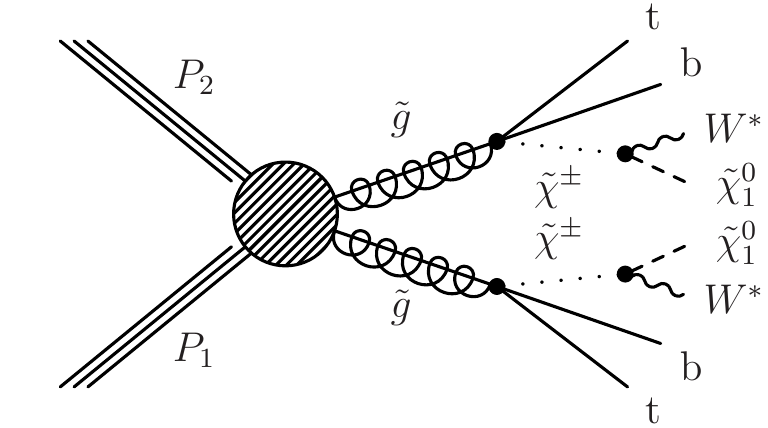} \\
\includegraphics[width=0.32\textwidth]{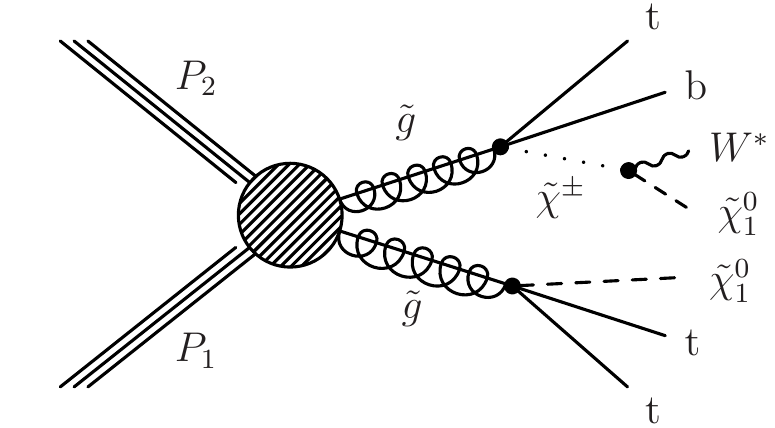}
\includegraphics[width=0.32\textwidth]{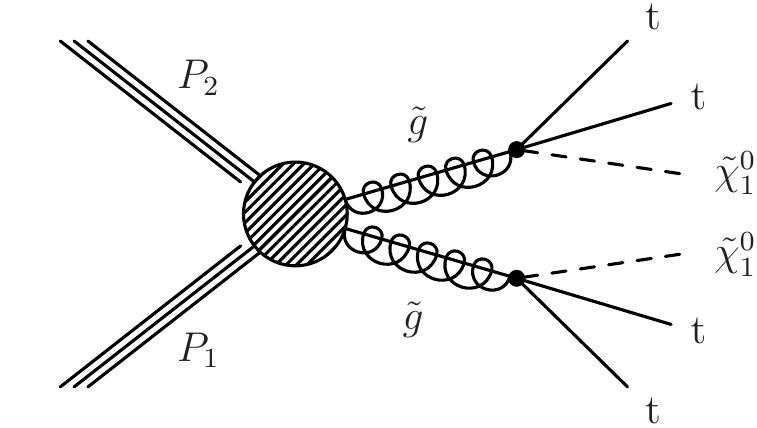} \\
\includegraphics[width=0.32\textwidth]{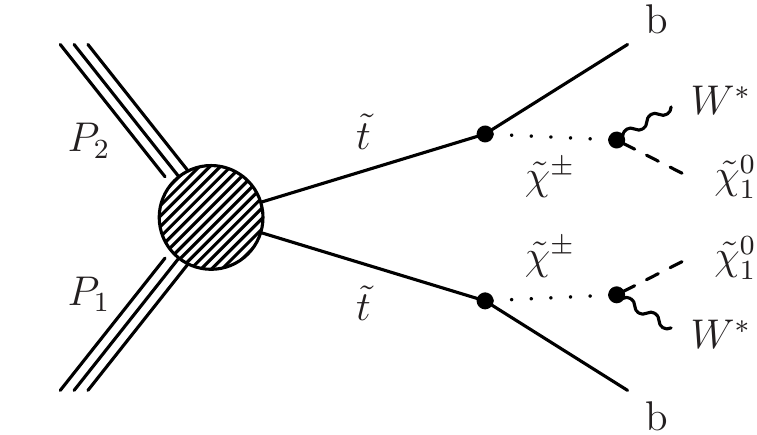}
\includegraphics[width=0.32\textwidth]{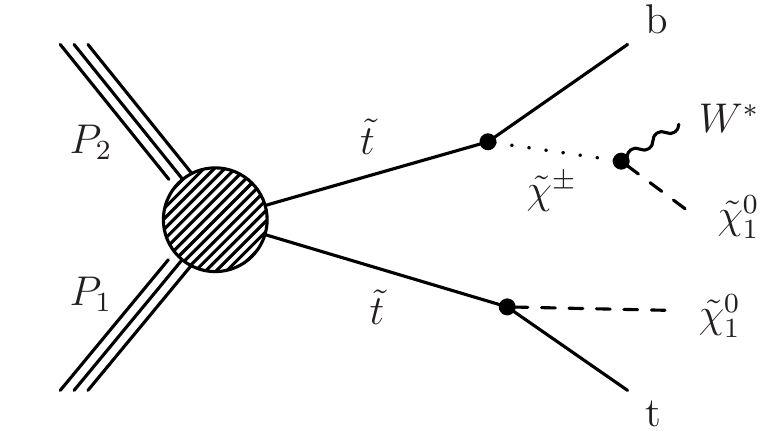}
\includegraphics[width=0.32\textwidth]{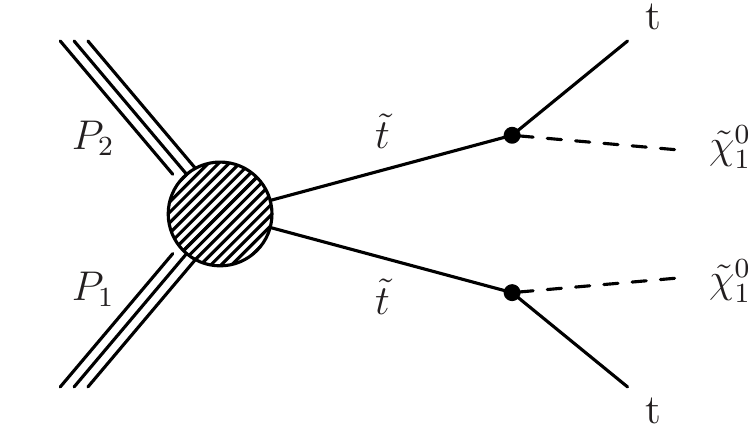}
\caption{Diagrams displaying the event topologies of gluino (upper 5
  diagrams) and top-squark (lower 3 diagrams) pair production
  considered in this paper.\label{fig:SMSGluinoTopology}}
\end{figure*}

In the context of this natural spectrum, five simplified
models~\cite{ArkaniHamed:2007fw,Alwall:2008ag,Alwall:2008va,Alves:2011sq,Alves:2011wf,Graesser:2012qy}
are considered for gluino pair production, based on three-body gluino decays~\cite{SUS-11-016}:
\begin{itemize}
\item \textbf{ T1bbbb}: pair-produced gluinos, each decaying with a 100\%
 branching fraction to a bottom quark-antiquark ($\bbbar$) pair and the LSP;
\item \textbf{ T1tbbb}: pair-produced gluinos, each decaying with a
 50\% branching fraction to a $\bbbar$ pair and the LSP or to a
 top quark (antiquark), a bottom antiquark (quark), and the NLSP;
\item \textbf{ T1ttbb}: pair-produced gluinos, decaying with a 100\%
  branching fraction to a top quark (antiquark), a bottom antiquark (quark), and the NLSP;
\item \textbf{ T1tttb}: pair-produced gluinos, each decaying with a
 50\% branching fraction to a top quark-antiquark ($\ttbar$) pair and the LSP or to a top
 quark (antiquark), a bottom antiquark (quark), and the NLSP;
\item \textbf{ T1tttt}: pair-produced gluinos, each decaying with a 100\%
 branching fraction to a $\ttbar$ pair and the LSP.
\end{itemize}
The corresponding Feynman diagrams are shown in Fig.~\ref{fig:SMSGluinoTopology}.

In addition, the following three simplified models are considered for
the production of top-squark pairs:
\begin{itemize}
\item \textbf{ T2bW$^{\ast}$}: pair-produced top squarks, each decaying
  with a 100\% branching fraction to a bottom quark and the NLSP;
\item \textbf{ T2tb}: pair-produced top squarks, each decaying with a 50\%
  branching fraction to a top quark and the LSP or to a bottom quark and
  the NLSP;
\item \textbf{ T2tt}: pair-produced top squarks, each decaying with a
  100\% branching fraction to a top quark and the LSP.
\end{itemize}
The corresponding Feynman diagrams are shown in
Fig.~\ref{fig:SMSGluinoTopology}.

Events for the eight simplified models are generated with the
\MADGRAPH V5 generator~\cite{Alwall:2011uj,Alwall:2014hca}, in
association with up to two partons. The SUSY particle decays are
treated with \PYTHIA V6.4.26 assuming a constant matrix element
(phase space decay). The parton showering is described by \PYTHIA and matched
to the matrix element kinematic configuration using the MLM
algorithm~\cite{Hoche:2006ph}, before being processed through a fast
simulation of the CMS detector~\cite{FastSim}. The SUSY
particle production cross sections are calculated to next-to-leading
order (NLO) plus next-to-leading-logarithm (NLL)
accuracy~\cite{NLONLL1,NLONLL2,NLONLL3,NLONLL4,NLONLL5}, assuming all
SUSY particles other than those in the relevant diagram to be too
heavy to participate in the interaction. The NLO+NLL cross section and
its associated uncertainty~\cite{NLONLLerr} are taken as a
reference to derive the exclusion limit on the SUSY particle masses.

\section{The CMS detector}
\label{sec:cms}

The central feature of the CMS detector is a
superconducting solenoid of 6\unit{m} internal diameter, providing a
magnetic field of 3.8\unit{T}. Within the superconducting solenoid
volume are a silicon pixel and a silicon strip tracker, a
lead-tungstate crystal electromagnetic calorimeter, and a
brass/scintillator hadron calorimeter, each composed of a barrel and
two endcap sections. Muons are measured in gas-ionization detectors
embedded in the magnet steel flux-return yoke outside the
solenoid. Extensive forward calorimetry complements the coverage
provided by the barrel and endcap detectors. Jets and leptons are
reconstructed within the pseudorapidity region $\abs{\eta}<3$, covered by the
electromagnetic and hadron calorimeters. Muons are reconstructed with
$\abs{\eta}<2.4$. Events are selected by a
two-level trigger system. The first level (L1) is based on a hardware
filter, followed by a software-based high level trigger (HLT). A more
detailed description of the CMS detector, together with a definition
of the coordinate system used and the relevant kinematic variables,
can be found in Ref.~\cite{Adolphi:2008zzk}.

\section{Event selection}
\label{sec:selection}
Events are selected at the L1 trigger level by requiring at least two
jets with $|\eta|<3$. At the HLT level, events are selected using
dedicated razor algorithms, consisting of a loose selection on \MR and
$\Rtwo$. Razor-specific triggers are used in the HLT in order to avoid
biases on the shapes of distributions from the SM background that are
introduced by requirements on more traditional selection variables
such as $\ETm$.  The razor triggers reject the majority of the SM
background, which mostly appears at low \Rtwo and low $\MR$, while
retaining events in the signal-sensitive regions of the ($\MR$,
$\Rtwo$) plane. Two types of triggers are used: i) a hadronic razor
trigger, which selects events that contain at least two jets with
transverse momentum $\pt>64\GeV$ by applying threshold requirements on
$\Rtwo$, $\MR$, and their product; ii) a muon and electron razor
trigger, which selects events with at least one isolated electron or
muon with $\pt>12\GeV$ in combination with looser requirements on
$\Rtwo$, $\MR$, and their product. The trigger efficiency, evaluated
using a dedicated trigger, is measured to be $(95 \pm 5)\%$ and is
independent of \Rtwo and $\MR$ for the events selected with the
baseline requirements described in Section~\ref{sec:razVar}.

Following the trigger selection, events are required to contain at
least one reconstructed interaction vertex. If more than one vertex is
found, the one with the highest $\pt^2$ sum of associated tracks is
chosen as the interaction point for event reconstruction. Algorithms are
used to remove events with detector- and beam-related noise that can
mimic event topologies with high energy and large $\pt$
imbalance~\cite{Chatrchyan:2011tn,Chatrchyan:2012lia,Khachatryan:2014gga}.

The analysis uses a global event description based on the CMS particle
flow (PF) algorithm~\cite{PF1,PF2}. Individual particles (PF
candidates) are reconstructed by combining the information from the inner
tracker, the calorimeters, and the muon system. Five categories of PF
candidates are defined: muons, electrons, photons (including their
conversions to $\Pep\Pem$ pairs), charged hadrons, and neutral
hadrons. The contamination from other proton-proton collisions in the
same or in neighboring bunch crossings is reduced by discarding the
charged PF candidates not compatible with the interaction point. When
computing lepton isolation and jet energy, the corresponding
contamination from neutral particles is subtracted on average by
applying an event-by-event correction based on the jet-area
method~\cite{jetarea_fastjet,jetarea_fastjet_pu,JME-JINST}.

A ``tight'' lepton identification is used for muons and electrons,
consisting of requirements on isolation and track reconstruction
quality. For electrons, the shape and position of the energy deposit
in the electromagnetic calorimeter is used to further reduce the contamination from
hadrons~\cite{Chatrchyan:2013iaa}. For events with one identified
tight lepton, additional muons or electrons are identified through a
``loose'' lepton selection, characterized by a relaxed isolation
requirement~\cite{Chatrchyan:2013mxa}. Tight leptons are
required to have $\pt>15$\GeV and loose leptons $\pt>10$\GeV.

Jets are reconstructed by clustering the PF candidates with the
\textsc{FastJet}~\cite{fastjet} implementation of the anti-\kt~\cite{antikt} algorithm with the distance parameter $R=0.5$. We
select events containing at least two jets with $\pt>80$\GeV and
$\abs{\eta}<2.4$, representing a tighter version of the L1 jet selection criterion. The $\pt$
imbalance in the event, $\ptvecmiss$, is the
negative of the sum of the $\ptvec$ of the PF candidates in the
event. Its magnitude is referred to as $\ETm$. For each event, the $\ptvecmiss$ and the
four-momenta of all the jets with $\pt>40$\GeV and $\abs{\eta}<2.4$ are
used to compute the razor variables, as described in
Section~\ref{sec:razVar}.

The medium working point of the combined secondary vertex
algorithm~\cite{btag8TeV} is used for b-jet tagging. The \PQb-tagging
efficiency and mistag probability are measured from data control
samples as a function of the jet $\pt$ and $\eta$. Correction factors
are derived for Monte Carlo (MC) simulations through comparison of the
measured and simulated \PQb-tagging efficiencies and mistag rates found
in these control samples~\cite{btag8TeV}.

Events with no \PQb-tagged jet are discarded, a criterion motivated by
the natural SUSY signatures described in Section~\ref{sec:sms}. A tighter
requirement ($\geq$2 \PQb-tagged jets) is imposed on events without an
identified tight lepton and fewer than four jets. This requirement reduces the
expected background from SM production of $\cPZ(\to\nu\bar\nu)$+jets
events to a negligible level.

\section{Box definitions}
\label{sec:razVar}

The selected events are categorized into the different razor boxes according to
their event content as shown in Table~\ref{tab:boxDef}. In the table,
the boxes are listed according to the filling order, from the first
(at the top of the table) to the last (at the bottom). If an event
satisfies the requirements of two or more boxes, the event is assigned
to the first listed box to ensure the boxes correspond to disjoint samples.

The events in the single-lepton and two-lepton boxes are recorded
using the electron and muon razor trigger. The remaining two boxes, generically
referred to as ``hadronic'' boxes, contain events recorded using the
hadronic razor trigger.

In the two-lepton boxes, the ($\MR$, $\Rtwo$)
distribution of events with at least one \PQb-tagged jet is studied. For
the other boxes, the data are binned according to the \PQb-tagged jet
multiplicity: 1 \PQb-tag, 2 \PQb-tags, and $\geq$3 \PQb-tags.

\begin{table*}[ht!]
\centering
 \topcaption{Kinematic and multiplicity requirements defining the nine
 razor boxes. Boxes are listed in order of event filling priority.
 \label{tab:boxDef}}
\resizebox{\textwidth}{!}{
\begin{scotch}{ccccc}
Box & Lepton & \PQb-tag & Kinematic & Jet \\
\hline
 \multicolumn{5}{c}{Two-lepton boxes}\\
\hline
\multirow{2}{*}{MuEle} & $\geq$1 tight electron and & \multirow{6}{*}{$\geq$1 \PQb-tag} & \multirow{2}{*}{} & \multirow{6}{*}{$\geq$2 jets}\\
& $\geq$1 loose muon & & & \\
\cline{1-2}
\multirow{2}{*}{MuMu} & $\geq$1 tight muon and & & ($\MR >300$\GeV and $\Rtwo > 0.15$) and & \\
& $\geq$1 loose muon & & ($\MR > 350$\GeV  or $\Rtwo > 0.2$) & \\
\cline{1-2}
\multirow{2}{*}{EleEle} & $\geq$1 tight electron and & & & \\
& $\geq$1 loose electron& & & \\
\hline
\multicolumn{5}{c}{Single-lepton boxes}\\
\hline
MuMultiJet & 1 tight muon & \multirow{4}{*}{$\geq$1 \PQb-tag} & & \multirow{2}{*}{$\geq$4 jets} \\
EleMultiJet &1 tight electron & & ($\MR > 300$\GeV and $\Rtwo > 0.15$) and & \\
\cline{5-5}
MuJet & 1 tight muon & & ($\MR > 350$\GeV or $\Rtwo > 0.2$) & \multirow{2}{*}{2 or 3 jets}\\
EleJet & 1 tight electron & & &  \\
\hline
\multicolumn{5}{c}{Hadronic boxes}\\
\hline
MultiJet & none & $\geq$1 \PQb-tag & ($\MR > 400$\GeV and $\Rtwo > 0.25$) and &$\geq$4 jets\\
$\geq$2 \PQb-tagged jet & none & $\geq$2 \PQb-tag &  ($\MR > 450$\GeV or $\Rtwo > 0.3$) & 2 or 3 jets\\
\end{scotch}}
\end{table*}

\begin{figure}[ht!]
\centering
\includegraphics[width=0.49\textwidth]{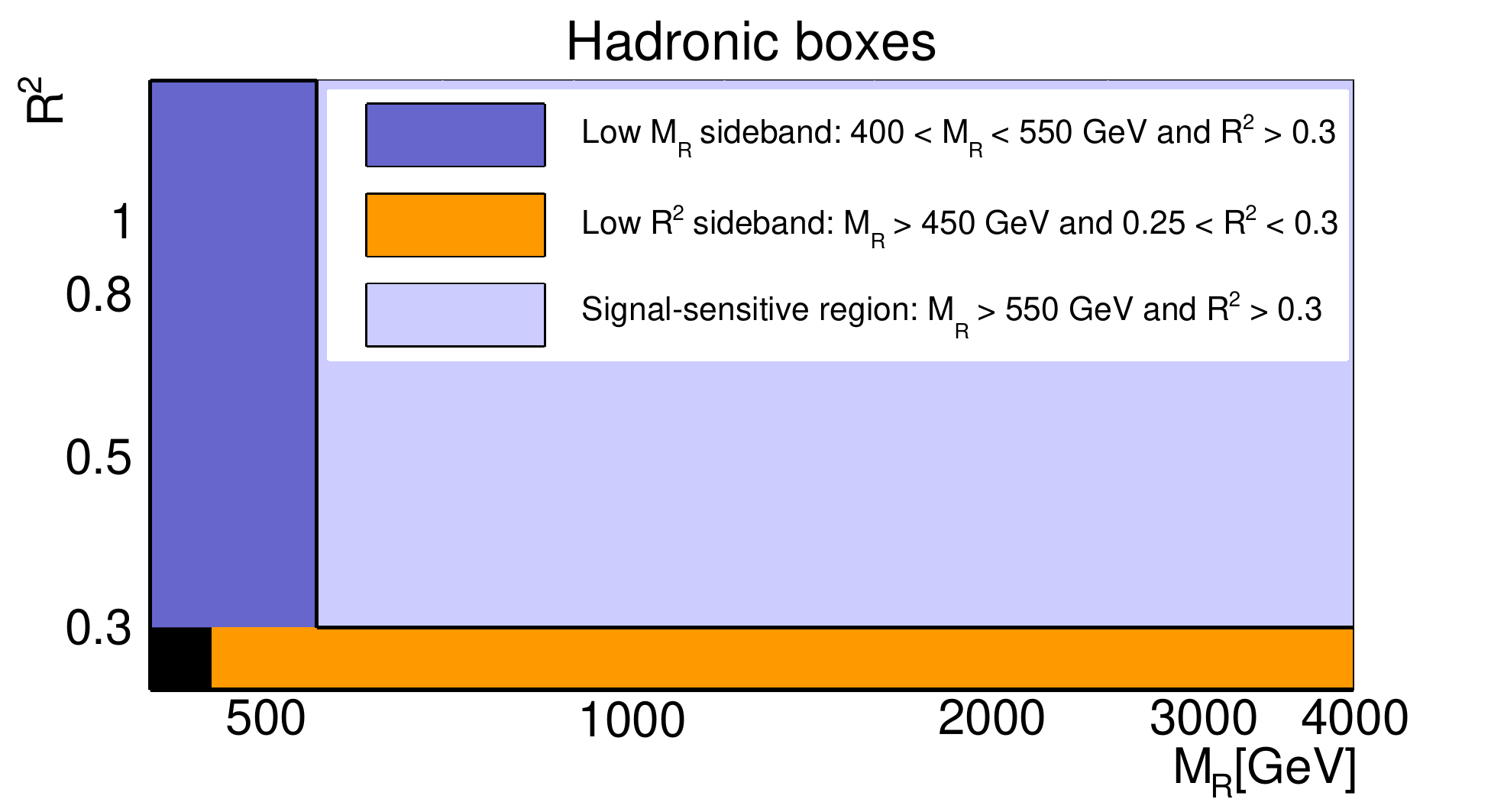}
\includegraphics[width=0.49\textwidth]{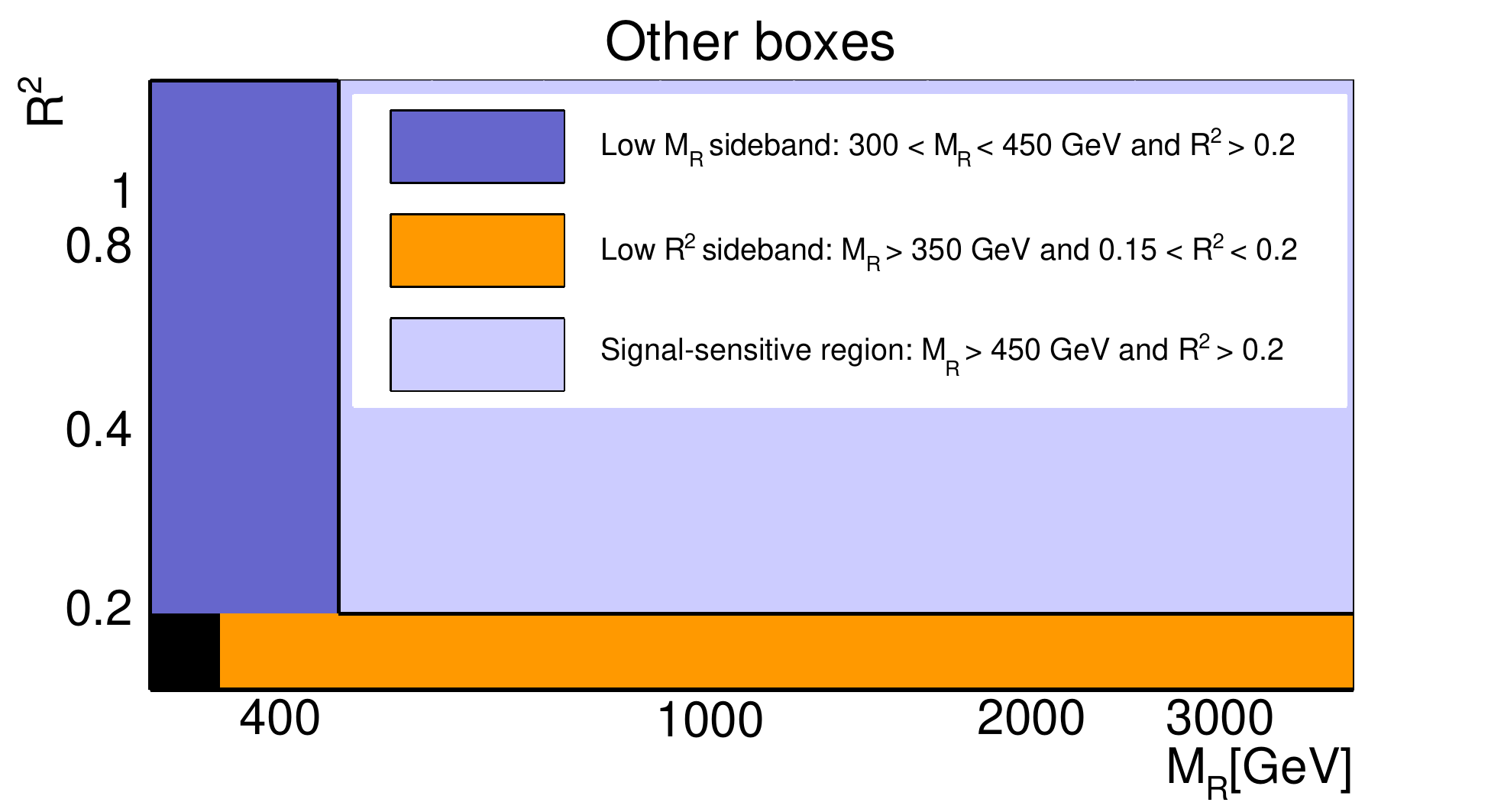}
\caption{\label{fig:regions} Definition of the sideband and the
 signal-sensitive regions used in the analysis, for (\cmsLeft) the hadronic
 boxes and (\cmsRight) the other boxes.}
\end{figure}

A baseline kinematic requirement is applied to define the region in
which we search for a signal:
\begin{itemize}
\item $\MR>400$\GeV and $\Rtwo>0.25$ for the hadronic boxes;
\item $\MR>300$\GeV and $\Rtwo>0.15$ for the other boxes.
\end{itemize}
The tighter baseline selection for the hadronic boxes is a consequence
of the tighter threshold used for the hadronic razor trigger. The
kinematic plane defined by the baseline selection is divided into three
regions (see Fig.~\ref{fig:regions}):
\begin{itemize}
\item Low \MR sideband: $400<\MR<550$\GeV
 and $\Rtwo>0.30$ for the hadronic boxes;
 $300<\MR<450$\GeV and $\Rtwo>0.20$ for the other
 boxes.
\item Low  \Rtwo sideband: $\MR>450$\GeV and
  $0.25<\Rtwo<0.30$ for the hadronic boxes;
  $\MR>350$\GeV and $0.15<\Rtwo<0.20$ for the other
  boxes.
\item Signal-sensitive region: $\MR>550$\GeV and
 $\Rtwo>0.30$ for the hadronic boxes; $\MR>450$\GeV
 and $\Rtwo>0.20$ for the other boxes.
\end{itemize}
The bottom left corner of the razor plane, not included in any of the
three regions, is excluded from the analysis. Given this selection,
the multijet background from quantum chromodynamics processes is
reduced to a negligible level due to the fact that these processes
typically peak at $\Rtwo\approx0$ and fall exponentially for
larger values of $\Rtwo$~\cite{razorPRL,razorPRD}.

\section{Modeling of the standard model backgrounds}
\label{sec:bmodel}
Under the hypothesis of no contribution from new-physics processes,
the event distribution in the considered portion of the
($\MR$, $\Rtwo$) plane can be described by the sum of
the contributions from SM $\cPV$+jets events (where
$\cPV$ indicates a $\PW$ or $\cPZ$ boson) and SM top quark-antiquark and
single-top events, where the events with a top quark are generically
referred to as the $\ttbar$ contribution. Based on MC studies, the
contributions from other processes are determined to be
negligible.

We study each of these processes using MC samples, generated with the
\MADGRAPH v5
simulation~\cite{Alwall:2011uj,Alwall:2014hca}. Parton shower and
hadronization effects are included by matching events to the \PYTHIA v6.4.26 simulation~\cite{Sjostrand:2006za} using the MLM
algorithm~\cite{Hoche:2006ph}. The events are processed by a
\GEANT-based~\cite{G4} description of the CMS apparatus in order to
account for the response of the detector.

Once normalized to the NLO inclusive cross
section and the integrated luminosity, the absolute yield of the
$\cPV$+jets events contribution satisfying the event selection is found
to be negligible in all of the two-lepton boxes. In the remaining boxes,
its contribution to the total SM background is found to be
approximately 25\%. The contribution of $\cPV$+jets events in
the $\geq$2 \PQb-tag and the $\geq$4 jet sample is found to be
negligible. The remainder of the background in each box originates
from $\ttbar$ events.

Based on the study of the data collected at $\sqrt{s}=7\TeV$ and the
corresponding MC samples~\cite{razorPRL,razorPRD}, the two-dimensional
probability density function
$P_\mathrm{SM}(\MR,\Rtwo)$ for each SM process is
found to be well described by the empirical function
\ifthenelse{\boolean{cms@external}}{
\begin{multline}
 f(\MR,\Rtwo) =  \bigl[b(\MR-{\MRz})^{1/n}(\Rtwo-{\Rtwoz})
  ^{1/n}-1\bigr]\times\\
  \re^{-bn(\MR-{\MRz})^{1/n}(\Rtwo-{\Rtwoz})
    ^{1/n}} ,
\label{eq:razFun}
\end{multline}
}{
\begin{equation}
 f(\MR,\Rtwo) =  \bigl[b(\MR-{\MRz})^{1/n}(\Rtwo-{\Rtwoz})
  ^{1/n}-1\bigr]\re^{-bn(\MR-{\MRz})^{1/n}(\Rtwo-{\Rtwoz})
    ^{1/n}} ,
\label{eq:razFun}
\end{equation}
}
where $b$, $n$, $\MRz$, and $\Rtwoz$ are free
parameters of the background model. For $n=1$, this function recovers
the two-dimensional exponential function used for previous
studies~\cite{razorPRL,razorPRD}. The shape of the empirical function
is determined through a \textsc{RooFit}-based extended and unbinned
maximum likelihood fit to the data~\cite{Verkerke:2003ir}. Two kinds
of fit are performed: (i)~a sideband-only fit, which is extrapolated
to the signal region in order to test for the presence of a signal
(discussed in the remainder of this section), and (ii)~a simultaneous
fit
to the signal and sideband regions, performed both under the
background-only and background-plus-signal hypotheses, which is used
for the interpretation of the results (Section~\ref{sec:limit}). In both cases, the empirical function is
found to adequately describe the SM background in each of the boxes,
for each \PQb-tagged jet multiplicity value.

The SM background-only likelihood function for the two-lepton boxes is written as:
\begin{equation}
\mathcal{L}(\text{data}|\Theta) = \frac{\re^{-N_\mathrm{SM}}}{N!} \prod_{i=1}^{N} N_\mathrm{SM}
 P_\mathrm{SM}({\MR}_{(i)},{\Rtwo}_{(i)}),
\label{eq:Lik1btag}
\end{equation}
where $P_\mathrm{SM}(\MR,\Rtwo)$ is the empirical function in
Eq.~(\ref{eq:razFun}) normalized to unity, $N_{SM}$ is the
corresponding normalization factor, $\Theta$ is the set of
background shape and normalization parameters, and the product runs
over the $N$ events in the data set. The same form of the
likelihood is used for the other boxes, for each \PQb-tagged jet
multiplicity. The total likelihood in these boxes is computed as the
product of the likelihood functions for each \PQb-tagged jet
multiplicity.

The fits are performed independently for each box and simultaneously
across the \PQb-tagged jet multiplicity bins. Common background shape
parameters ($b$, ${\MR}^0$, $\Rtwoz$, and $n$) are used
for the 2 \PQb-tag and $\geq$3 \PQb-tag bins, since no substantial
difference between the two distributions is observed on large samples
of $\ttbar$ and $\cPV$+jets MC events. A difference is observed
between 1 \PQb-tag and $\geq$2 \PQb-tag samples, due to the observed
dependence of the \PQb-tagging efficiency on the jet $\pt$. Consequently,
the shape parameters for the 1 \PQb-tag bins are allowed to differ
from the corresponding parameters for the $\geq$2 \PQb-tag bins. The
background normalization parameters for each \PQb-tagged jet multiplicity
bin are also treated as independent parameters.

\begin{figure}[tb!]
\centering
\includegraphics[width=0.49\textwidth]{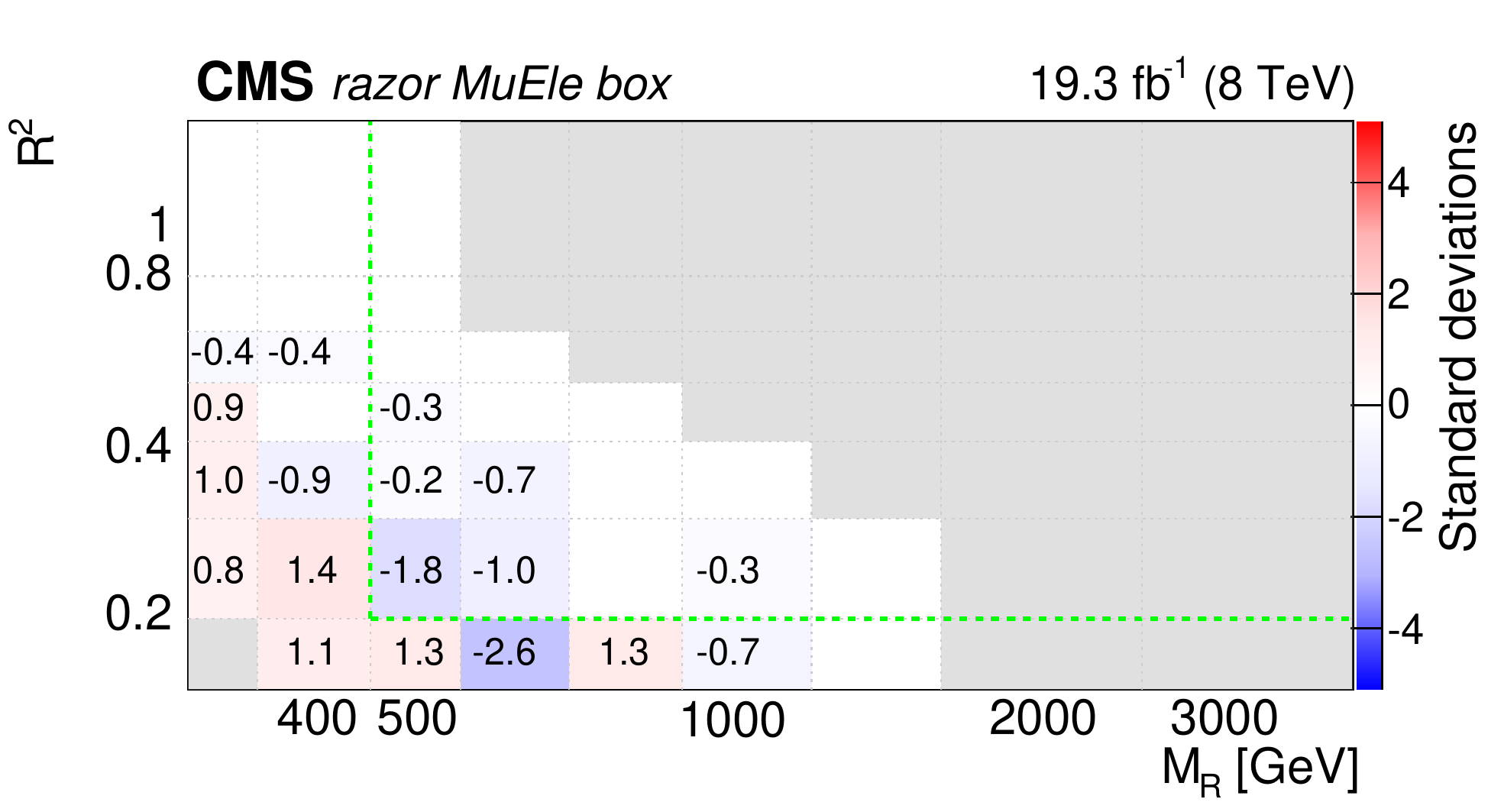}
\includegraphics[width=0.49\textwidth]{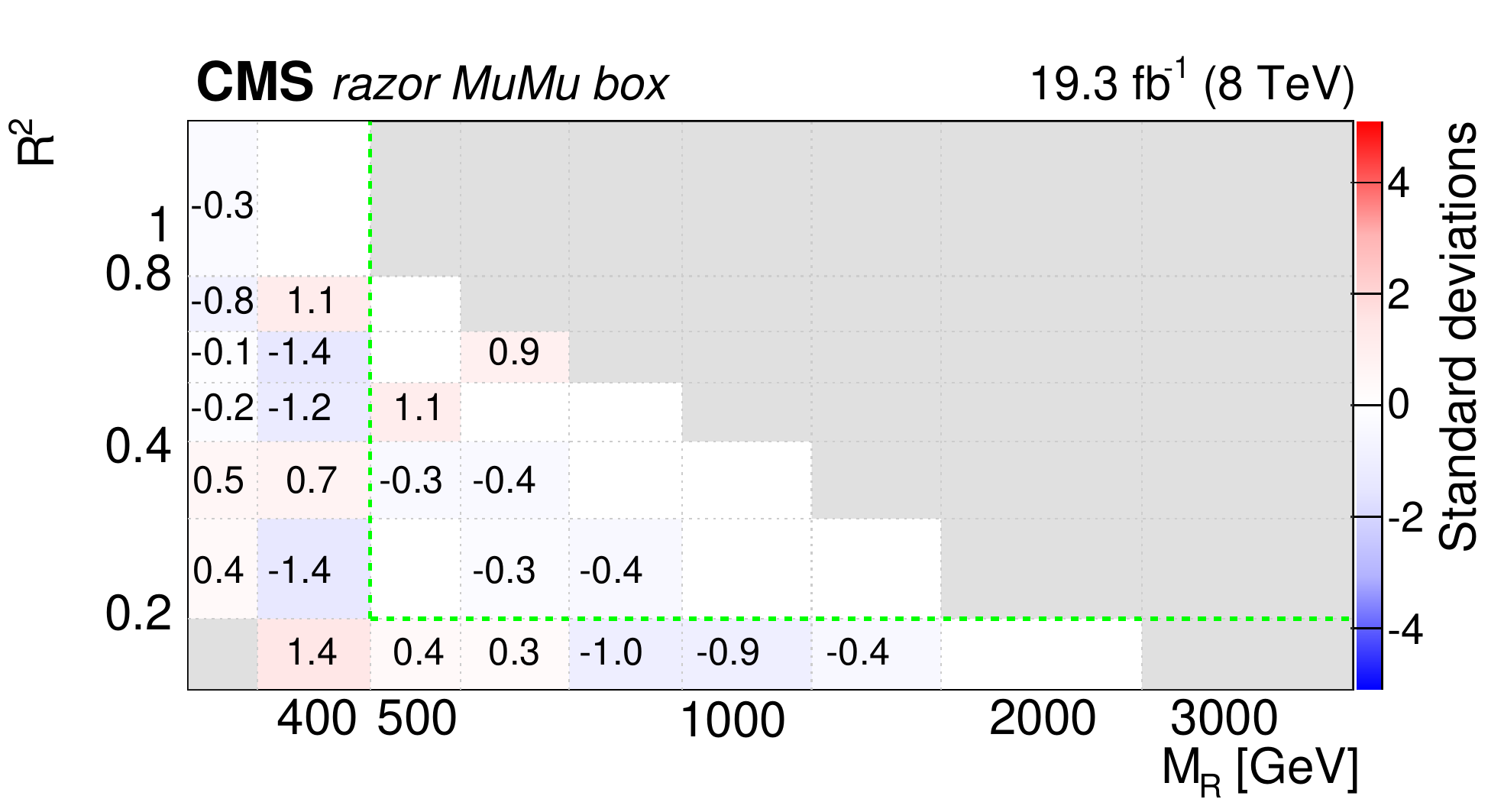}
\includegraphics[width=0.49\textwidth]{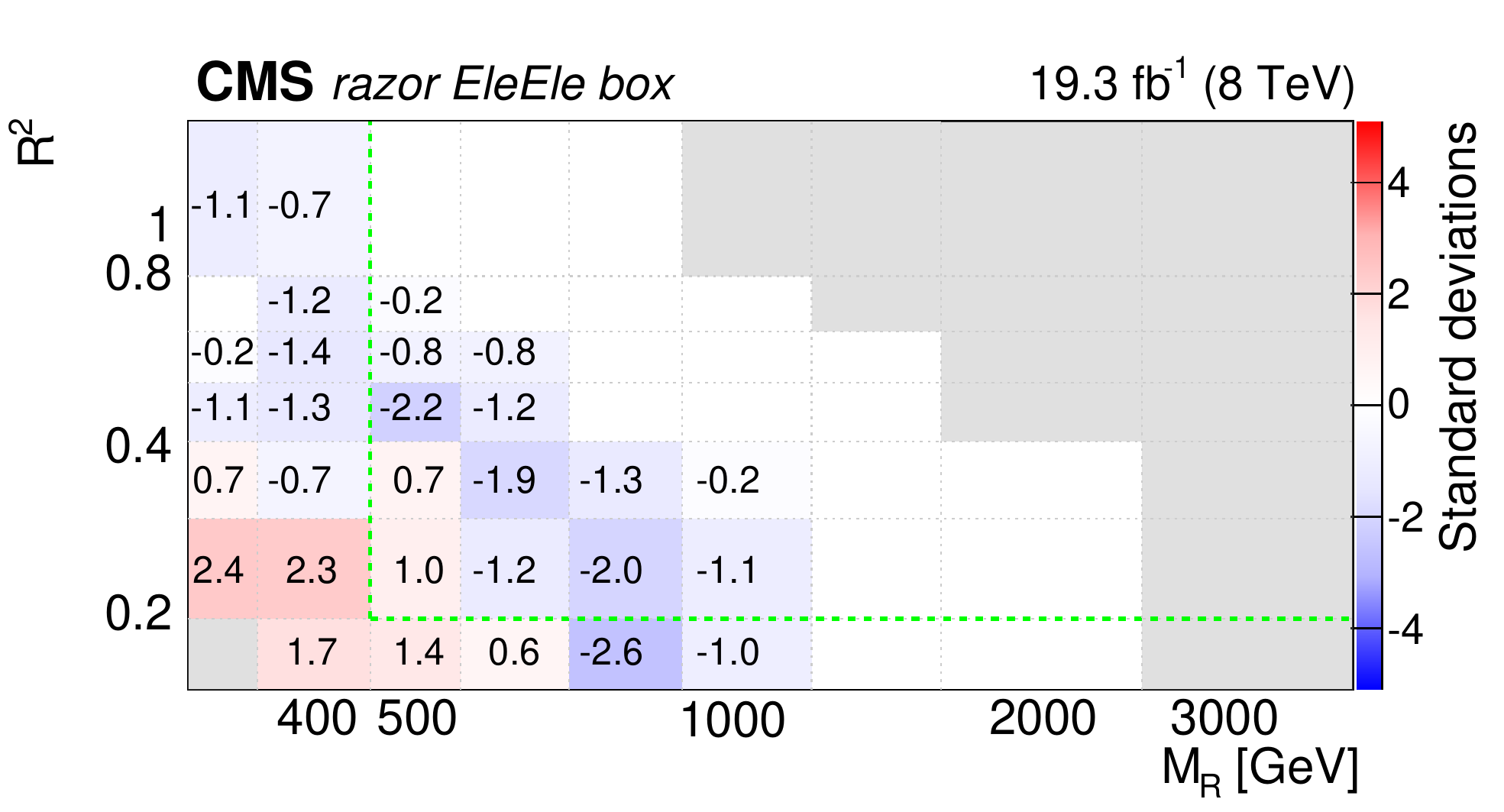}
\caption{Comparison of the expected background and the observed yield
  in the (\cmsUpperLeft) MuEle, (\cmsUpperRight) MuMu, and (bottom)
  EleEle boxes. A probability density function is derived for the
  bin-by-bin yield using pseudo-experiments, sampled from the output
  of the corresponding sideband fit. A two sided p-value is computed
  comparing the observed yield to the distribution of background yield
  from pseudo-experiments. The p-value is translated into the
  corresponding number of standard deviations, quoted in each bin and
  represented by the bin-filling color. Positive and negative
  significance correspond to regions where the observed yield is
  respectively larger and smaller than the predicted one. The white areas
  correspond to bins in which a difference smaller than 0.1 standard
  deviations is observed. The gray areas correspond to empty bins with
  less than one background event expected on average. The dashed lines
  represent the boundaries between the sideband and the signal
  regions.\label{fig:FrenchFlagDilep}}

\end{figure}

\begin{figure*}[tb!]
\centering
\includegraphics[width=0.49\textwidth]{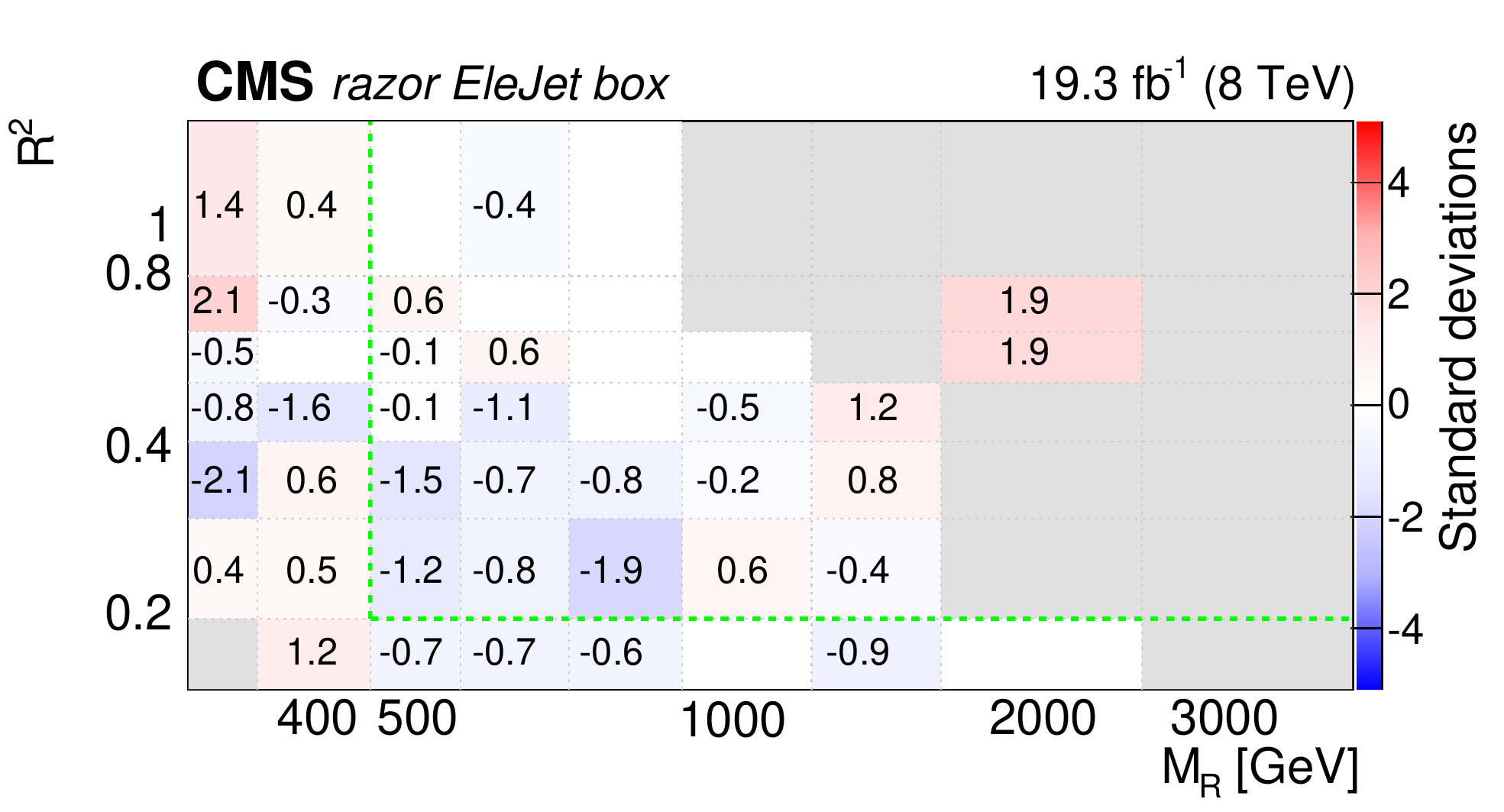}
\includegraphics[width=0.49\textwidth]{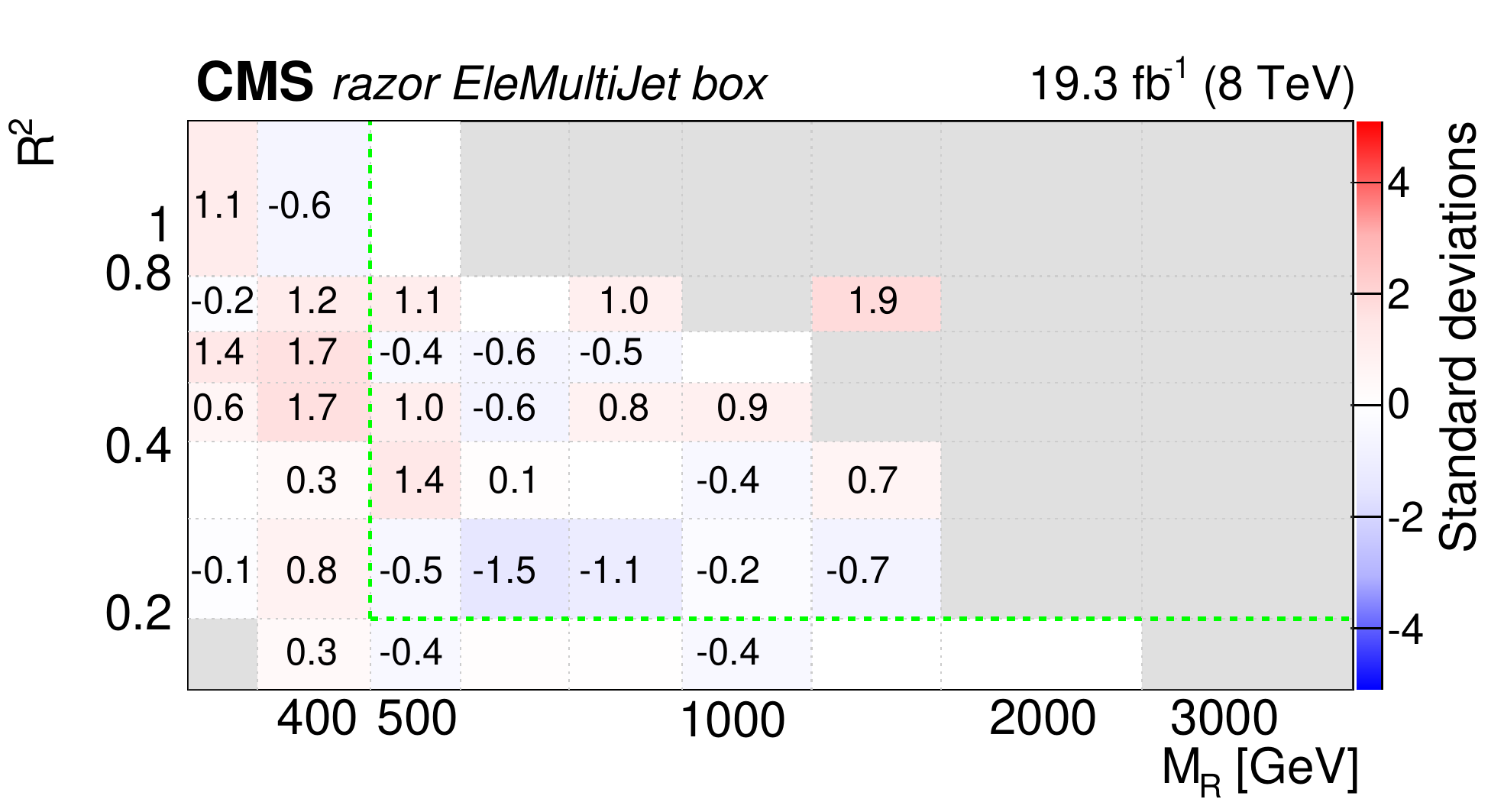}
\includegraphics[width=0.49\textwidth]{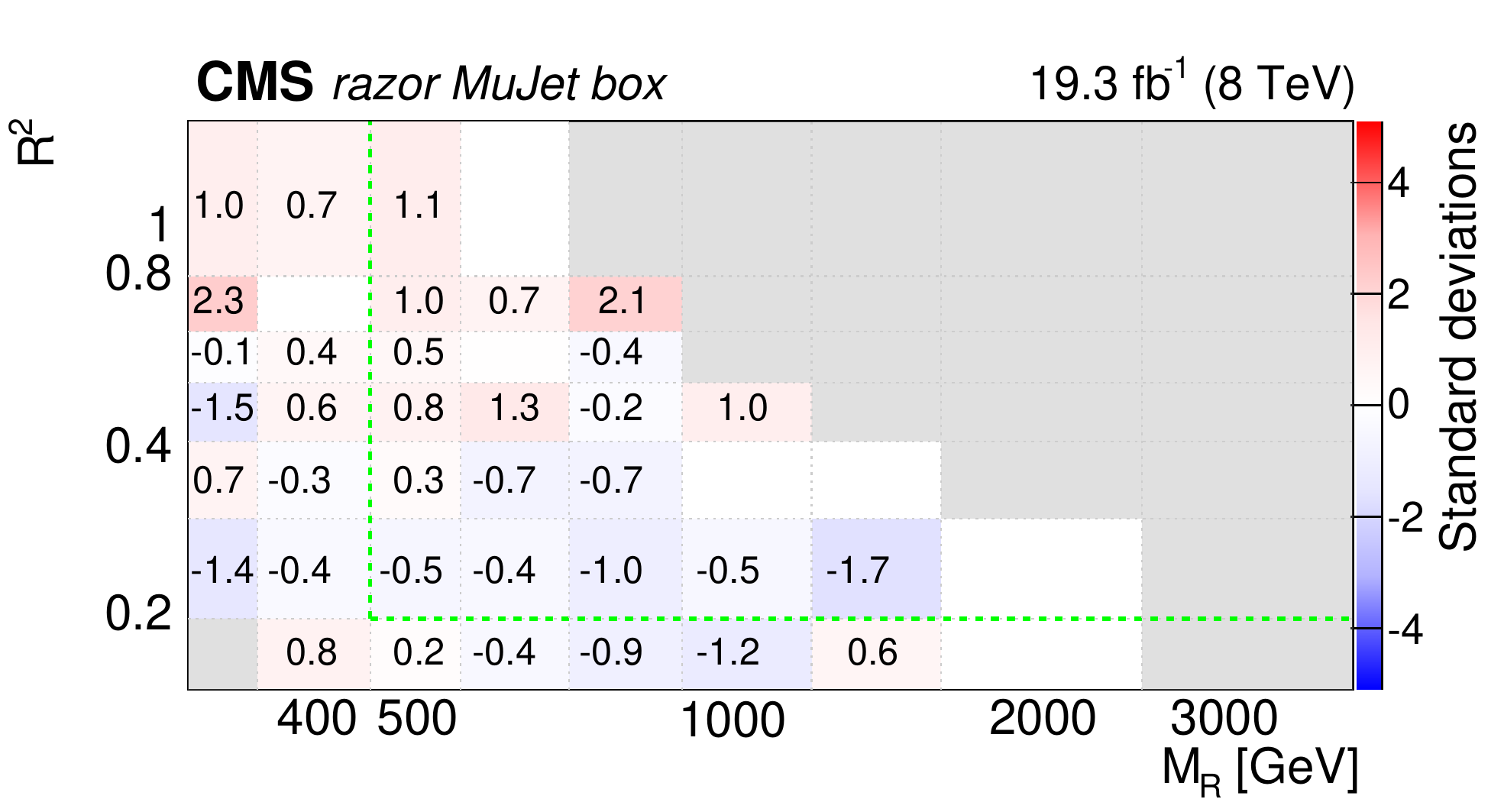}
\includegraphics[width=0.49\textwidth]{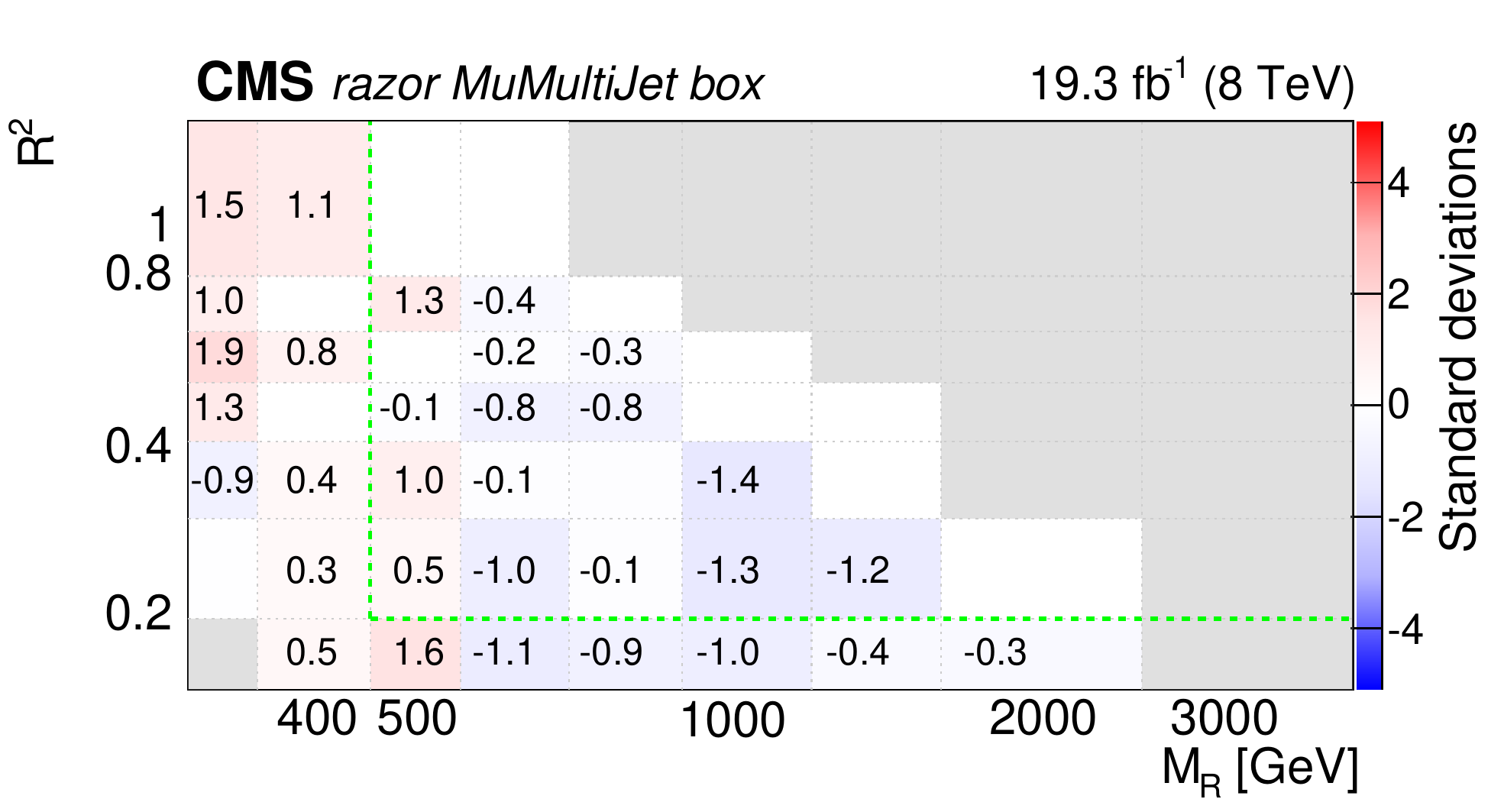}
\caption{Comparison of the expected background and the observed yield
  in (upper left) the EleJet, (upper right) the EleMultiJet, (lower left) the MuJet, and (lower right) the MuMultiJet
  boxes. A detailed explanation is given in the caption of
  Fig.~\ref{fig:FrenchFlagDilep}.\label{fig:FrenchFlagLep}}
\end{figure*}

\begin{figure}[tb!]
\centering
\includegraphics[width=0.49\textwidth]{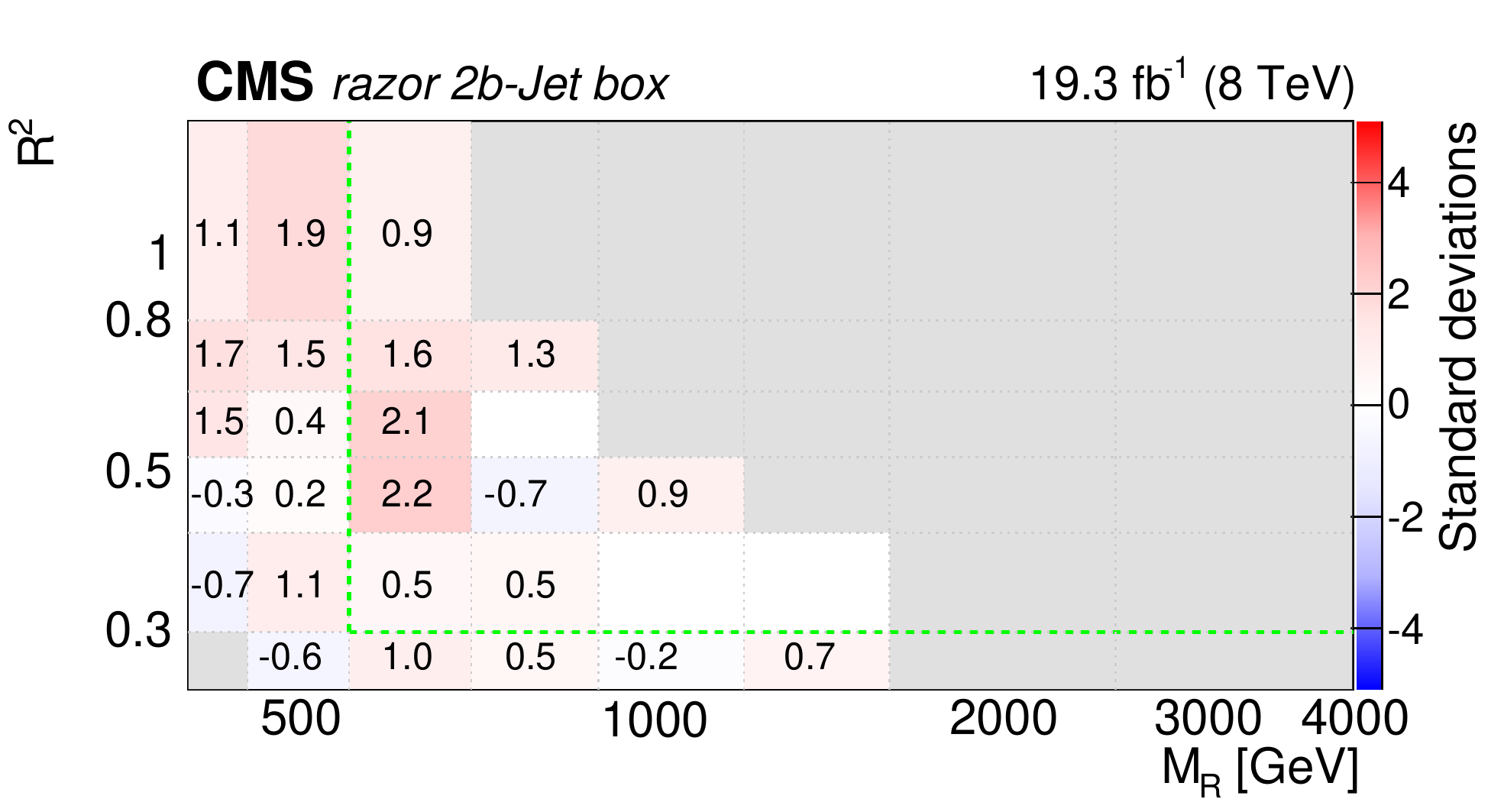}
\includegraphics[width=0.49\textwidth]{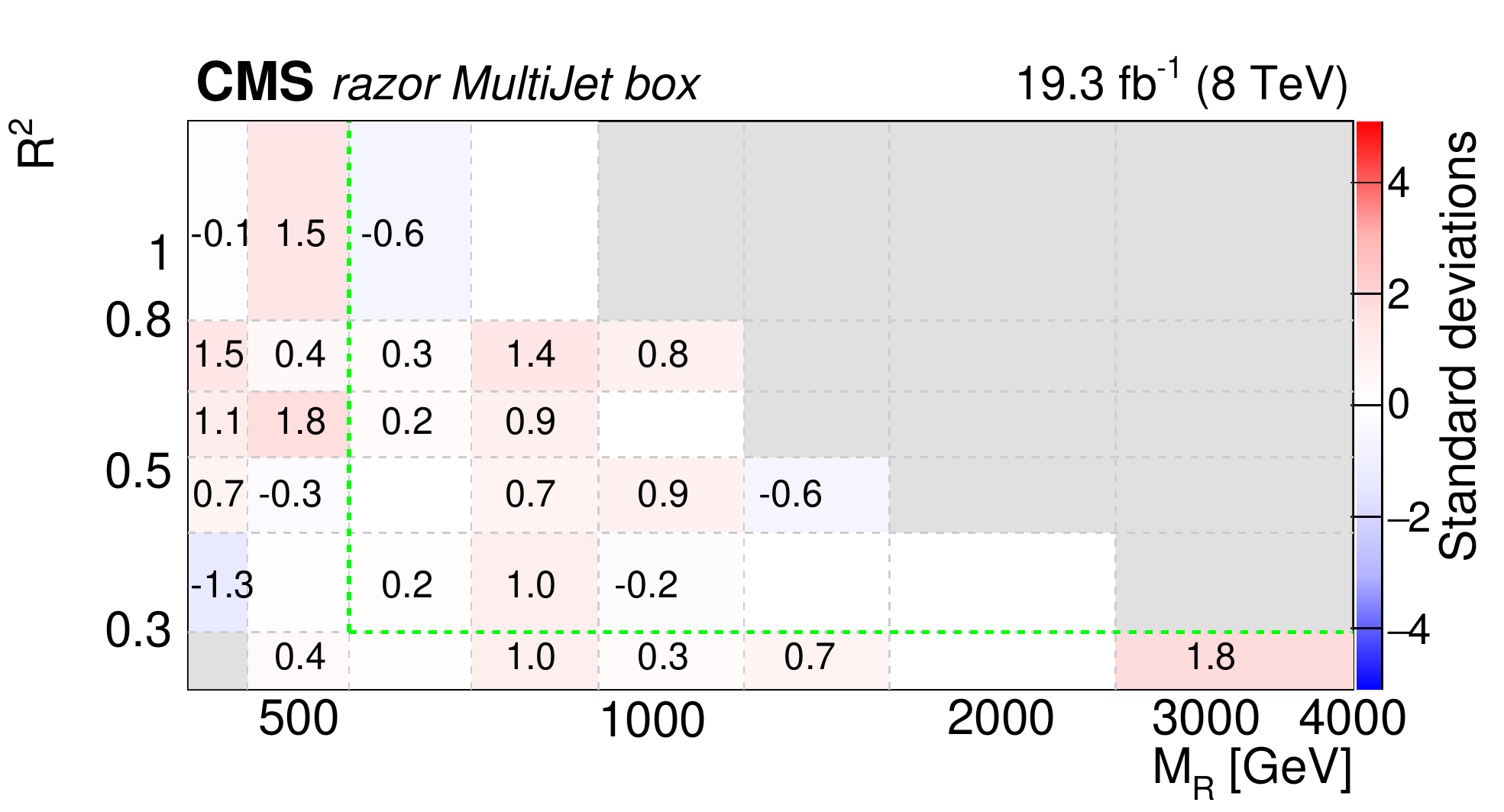}
\caption{Comparison of the expected background and the observed yield
  in the $\geq$2 \PQb-tagged jet box (\cmsLeft) and the MultiJet box
  (\cmsRight). A detailed explanation is given in the caption of
  Fig.~\ref{fig:FrenchFlagDilep}.\label{fig:FrenchFlagHad}}

\end{figure}

\begin{figure*}[tb!]
\centering
\includegraphics[width=0.49\textwidth]{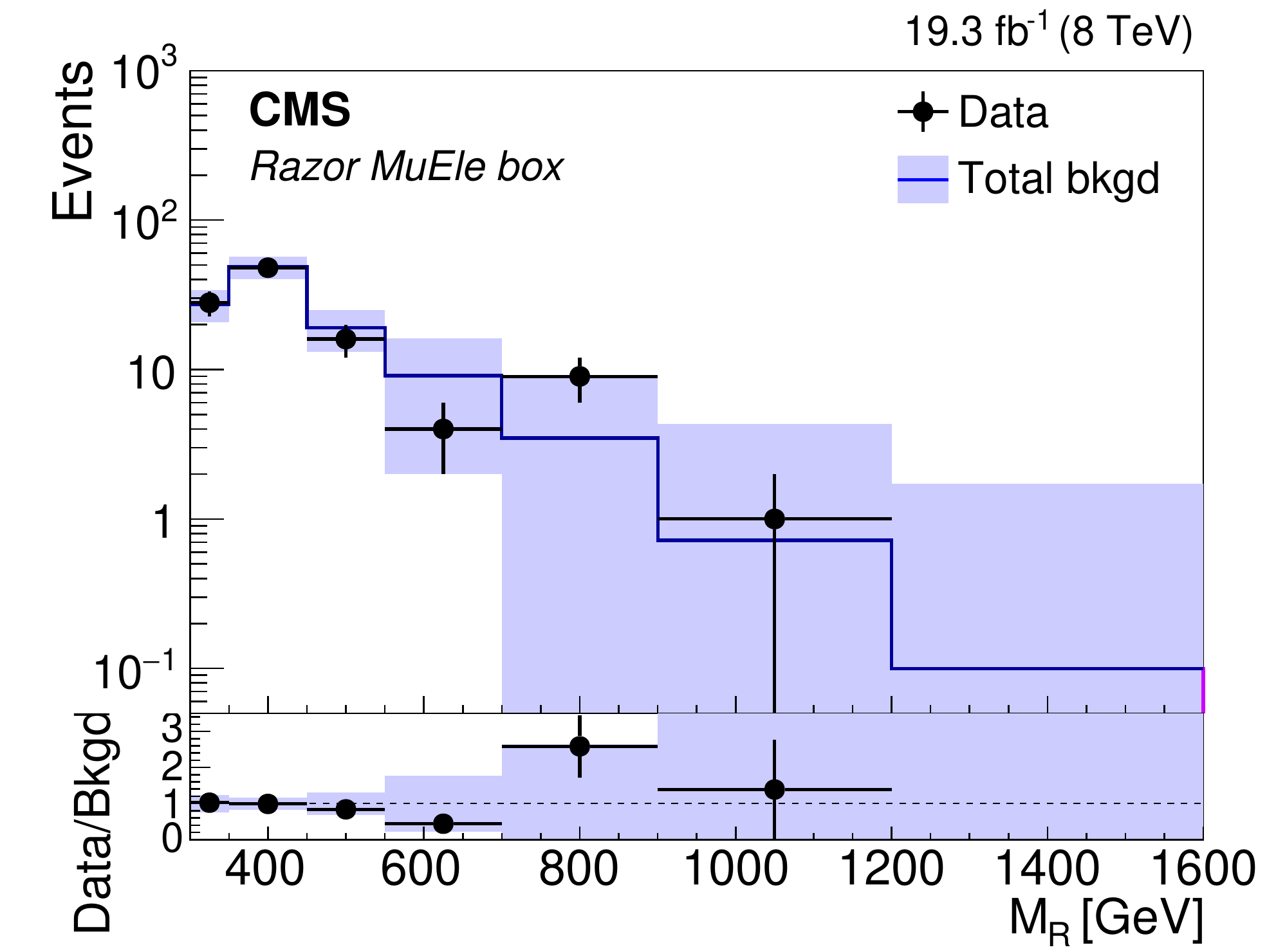}
\includegraphics[width=0.49\textwidth]{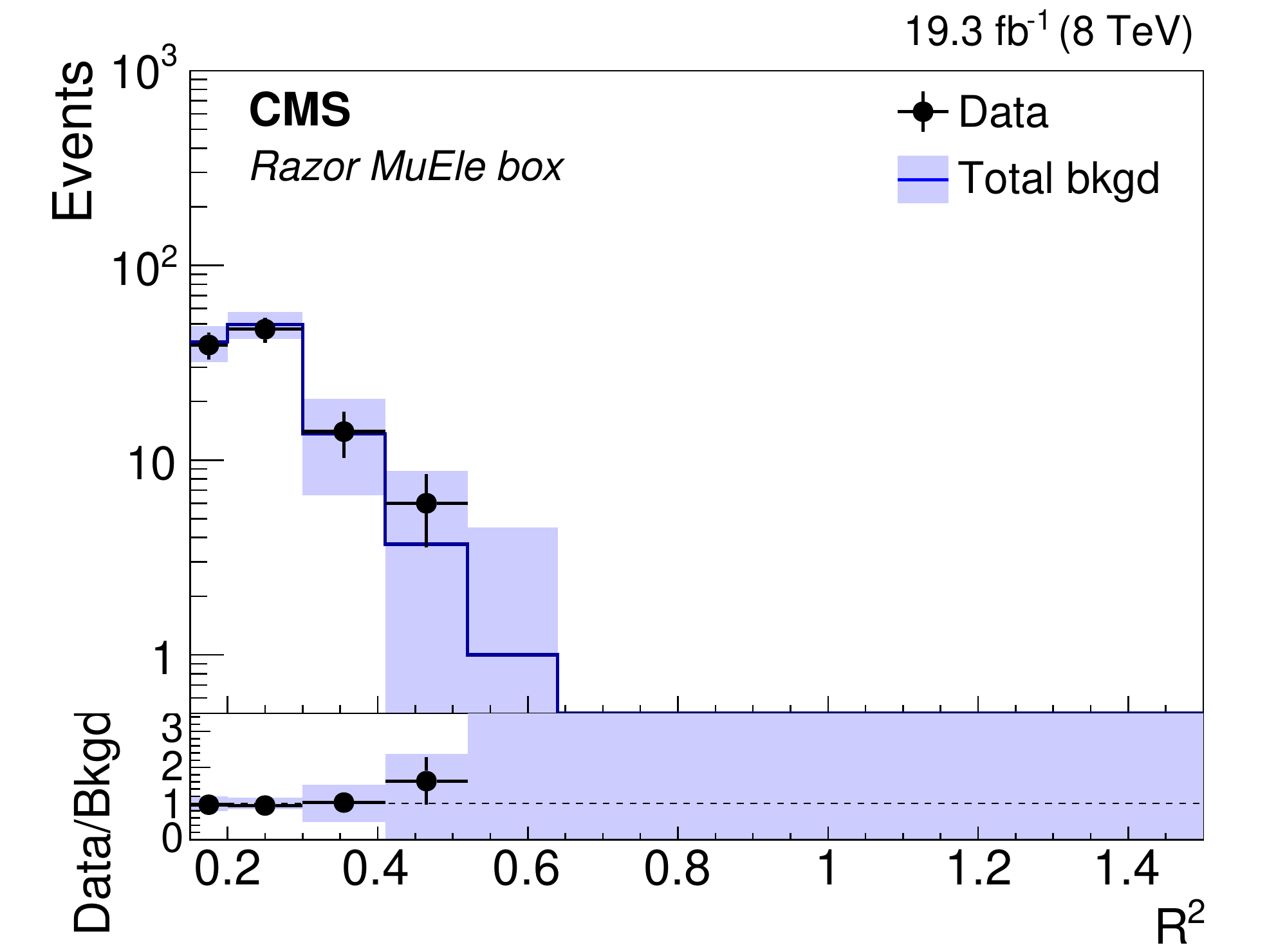}
\includegraphics[width=0.49\textwidth]{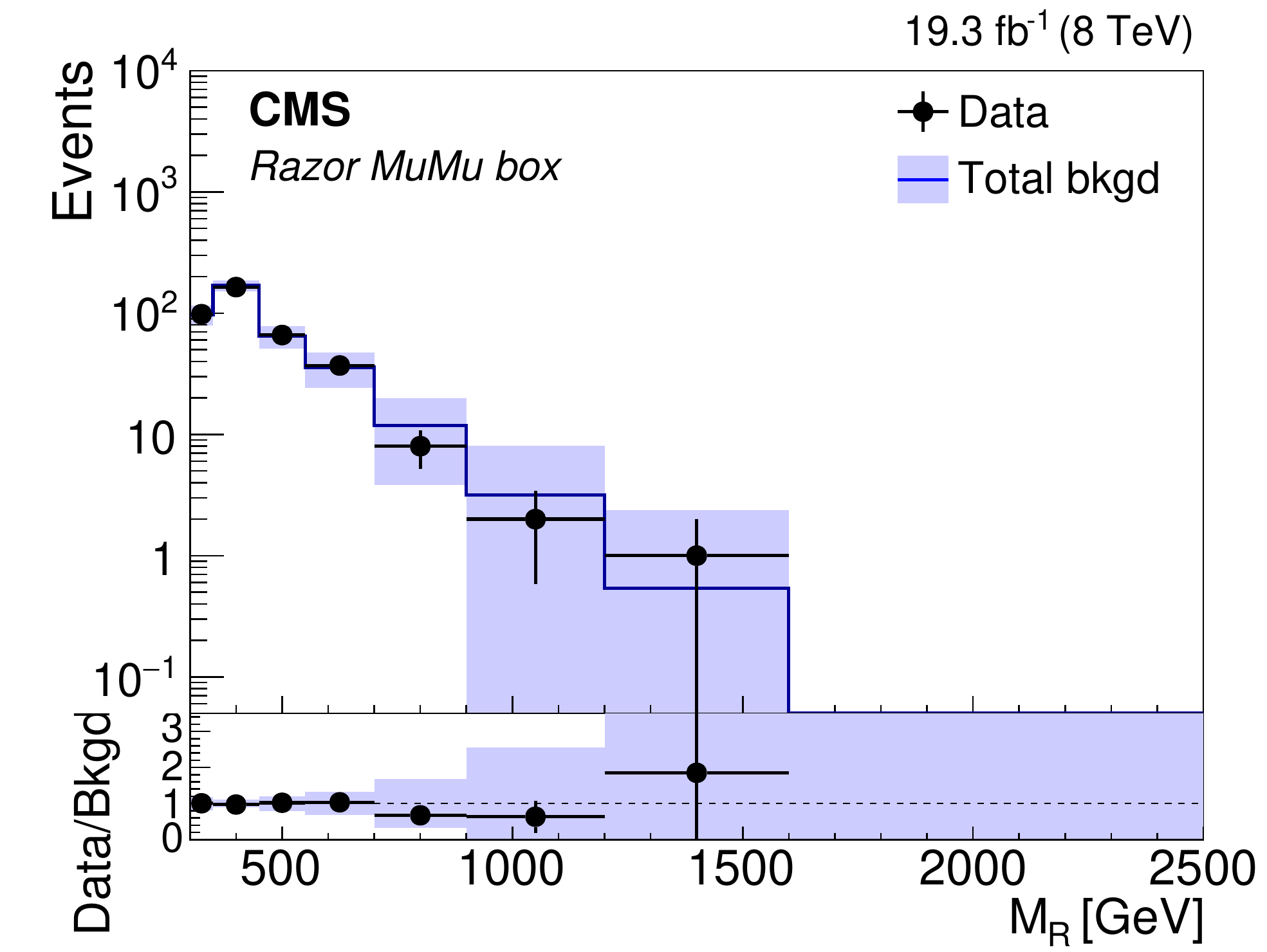}
\includegraphics[width=0.49\textwidth]{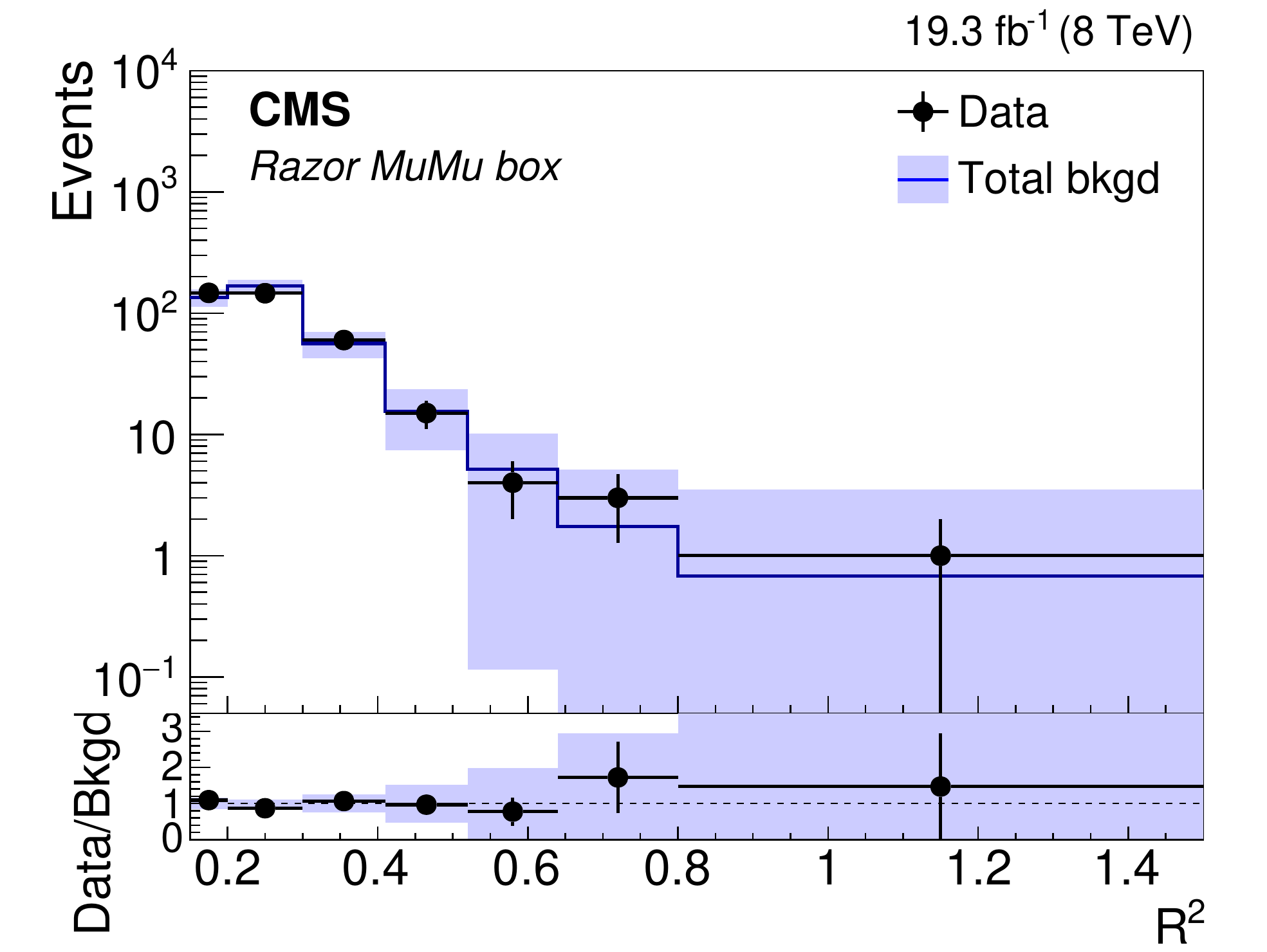}
\includegraphics[width=0.49\textwidth]{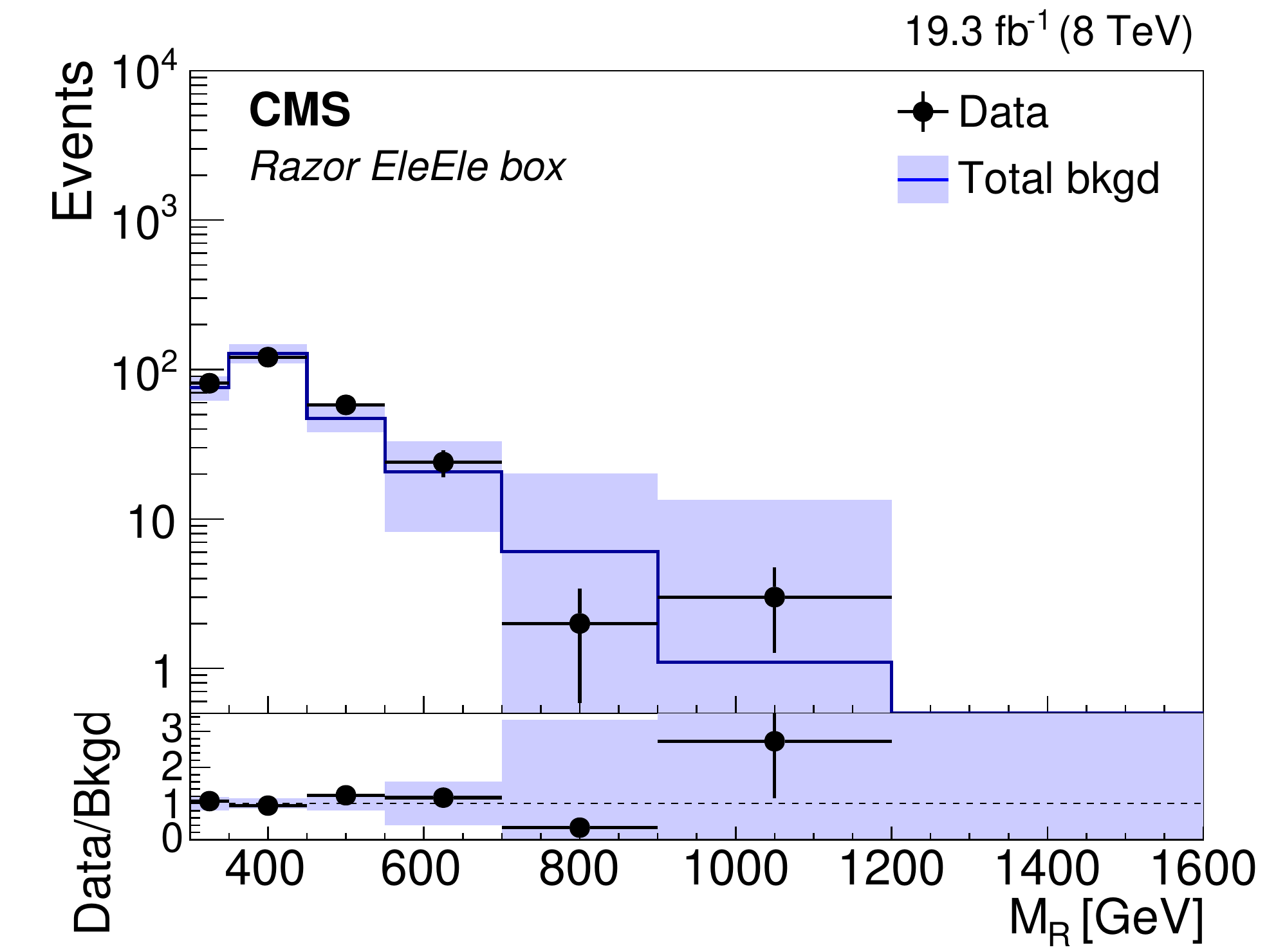}
\includegraphics[width=0.49\textwidth]{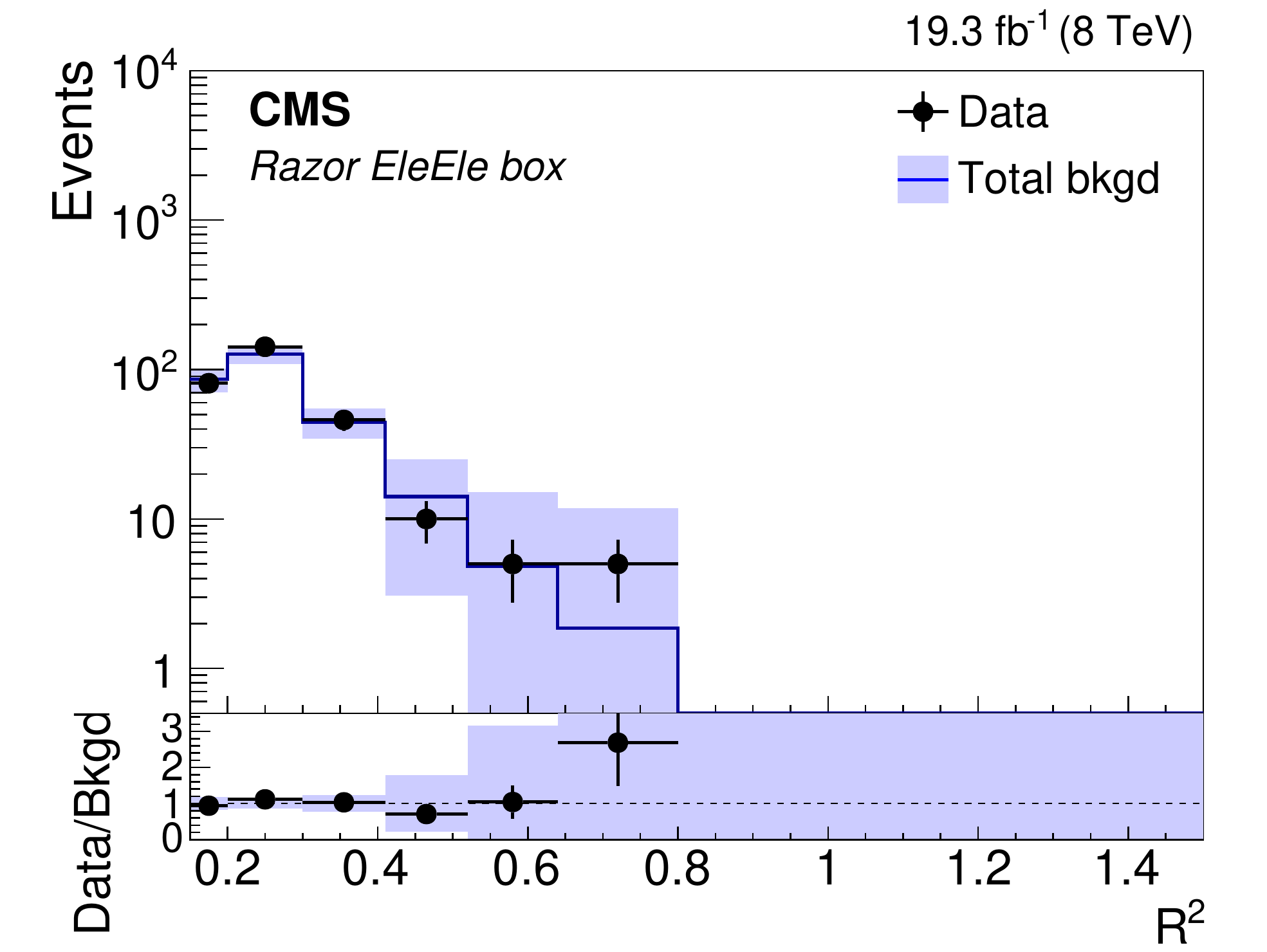}
\caption{Projection of the sideband fit result in the (upper row) MuEle, (middle row)
  MuMu, and (lower row) EleEle boxes on \MR (left) and
   \Rtwo (right), respectively. The fit is performed
  in the sideband regions and extrapolated to the signal-sensitive
  region. The solid line and the filled band represent the total
  background prediction and its uncertainty. The points and the band
  in the bottom panel represent the data-to-prediction ratio and the
  prediction uncertainty, respectively.\label{fig:Proj1DDilep}}
\end{figure*}

\begin{figure*}[tb!]
\centering
\includegraphics[width=0.49\textwidth]{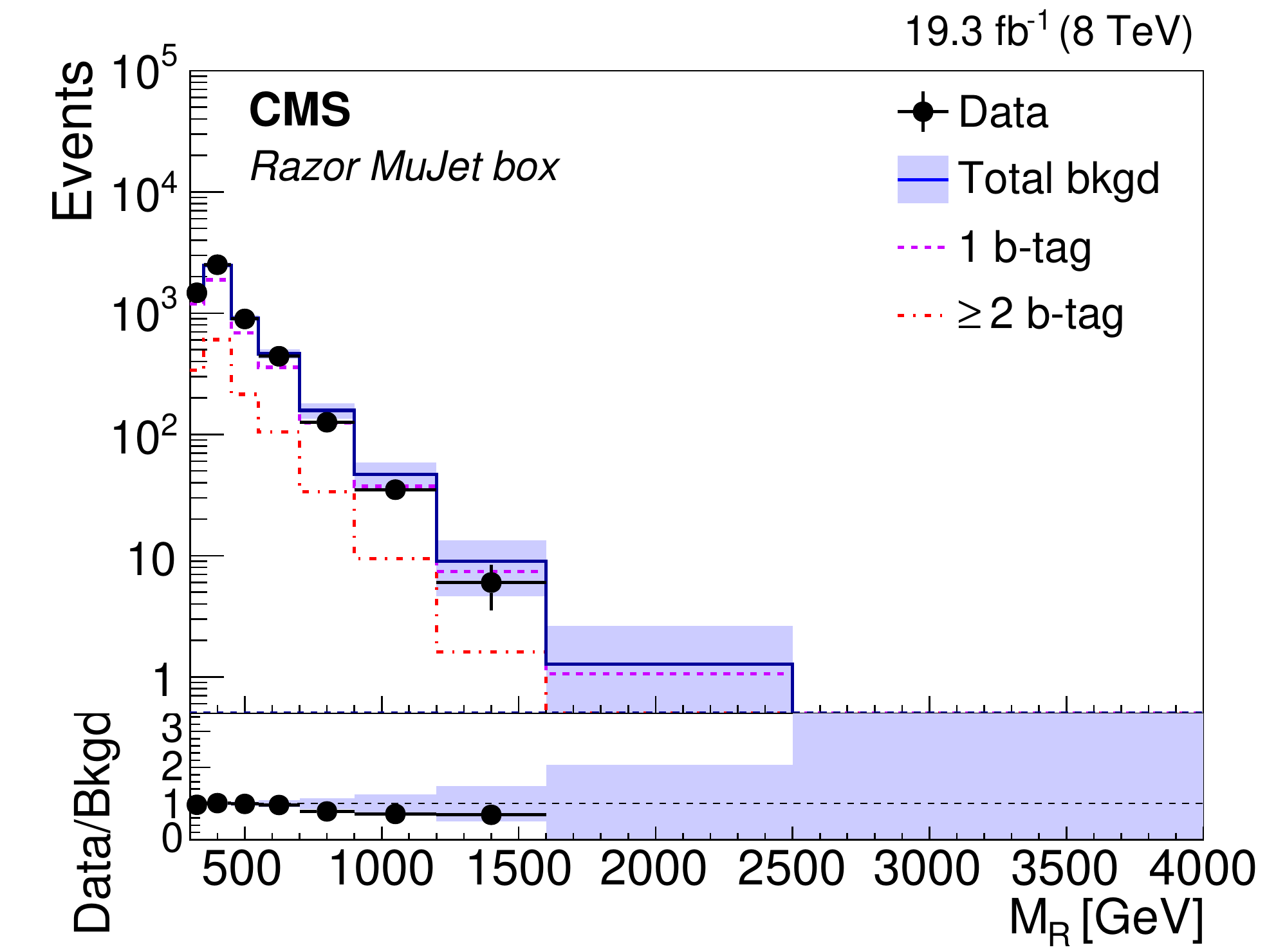}
\includegraphics[width=0.49\textwidth]{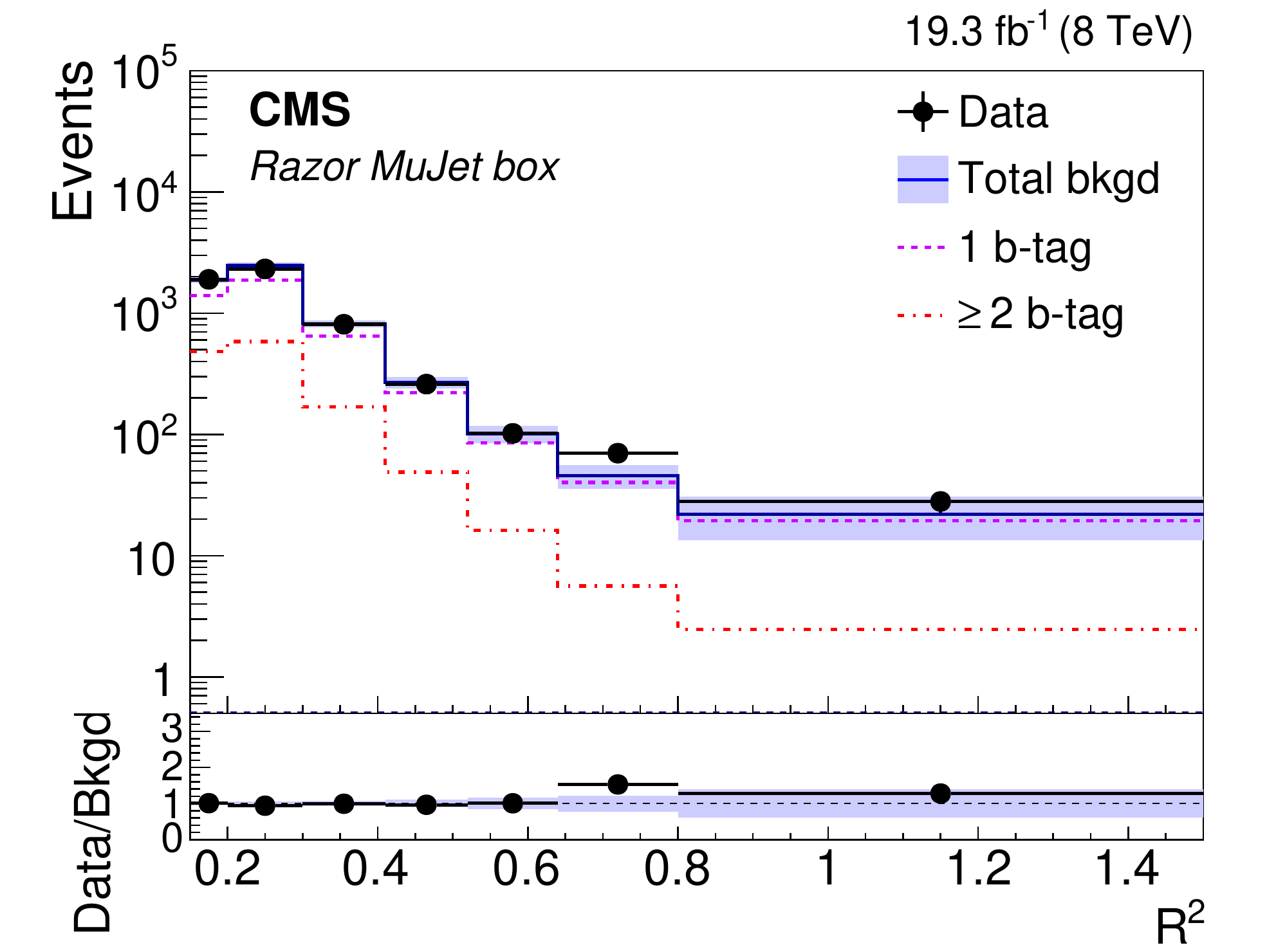}
\includegraphics[width=0.49\textwidth]{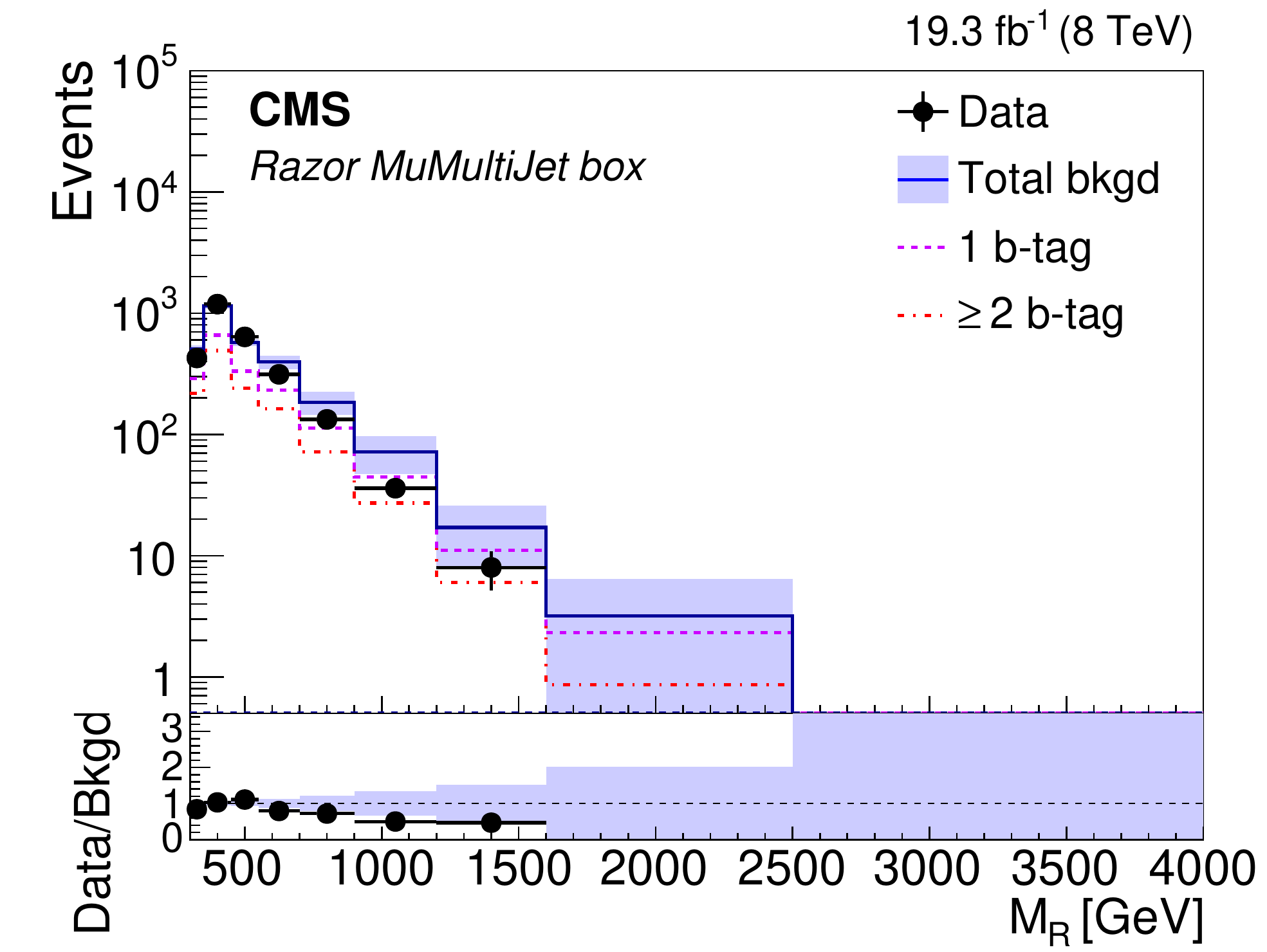}
\includegraphics[width=0.49\textwidth]{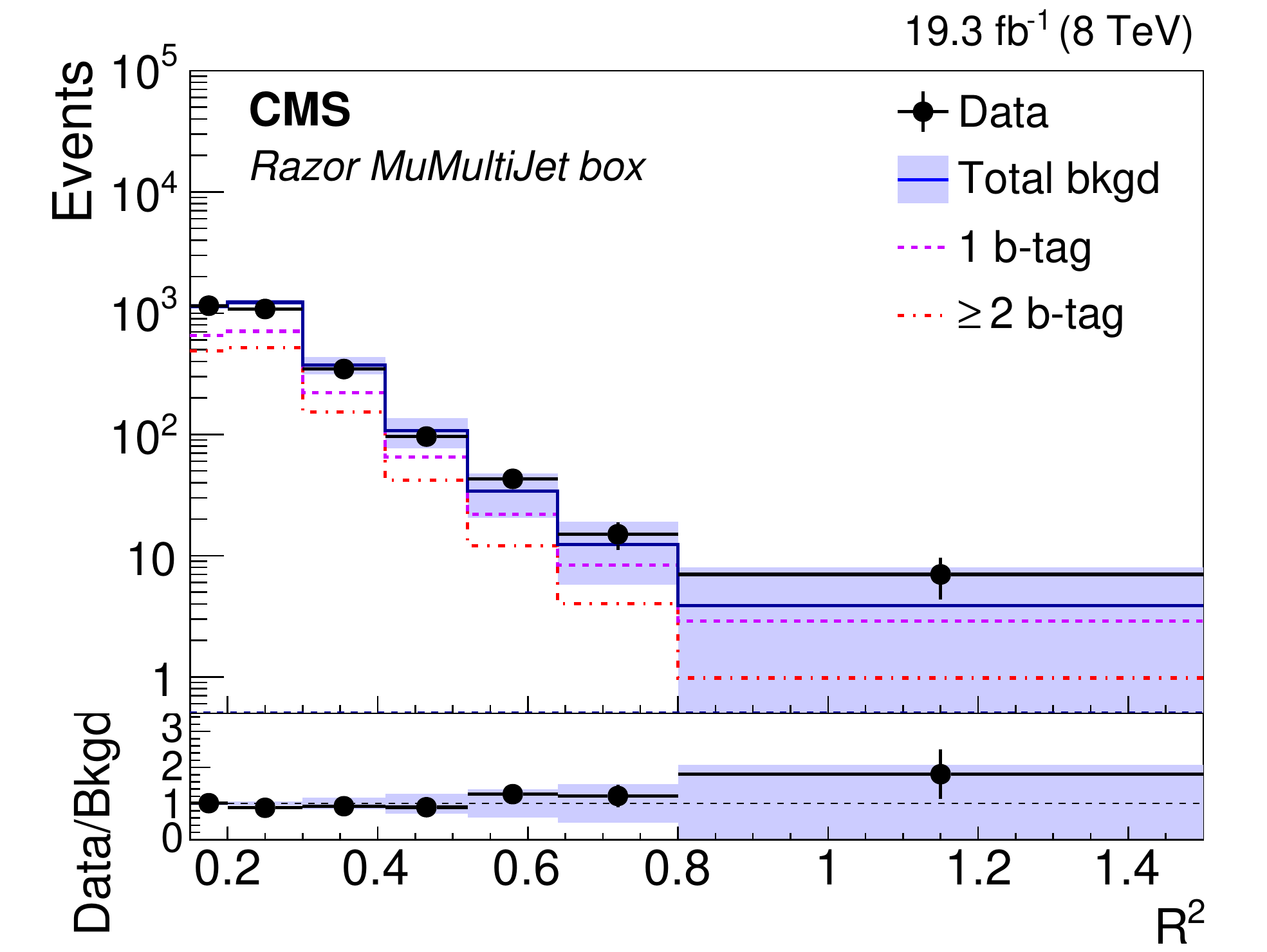}
\caption{Projection of the sideband fit result in the MuJet box on (upper left)
  \MR and (upper right) $\Rtwo$, and of the sideband fit
  result in the MuMultiJet box on (lower left) \MR and (lower right)
  $\Rtwo$. The fit is performed in the sideband regions and
  extrapolated to the signal-sensitive region. The solid line and the
  filled band represent the total background prediction and its
  uncertainty. The dashed and dot-dashed lines represent the
  background shape for 1 \PQb-tag and $\geq$2 \PQb-tag events,
  respectively. The points and the band in the bottom panel represent
  the data-to-prediction ratio and the prediction uncertainty,
  respectively.\label{fig:Proj1DMu}}

\end{figure*}

\begin{figure*}[tb!]
\centering
\includegraphics[width=0.49\textwidth]{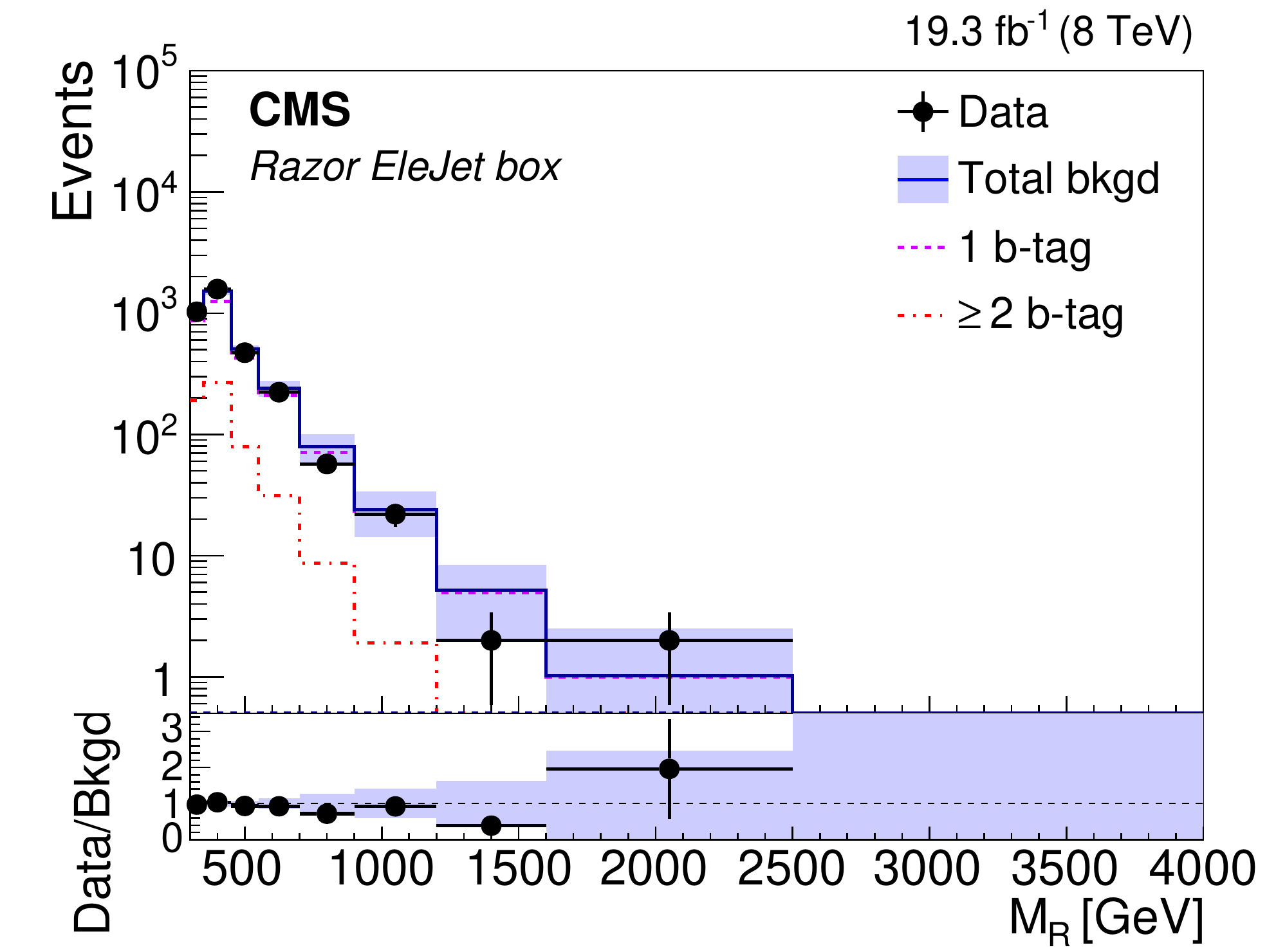}
\includegraphics[width=0.49\textwidth]{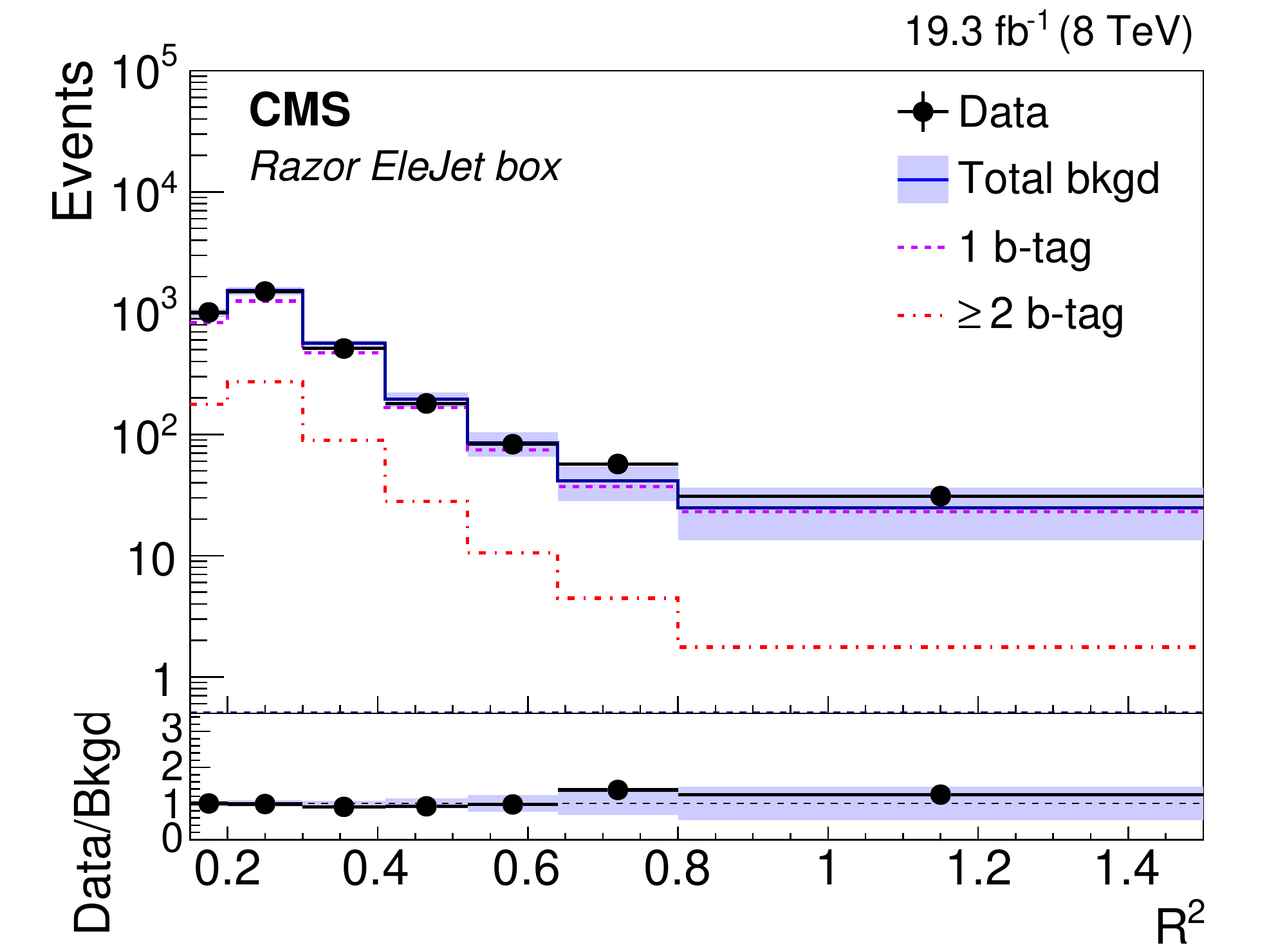}
\includegraphics[width=0.49\textwidth]{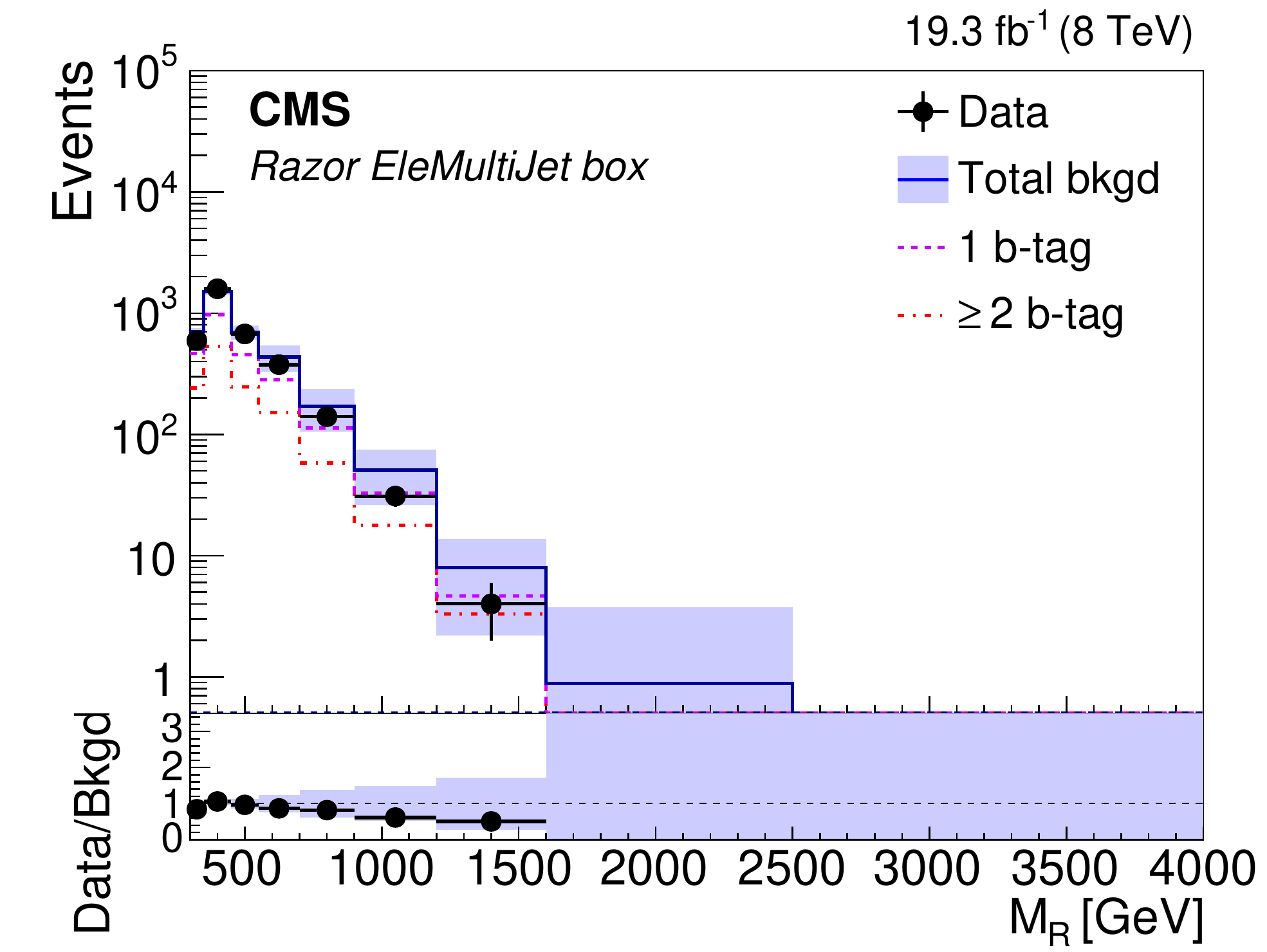}
\includegraphics[width=0.49\textwidth]{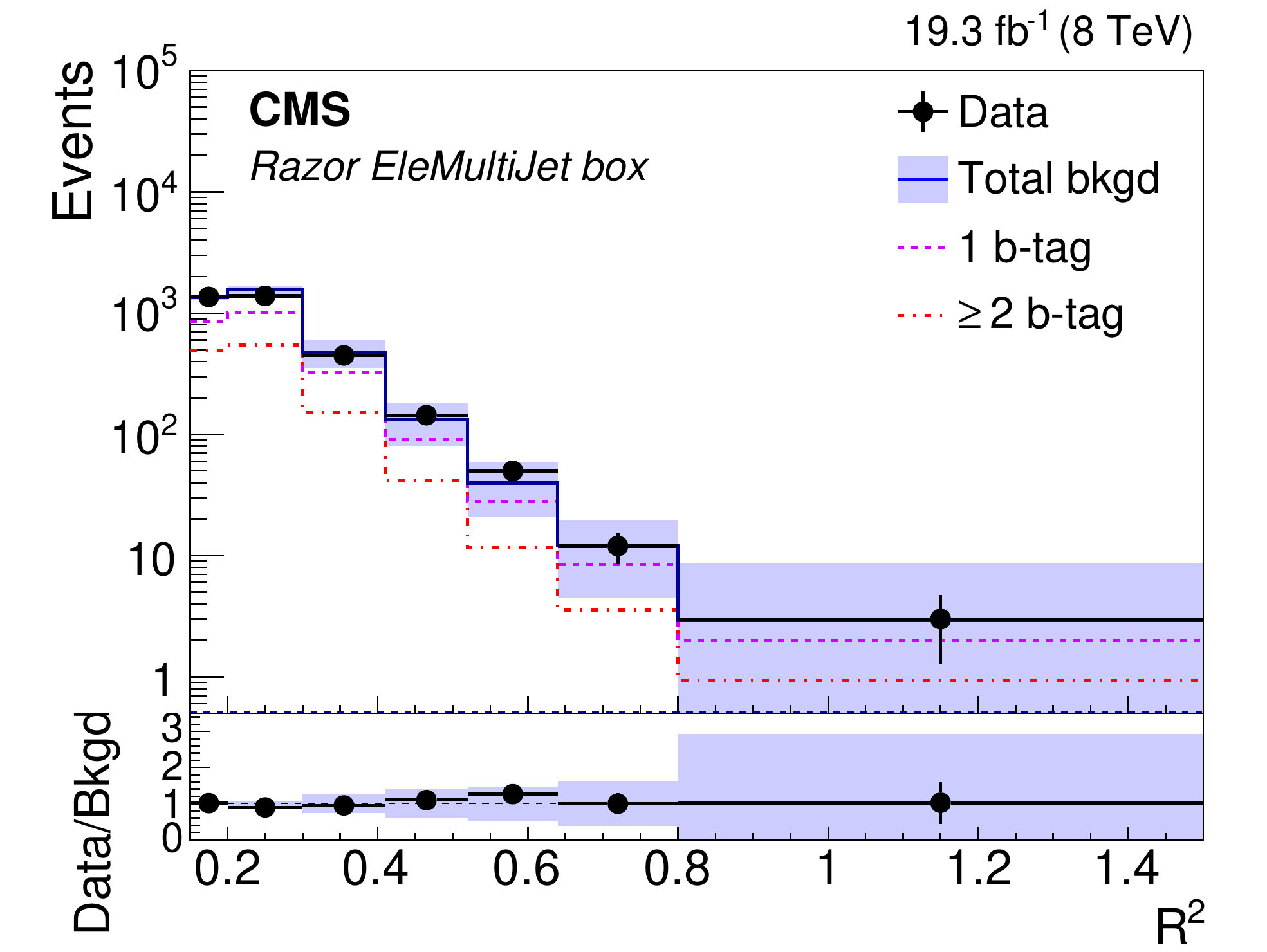}
\caption{Projection of the sideband fit result in the EleJet box on
  (upper left) \MR and (upper right) $\Rtwo$, and projection of the
  sideband fit result in the EleMultiJet box on (lower left) \MR and
  (lower right) $\Rtwo$. A detailed explanation is given in the caption
  of Fig.~\ref{fig:Proj1DMu}.\label{fig:Proj1DEle}}
\end{figure*}

\begin{figure*}[tb!]
\centering
\includegraphics[width=0.49\textwidth]{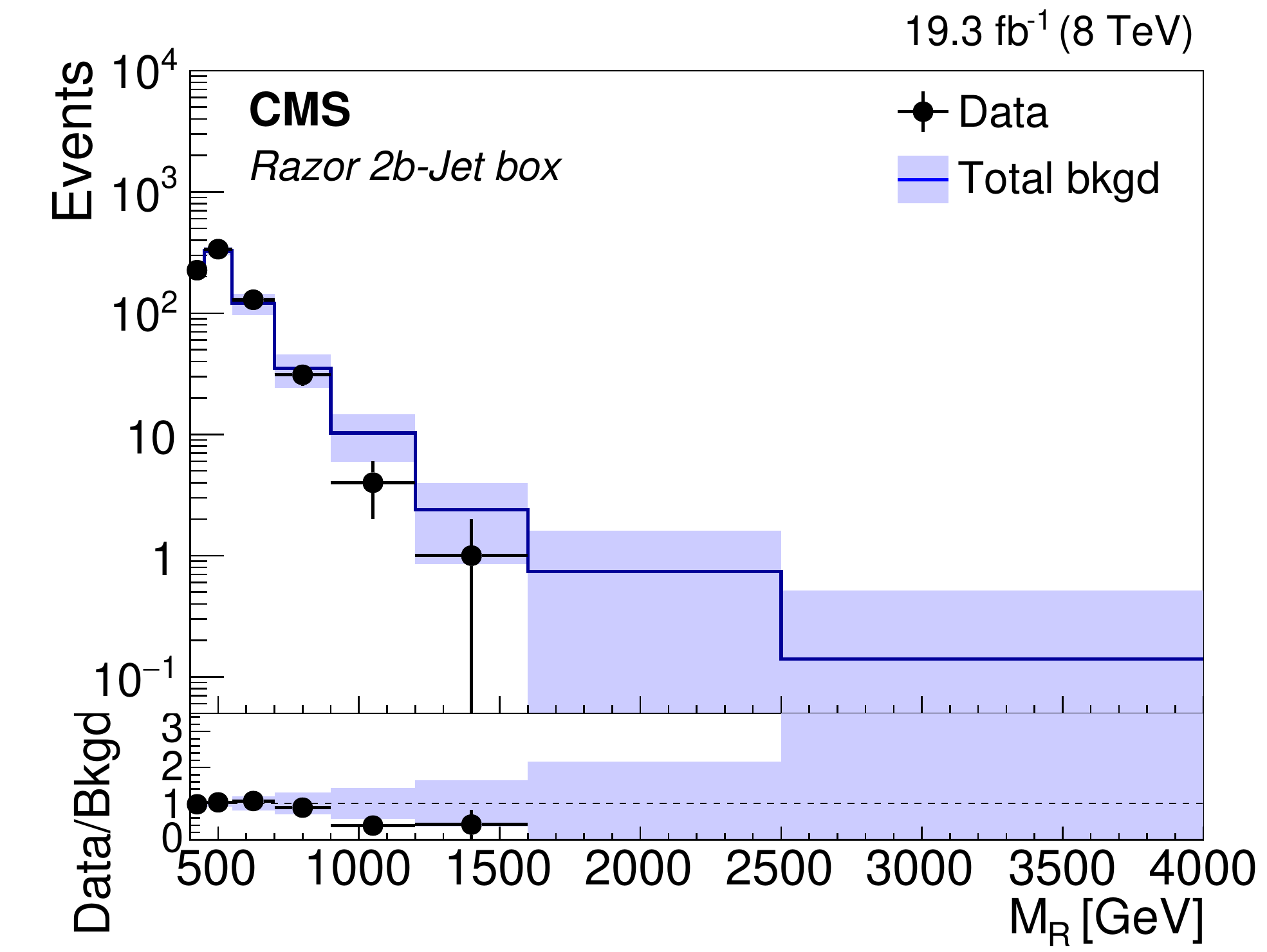}
\includegraphics[width=0.49\textwidth]{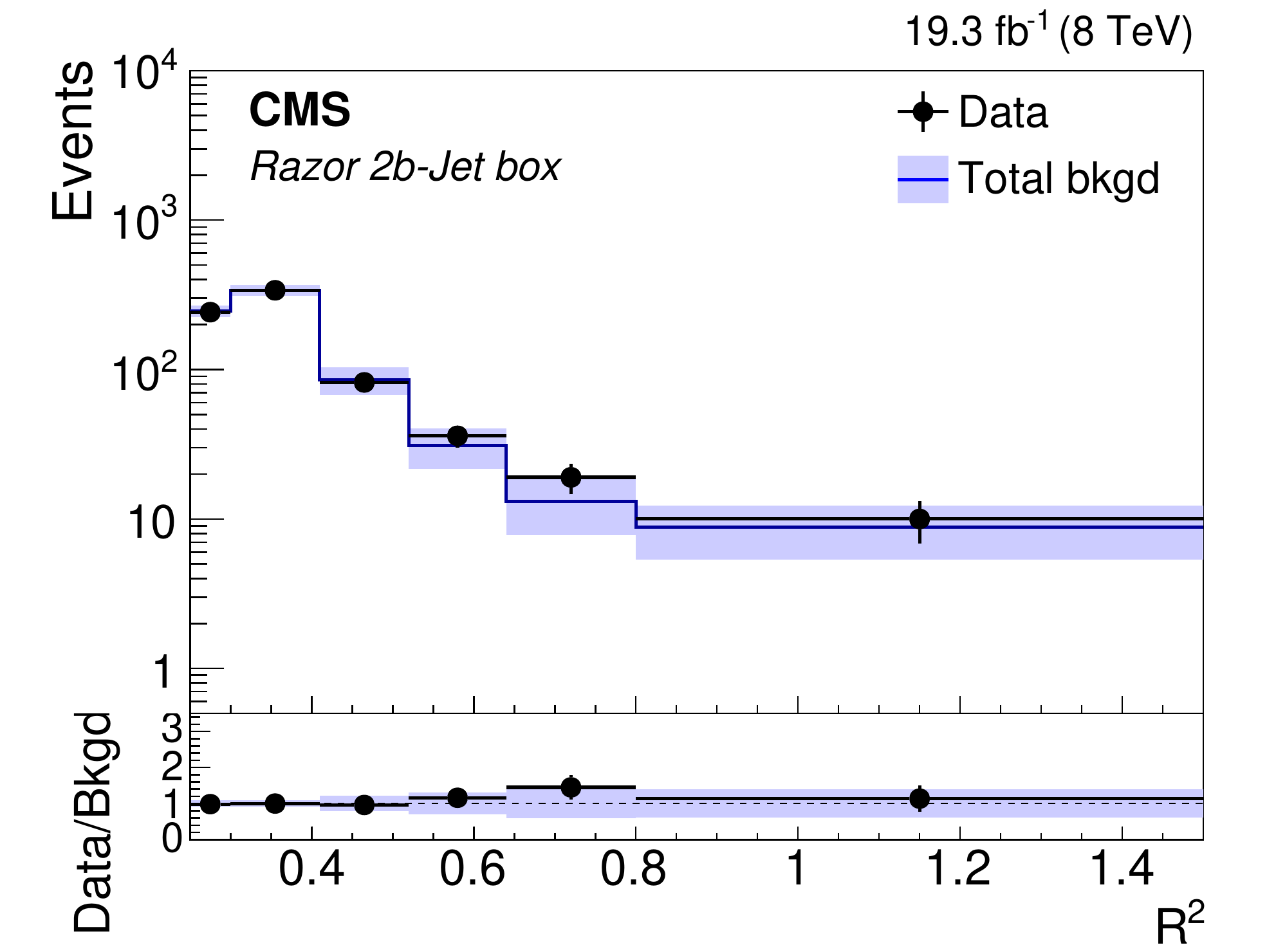}
\includegraphics[width=0.49\textwidth]{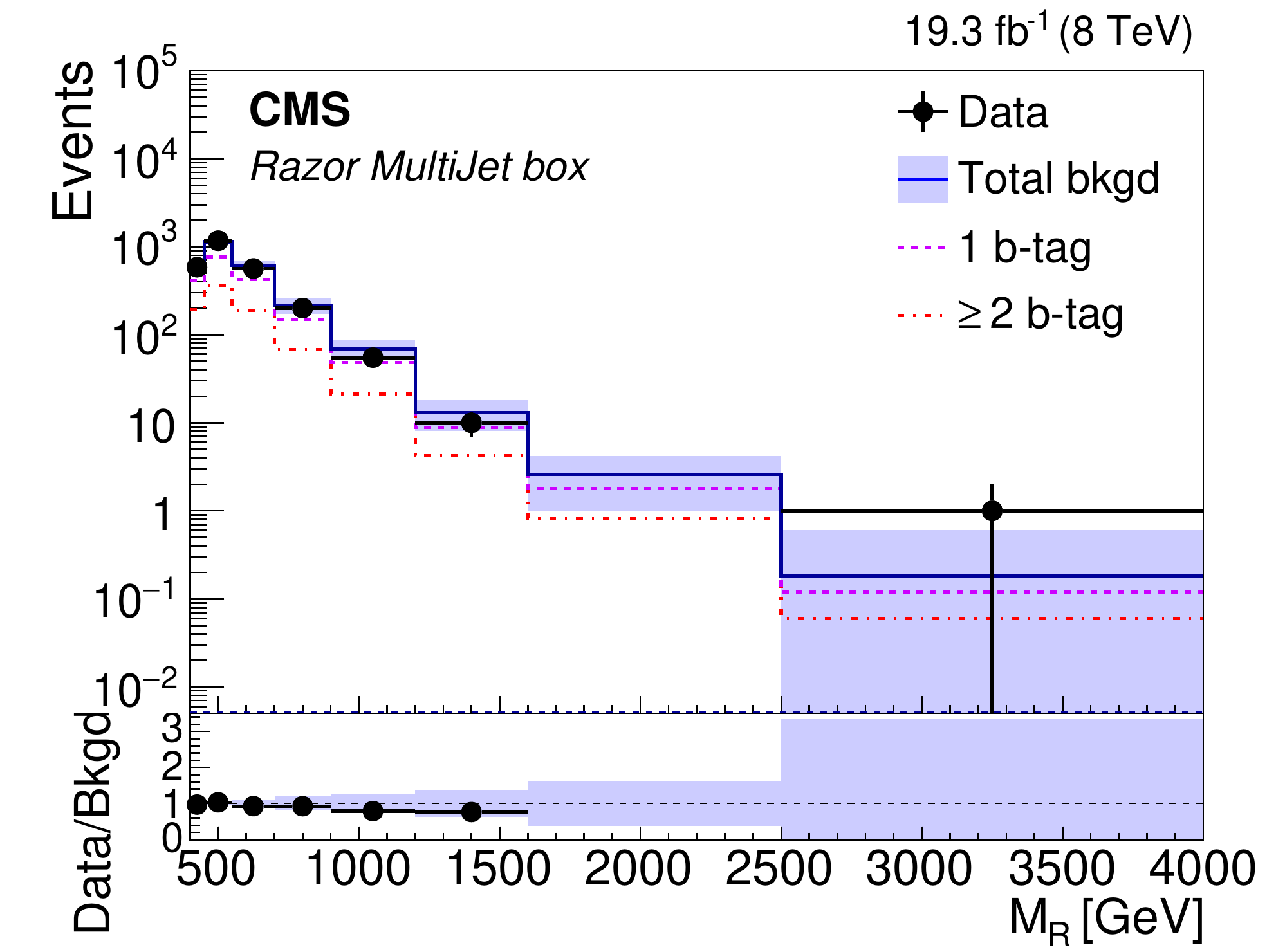}
\includegraphics[width=0.49\textwidth]{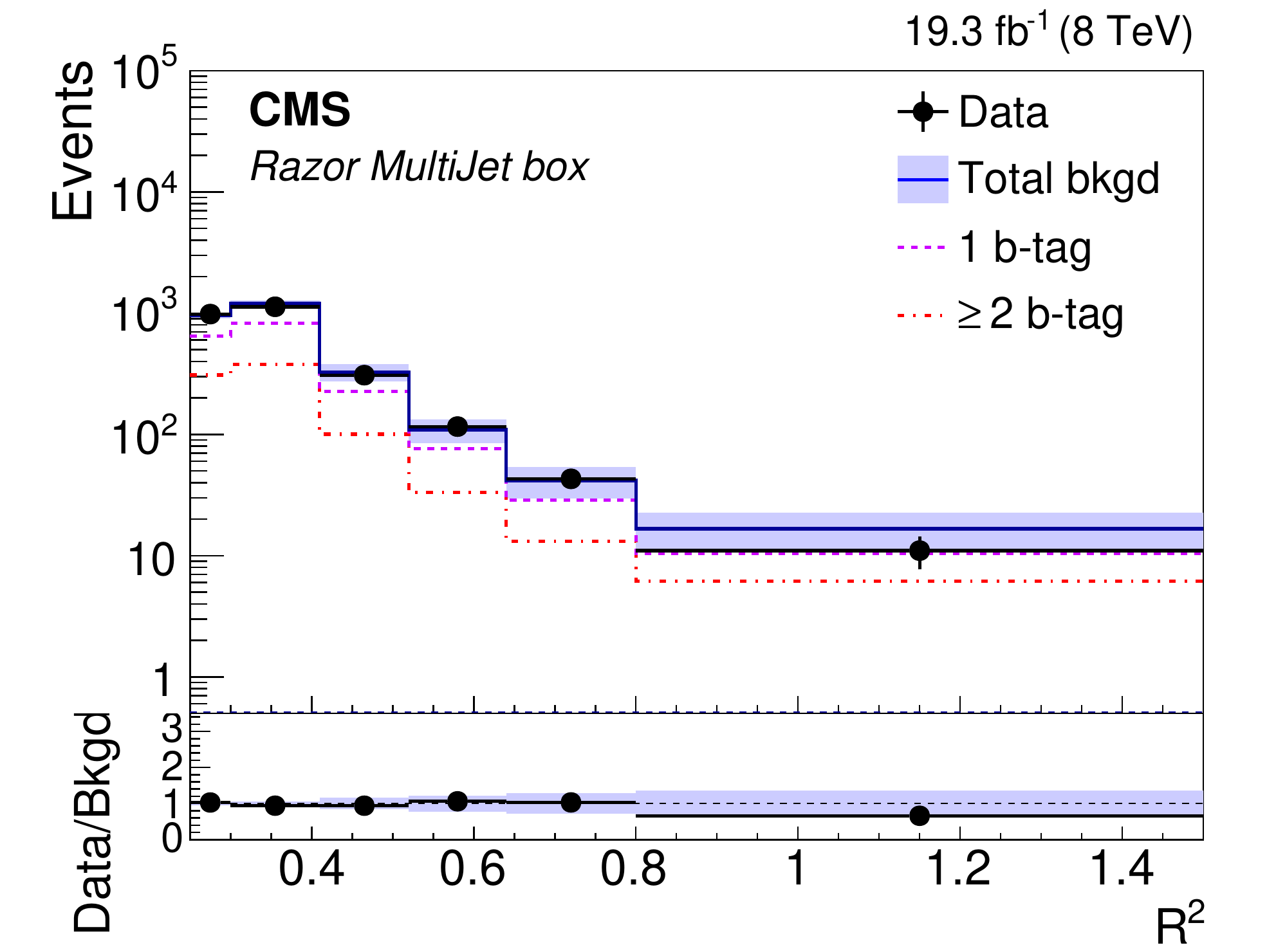}
\caption{Projection of the sideband fit result in the $\geq$2 \PQb-tagged jet
  box on (upper left) \MR and (upper right) $\Rtwo$, and projection of
  the sideband fit result in the MultiJet box on (lower left) \MR   and (lower right) $\Rtwo$. A detailed explanation is given in the
  caption of Fig.~\ref{fig:Proj1DMu}.\label{fig:Proj1DHad}}
\end{figure*}

The background shape parameters are estimated from the events in the
two sidebands (Section~\ref{sec:razVar}). This shape is then used to
derive a background prediction in the signal-sensitive region:
$30\,000$ alternative sets of background shape parameters are generated
from the covariance matrix returned by the fit. An ensemble of
pseudo-experiment data sets is created, generating random
($\MR$, $\Rtwo$) pairs distributed according to each
of these alternative shapes. For each bin of the signal-sensitive
region, the distribution of the predicted yields in each
pseudo-experiment is compared to the observed yield in data in order
to quantify the agreement between the background model and the
observation. The agreement, described as a two-sided p-value, is then
translated into the corresponding number of standard deviations for a
normal distribution. The p-value is computed using the probability
density as the ordering principle. The observed numbers of standard
deviations in the two-lepton boxes are shown in
Fig.~\ref{fig:FrenchFlagDilep}, as a function of \MR and
$\Rtwo$. Positive and negative significance correspond to
regions where the observed yield is respectively larger and smaller
than the predicted one. Light gray areas correspond to empty bins with
less than one event expected on average. Similar results for the
one-lepton and hadronic boxes are shown in
Figs.~\ref{fig:FrenchFlagLep} and
\ref{fig:FrenchFlagHad}. Figures~\ref{fig:Proj1DDilep}--\ref{fig:Proj1DHad}
illustrate the extrapolation of the fit results to the full
($\MR$, $\Rtwo$) plane, projected onto  \Rtwo and \MR and summed over the \PQb-tagged jet multiplicity
bins. No significant deviation of data from the SM background
predictions is observed.

\begin{figure}[htb!]
\centering
\includegraphics[width=0.49\textwidth]{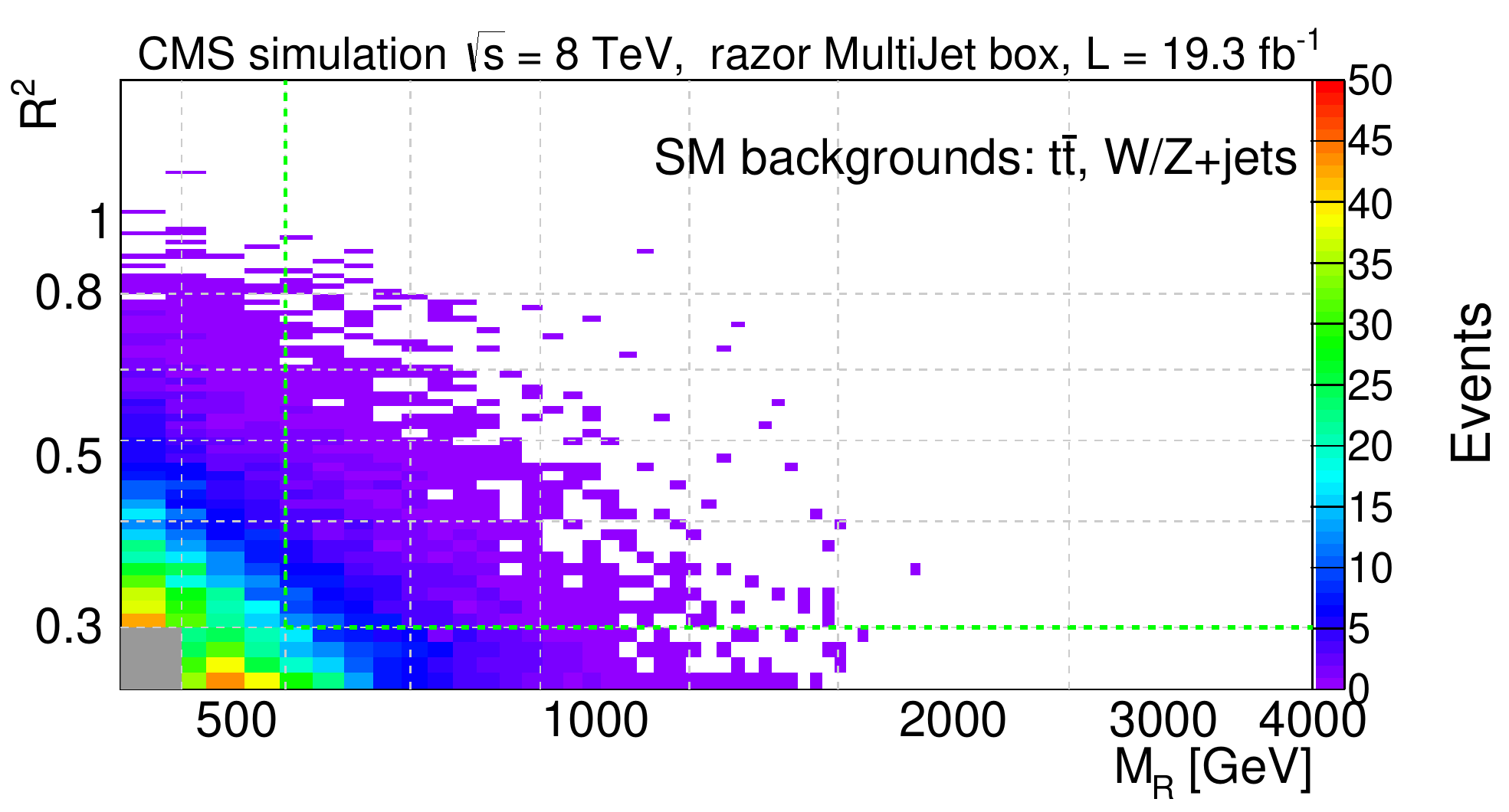}
\includegraphics[width=0.49\textwidth]{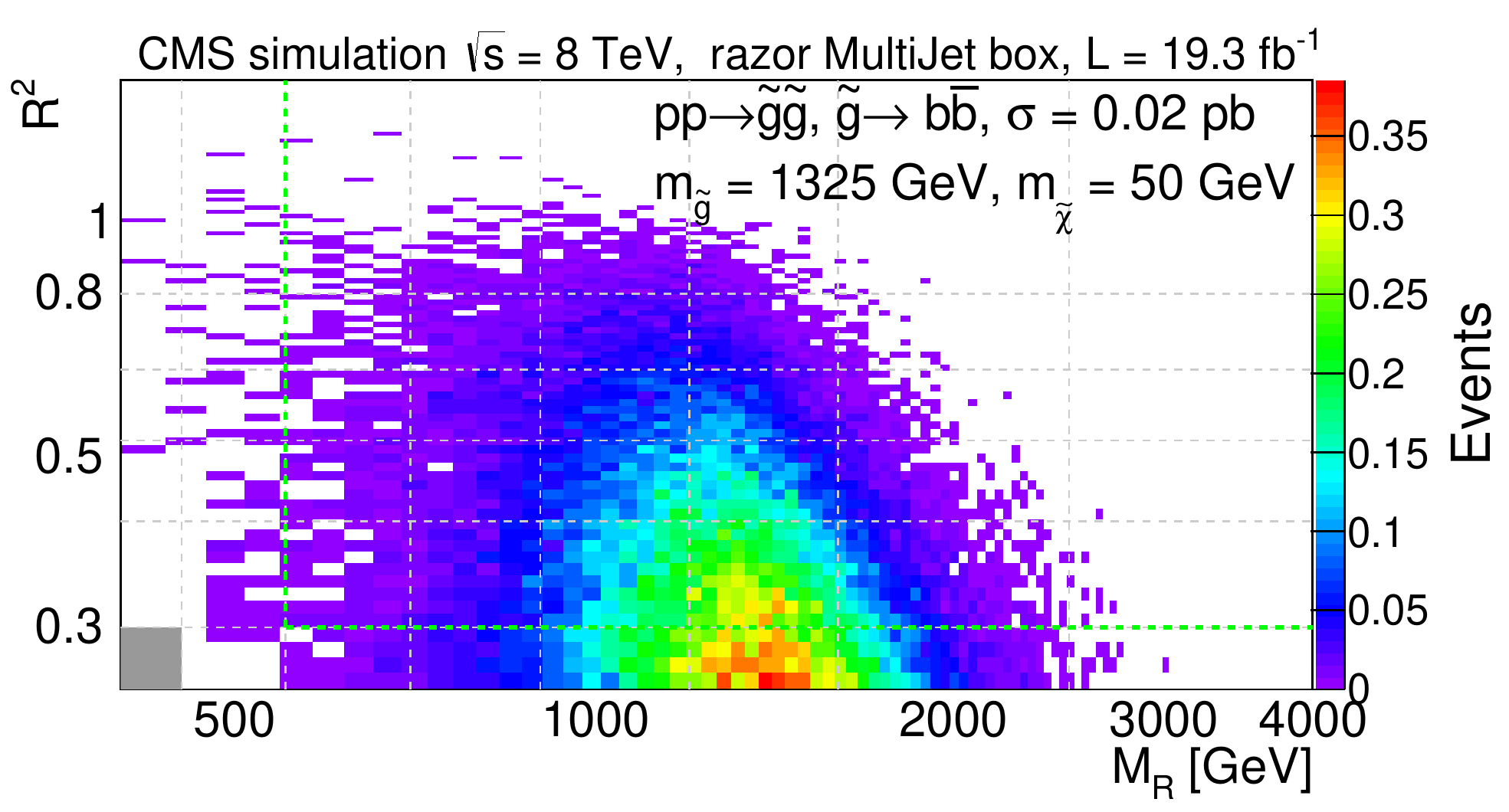}
\caption{Distribution of (\cmsLeft) simulated SM background events and (\cmsRight)
  T1bbbb gluino-gluino events in the MultiJet box. Each $\sGlu$ is
  forced to decay to a \bbbar pair and a $\chiz_1$,
  assumed to be the stable LSP. The $\sGlu$ and $\chiz_1$ masses are
  fixed to 1325\GeV and 50\GeV,
  respectively.\label{fig:T1bbbbsignalinj}}

\end{figure}

\begin{figure}[htb!]
\centering
\includegraphics[width=0.49\textwidth]{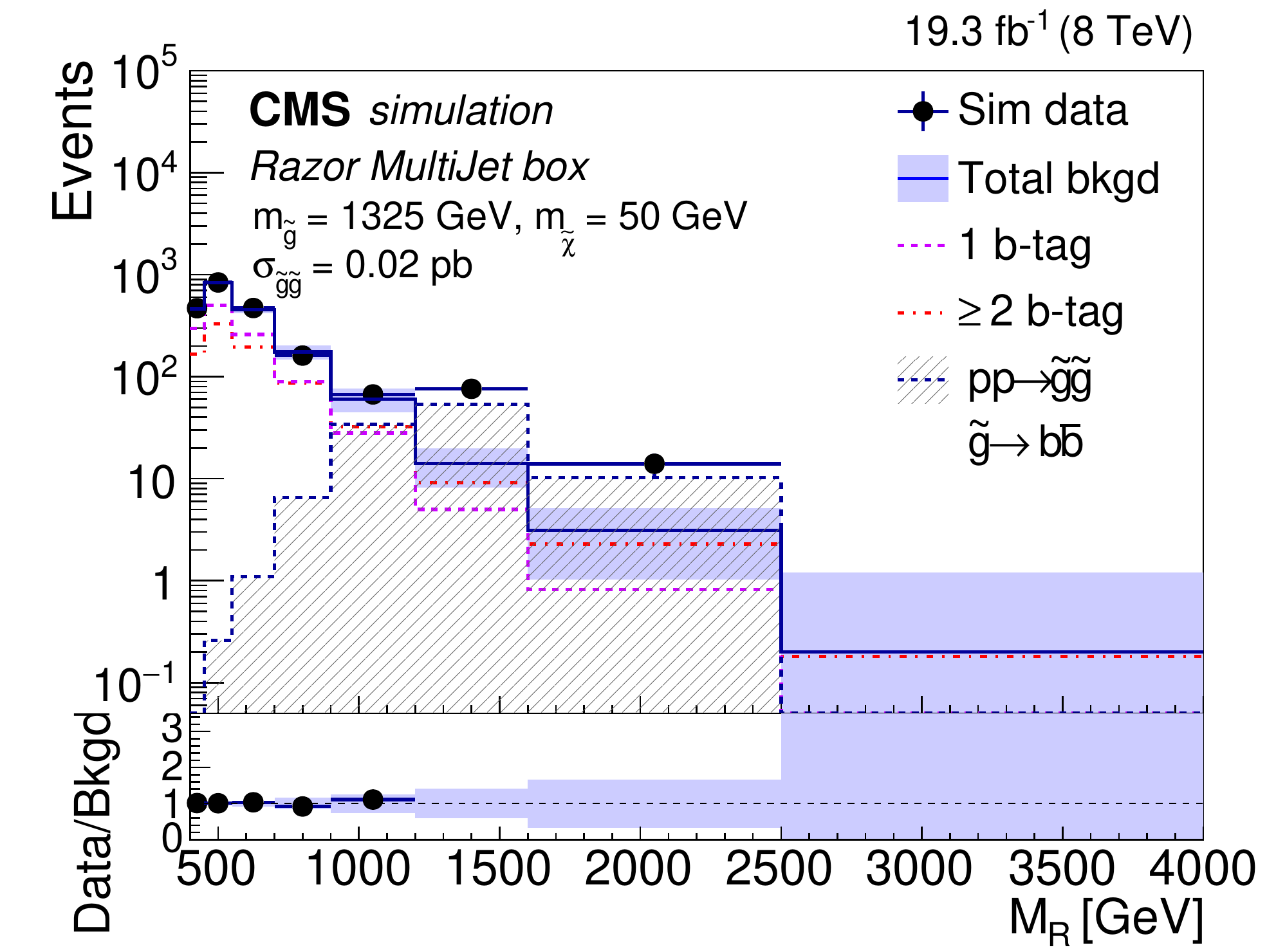}
\includegraphics[width=0.49\textwidth]{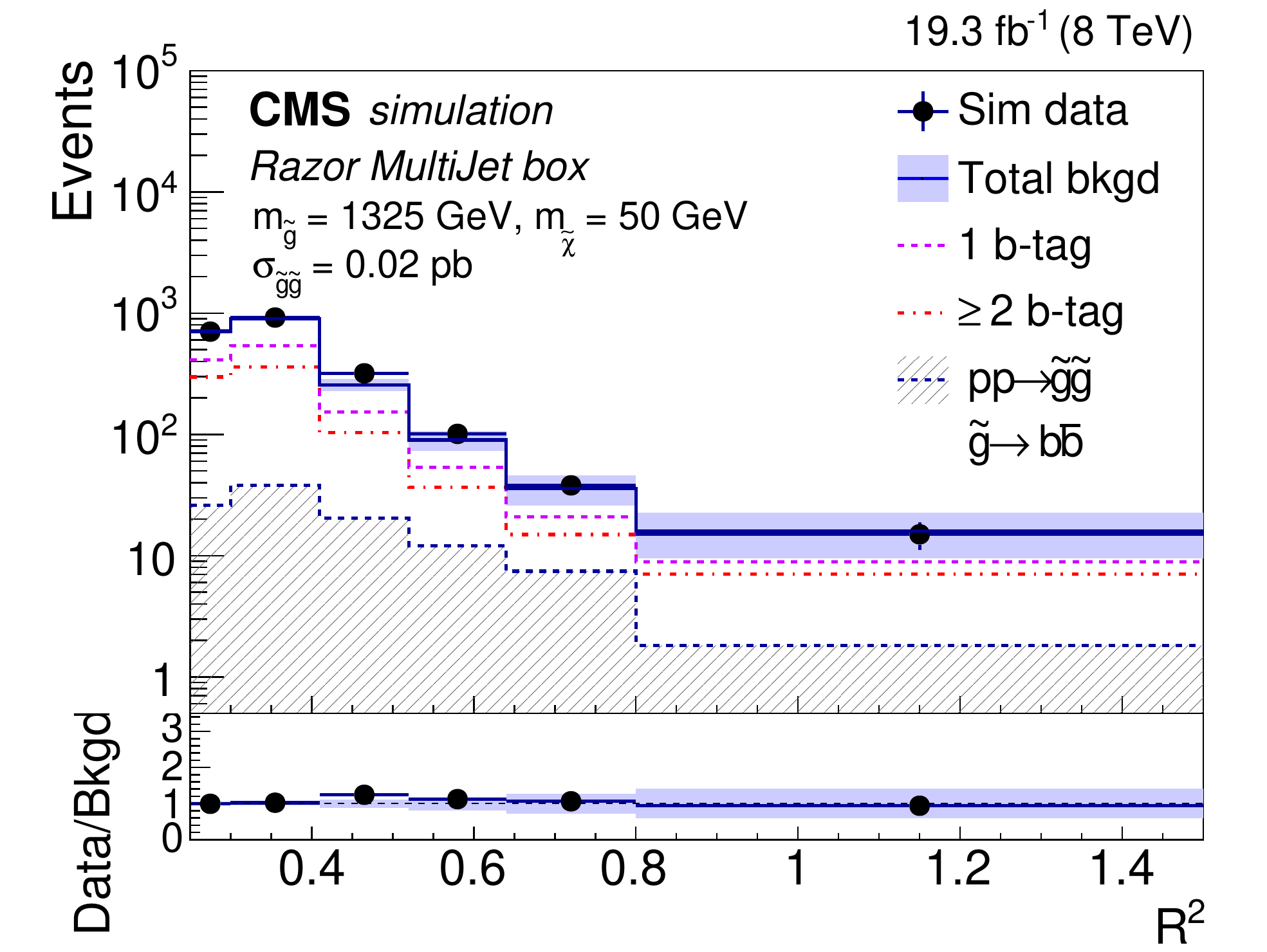}
\includegraphics[width=0.49\textwidth]{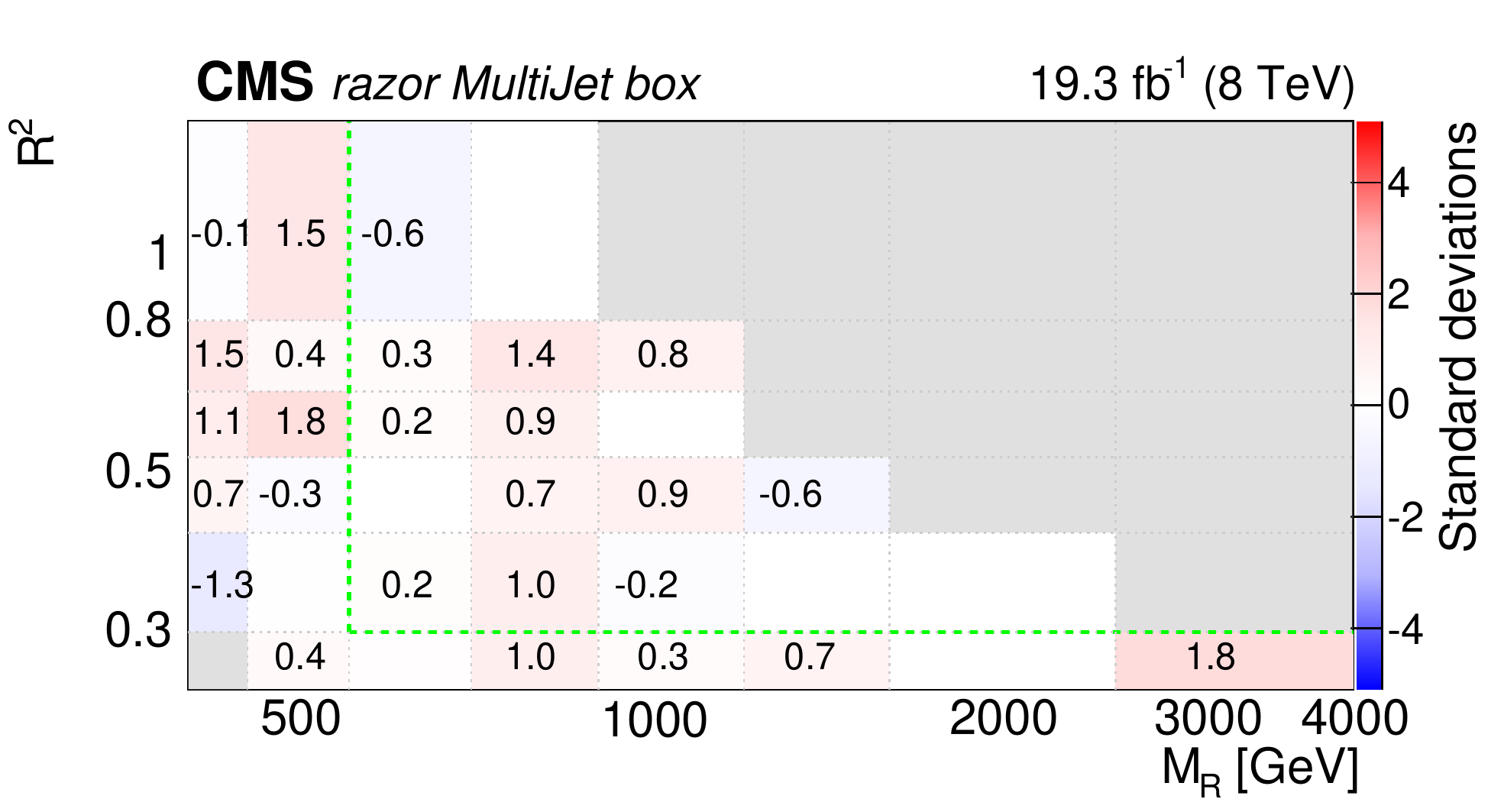}
\caption{Result of the fit to the sideband events of a
  signal-plus-background MC sample, corresponding to the gluino model
  whose distribution is shown in Fig.~\ref{fig:T1bbbbsignalinj}. A
  gluino-gluino production cross section of 0.02\unit{pb} is assumed. The
  one-dimensional projections on (\cmsUpperLeft) \MR and (\cmsUpperRight)
   \Rtwo are shown, together with (bottom) the agreement between
  the observed yield and the prediction from the sideband fit as a
  function of  \Rtwo and $\MR$. This agreement is
  evaluated from a two-sided p-value using an ensemble of
  background-only pseudo-experiments as described in
  Section~\ref{sec:bmodel}.\label{fig:FFsigma0p02}}
\end{figure}

To demonstrate the discovery potential of this analysis, we apply the
background-prediction procedure to a simulated signal-plus-background
MC sample. Figure~\ref{fig:T1bbbbsignalinj} shows the \MR and  \Rtwo distributions of SM background events and T1bbbb
events (Section~\ref{sec:sms}). The gluino and LSP masses are set
respectively to 1325\GeV and 50\GeV, representing a new-physics
scenario near the expected sensitivity of the analysis. A
signal-plus-background sample is obtained by adding the two
distributions of Fig.~\ref{fig:T1bbbbsignalinj}, assuming an
integrated luminosity of 19.3\fbinv and a gluino-gluino production
cross section of 0.02\unit{pb}, corresponding to 78 expected signal events
in the signal-sensitive region. The agreement between the background
prediction from the sideband fit and the yield of the
signal-plus-background pseudo-experiments is displayed in
Fig.~\ref{fig:FFsigma0p02}. The contribution of signal events to the
sideband region has a negligible impact on the determination of the
background shape, while a disagreement is observed in the
signal-sensitive region, characterized as an excess of events
clustered around $\MR\approx1300$\GeV. The excess indicates
the presence of a signal, and the position of the excess in the
$\MR$ variable provides information about the underlying SUSY
mass spectrum.

\section{Limit-setting procedure}
\label{sec:limit}

We interpret the results of the searches by determining the 95\%
confidence level (\CL) upper limits on the production cross sections of
the SUSY models presented in Section~\ref{sec:sms}, using the LHC
\CLs procedure~\cite{LHCCLs} and a global likelihood
determined by combining the likelihoods of the different search boxes
and sidebands.  To reduce computational requirements, a binned
likelihood is used.

For the razor search boxes, the signal contribution is modeled by a
template function, for a given signal hypothesis in a specific box and
a given \PQb-tagged jet multiplicity. The template function, normalized
to unit probability, is multiplied by the expected signal yield in
each bin ($\sigma_\mathrm{NLO+NLL} L
\epsilon^{\text{box}}_{\PQb\text{-tag}}$). Here $\sigma_\mathrm{NLO+NLL}$ is
the SUSY signal cross section, $L$ is the integrated luminosity
corresponding to the size of the data set, and
$\epsilon^{\text{box}}_{\PQb\text{-tag}}$ is the signal selection
efficiency for a given box and, in case of the single-lepton and
hadronic boxes, for a given \PQb-tagged jet multiplicity.

Each systematic uncertainty is incorporated in the likelihood with a
dedicated nuisance parameter, whose value is not known a priori but
rather must be estimated from the data. The set of nuisance parameters
may be divided into three distinct classes (though their statistical
treatment is the same): those related to the signal normalization,
those related to the signal shape, and those related to the background
normalization and shape.

We consider the following systematic uncertainties associated with the
signal normalization, with the size of the uncertainty indicated in
parentheses:
\begin{itemize}
\item integrated luminosity (2.6\%)~\cite{CMS:2013gfa};
\item trigger efficiency (5\%);
 \item lepton reconstruction and identification efficiencies (3\%
  per lepton), measured from an inclusive $\cPZ\to \ell^+\ell^-$ event
  sample ($\ell=\Pe,\Pgm$) as a function of the lepton $\pt$ and $\eta$
  values~\cite{Chatrchyan:2013iaa,Chatrchyan:2013mxa}.
\end{itemize}

In addition, four signal-shape systematic uncertainties are considered, whose
sizes vary with $\Rtwo$, $\MR$, and the \PQb-tagged jet
multiplicity:
\begin{itemize}
\item The uncertainty in the jet \PQb-tagging and mistagging efficiencies
  (up to 20\% depending on the signal model), evaluated for each
  ($\MR$, $\Rtwo$) and \PQb-tagged jet multiplicity
  bin. The uncertainty is evaluated by propagating the uncertainty in
  data-to-simulation scale factors~\cite{btag8TeV}.
\item the uncertainty in the modeling of the parton distribution
  functions (PDFs) (up to 10\% depending on the signal model),
  evaluated for each bin in the ($\MR$, $\Rtwo$) plane
  and for each box and \PQb-tag multiplicity following the
  PDF4LHC~\cite{Bourilkov:2006cj,Alekhin:2011sk,Botje:2011sn}
  prescription, using the CTEQ-6.6~\cite{Nadolsky:2008zw} and
  MRST-2006-NNLO~\cite{Martin:2007bv} PDF sets.
\item The uncertainty in the jet energy scale and resolution (up to
  5\% depending on the signal model), evaluated from a set of data
  control samples and MC simulations~\cite{JME-JINST}.
\item The uncertainty in the modeling of the associated jet production
  by the \MADGRAPH simulation (up to 20\% depending on the
  signal model), studied using $\cPZ$+jets and $\ttbar$ data events
  and parameterized by an MC-to-data scale factor as a function of the
  magnitude of the vector sum of the $\pt$ values of the two produced
  SUSY particles~\cite{1LepMVA}.
\end{itemize}
The impact of each of these uncertainties on the SUSY signal shape is
taken into account by varying each effect up or down by one standard
deviation.

The uncertainty in the knowledge of the background distributions
is taken into account by maximizing the likelihood with respect to the background shape and
normalization parameters using the data in the two
sidebands and the signal-sensitive region. The background parameterization
is able to accommodate several sources of
systematic uncertainties defined below:
\begin{itemize}
\item dependence of the background shape on the \PQb-tag multiplicity;
\item dependence of the background shape on the lepton and jet
 multiplicities;
\item deviation of the two-dimensional shape from an exponentially falling
 distribution, through the background empirical function parameter $n$,
 which modifies the tail in \MR and $\Rtwo$;
\item shape bias induced by the dependence of the \PQb-tagging efficiency and
 mistag rate on the jet $\pt$;
\item deviation of the \PQb-tagging and mistagging efficiencies from the
 MC prediction, through independent normalization factors in each
 \PQb-tagged jet multiplicity bin.
\end{itemize}

The combination of razor and exclusive single-lepton~\cite{1LepMVA}
searches is performed using the same procedure, taking into account
the systematic uncertainties associated with the five following
effects:
\begin{itemize}
\item the PDFs;
\item the jet energy scale correction;
\item the integrated luminosity;
\item the b-jet tagging efficiency;
\item the associated jet production.
\end{itemize}
The uncertainties in the background predictions are taken to be
uncorrelated, being derived from independent data control samples with
different techniques. We verified that the correlation model for the
systematics has a negligible impact on the combination, since similar
results are obtained when neglecting any correlation between the
systematic uncertainties of the two searches.

\section{Interpretation}
\label{sec:interp}

The results of this search are interpreted in the context of the natural
SUSY simplified models presented in Section~\ref{sec:sms}.

\subsection{Limits on gluino pair production}
\label{sec:interp:gluino}

\begin{figure*}[t]
\centering
\includegraphics[width=0.40\textwidth]{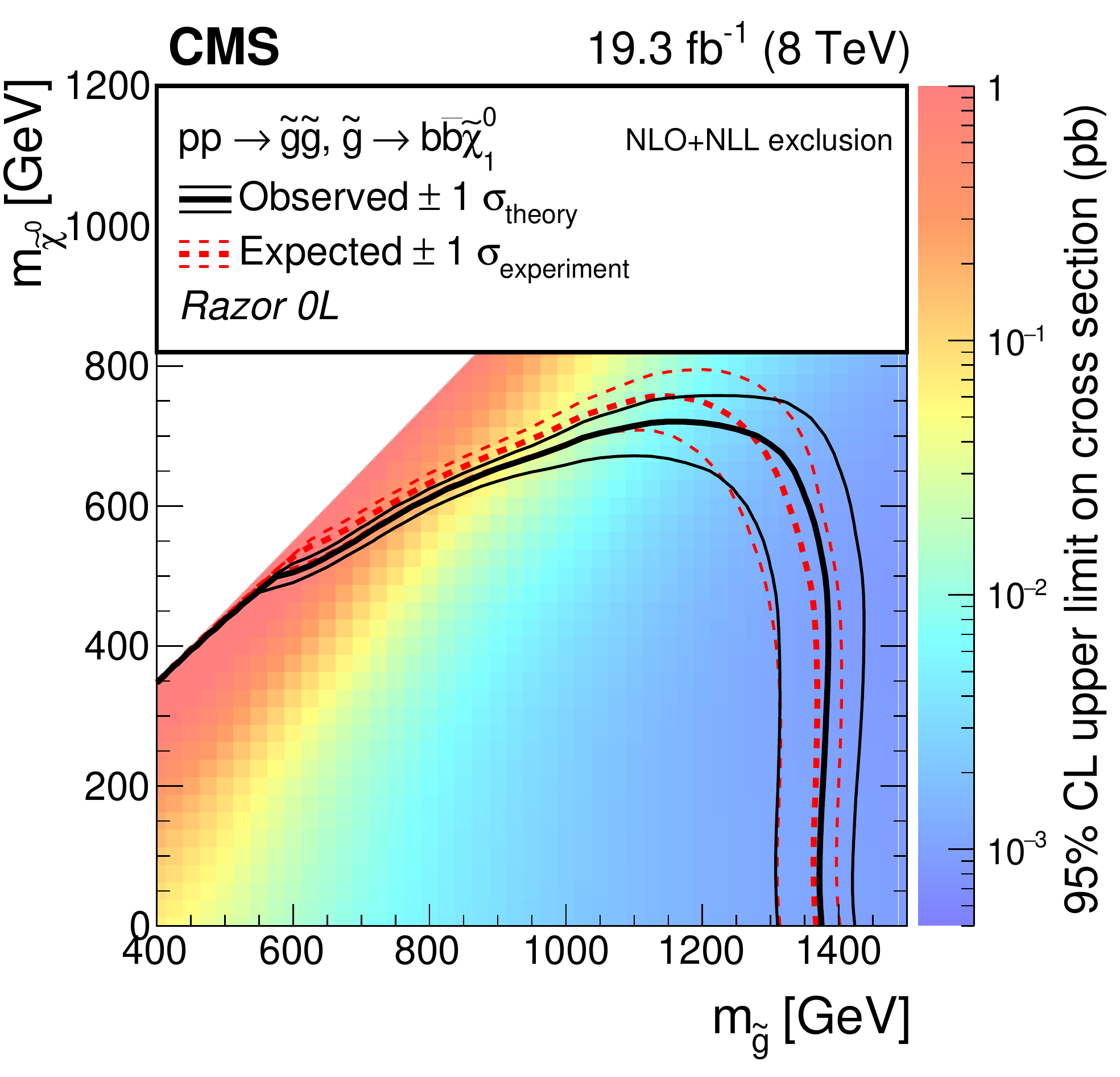}
\includegraphics[width=0.4\textwidth]{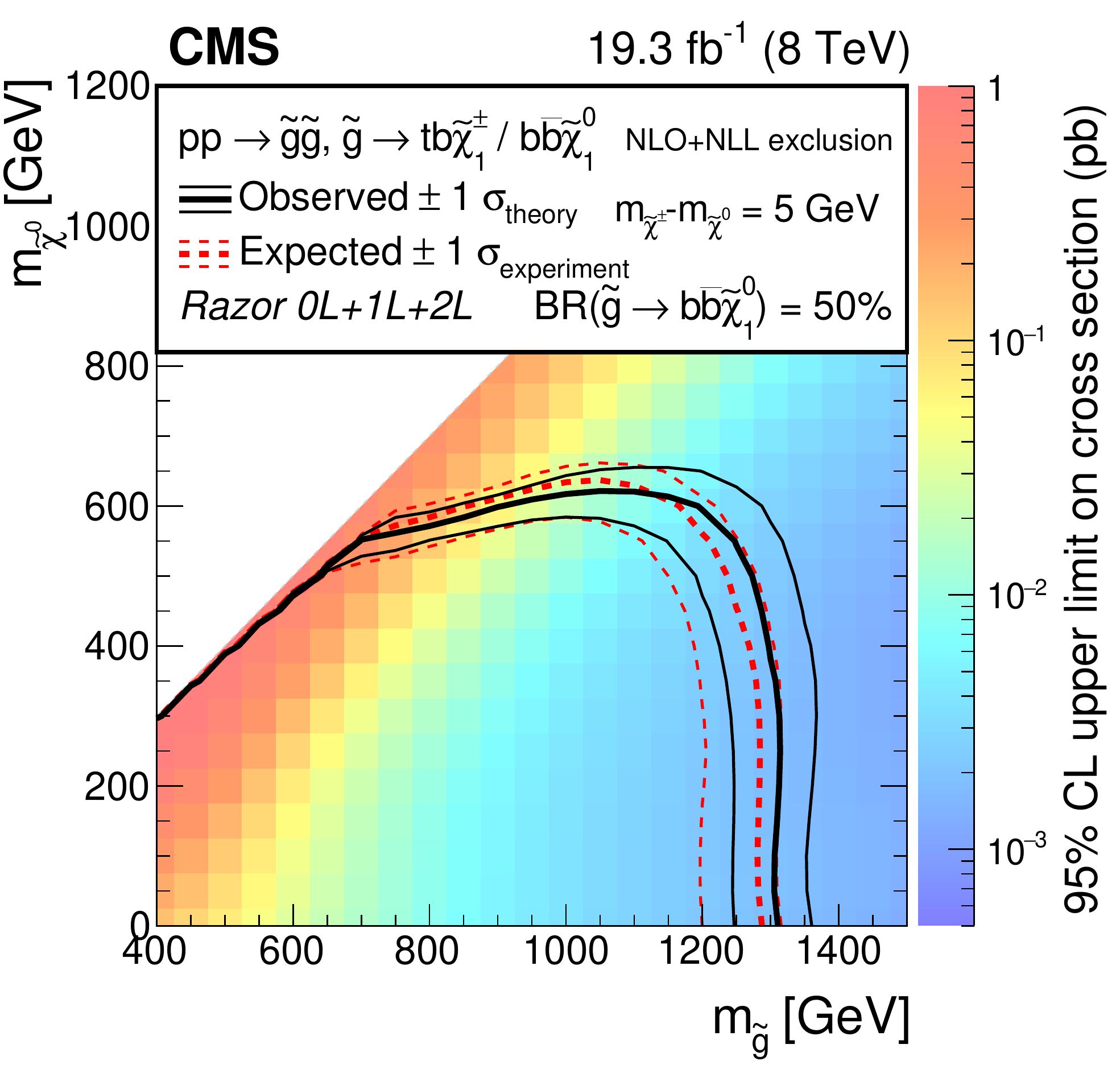}
\includegraphics[width=0.4\textwidth]{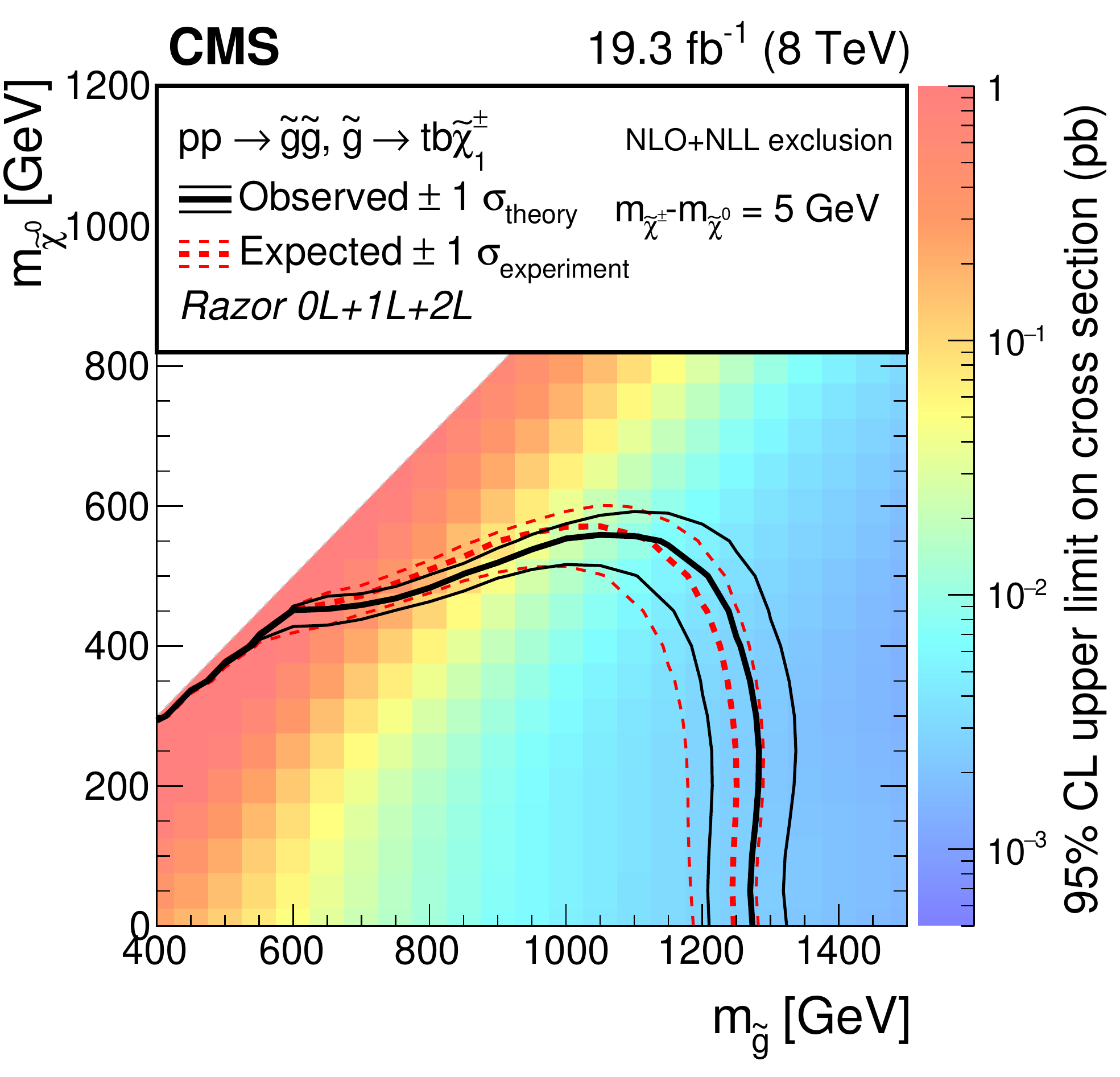}
\includegraphics[width=0.4\textwidth]{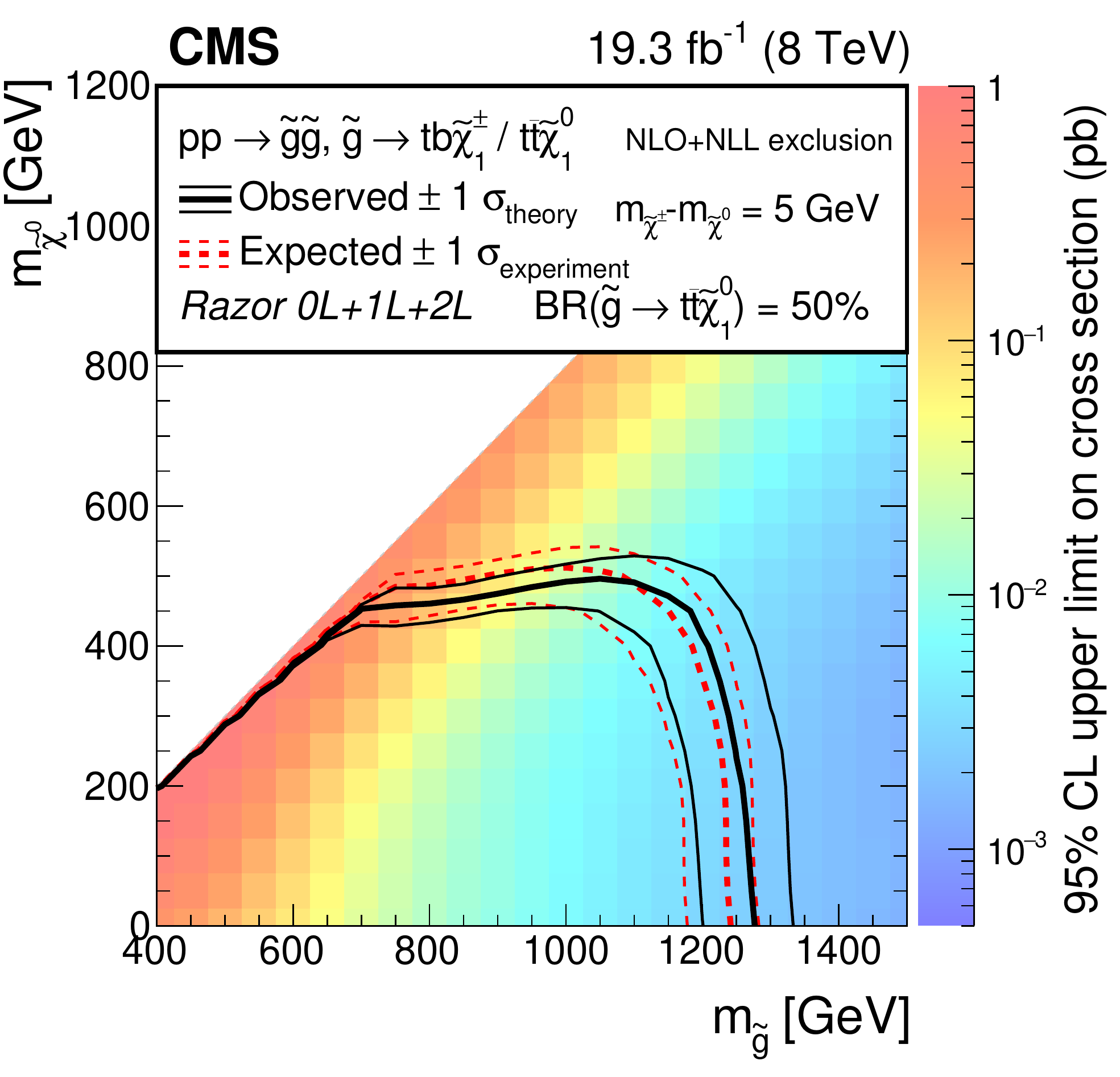}
\includegraphics[width=0.4\textwidth]{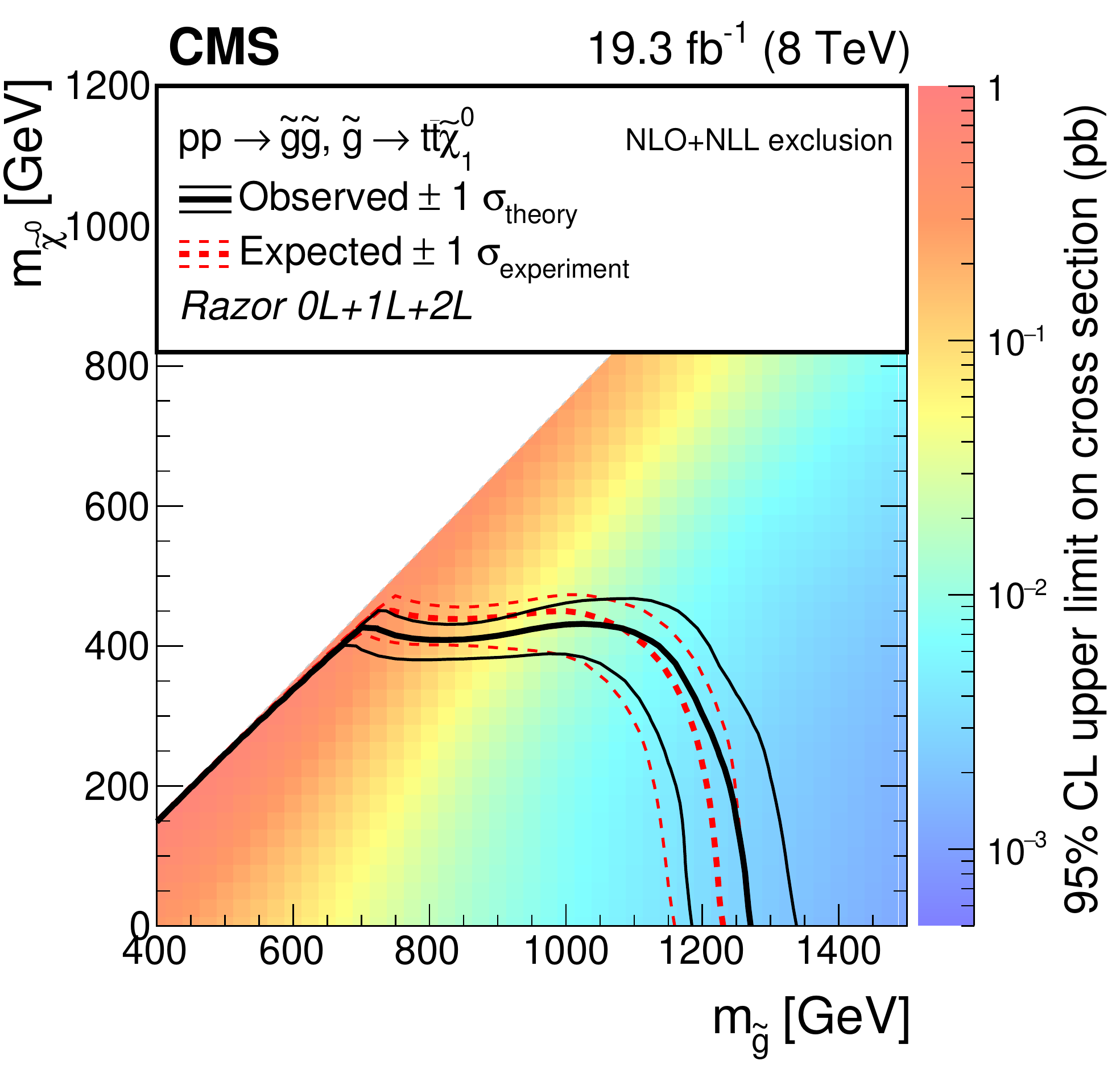}
\caption{Interpretation of the inclusive search with razor variables
  in the context of gluino pair production models: (upper left) T1bbbb, (upper right)
  T1tbbb, (middle left) T1ttbb, (middle right) T1tttb, and (bottom) T1tttt. The limit for T1bbbb
  is derived using only the hadronic boxes, while the limits for the
  remaining models are derived using all nine boxes. The color coding
  indicates the observed 95\% \CL upper limit on the signal cross
  section. The dashed and solid lines represent the expected and
  observed exclusion contours at a 95\% \CL, respectively. The dashed
  contours around the expected limit and the solid contours around the
  observed one represent the one standard deviation theoretical
  uncertainties in
  the cross section and the combination of the statistical and
  experimental systematic uncertainties,
  respectively.\label{fig:gluino2D}}
\end{figure*}

\begin{figure}[tb!]
\centering
\includegraphics[width=\cmsFigWidth]{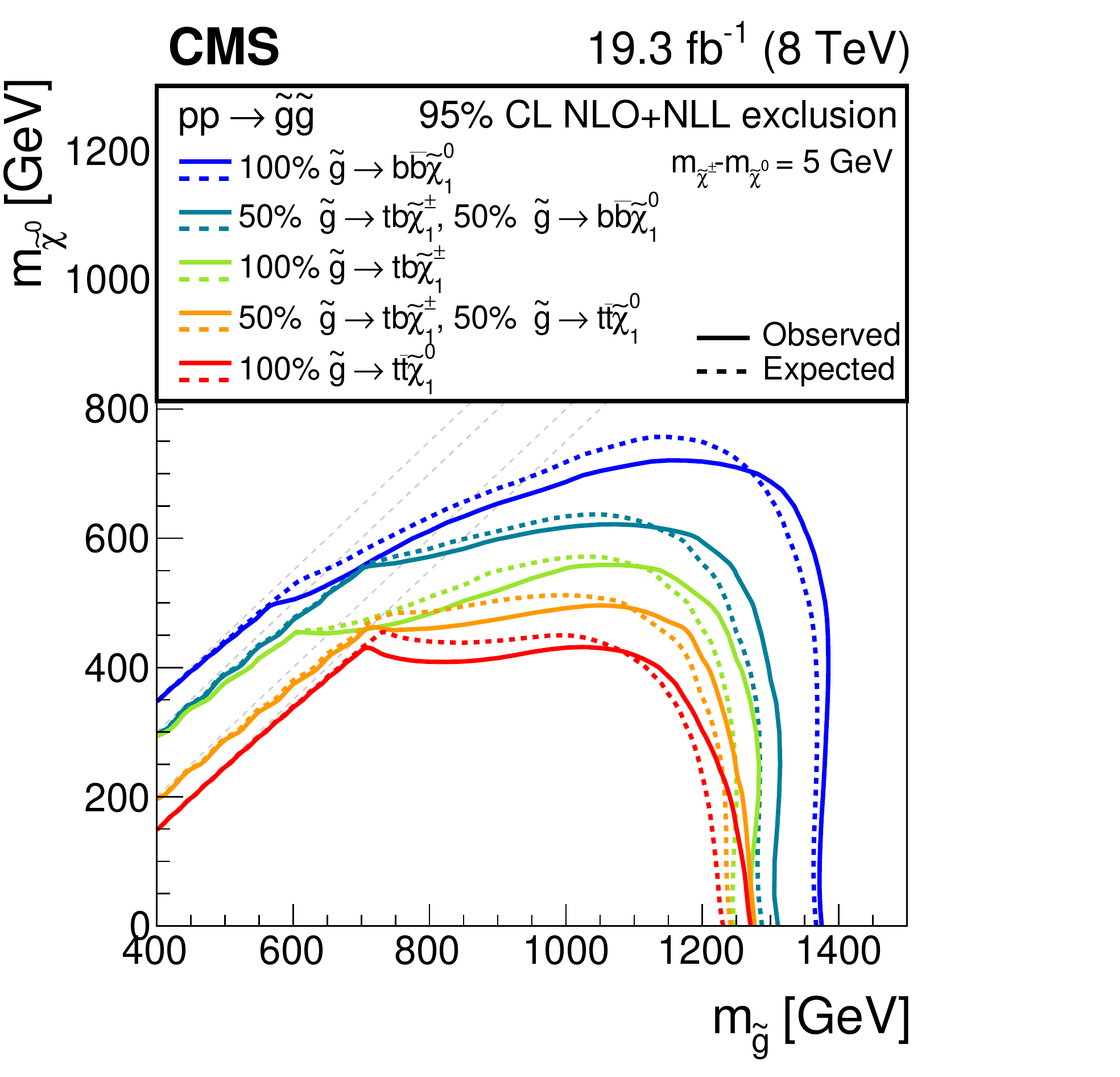}
\caption{Gluino mass limit at a 95\% \CL, obtained for different gluino
  pair production models with the inclusive razor analysis in the
  context of the natural SUSY spectrum of
  Fig.~\ref{fig:spectrum}.\label{fig:gluinogluinoSUMMARY}}

\end{figure}

Derived limits on gluino pair production in the T1bbbb, T1tbbb,
T1ttbb, T1tttb, and T1tttt scenarios are presented in
Fig.~\ref{fig:gluino2D}.  A comparison of the simplified natural SUSY
gluino-gluino exclusions, obtained for the different decay-mode
combinations of third generation quarks, is shown in
Fig.~\ref{fig:gluinogluinoSUMMARY}. The limits corresponding to
gluino-gluino topologies with mixed branching fractions lie within the
band defined by the T1bbbb and the T1tttt contours. As an example,
gluino masses smaller than 1175\GeV for T1tttt and 1310\GeV for T1bbbb
are excluded, for an LSP mass of 100\GeV. For any LSP mass value, a
larger number of top quarks in the decay topology corresponds to a
weaker limit, mainly due to a reduced total signal efficiency with
respect to the four-bottom-quark final state and a worse \MR and
$\Rtwo$ resolution for events with higher jet multiplicity in the
final state. Given this fact and the inclusive nature of the analysis,
the T1tttt limit can be considered to represent a conservative
estimate of a branching-fraction-independent limit, generically valid
for gluino-gluino production within the context of the natural SUSY
spectrum shown in Fig.~\ref{fig:spectrum}.

\subsection{Limits on top-squark pair production}
\label{sec:interp:stop}

\begin{figure*}[tb!]
\centering
\includegraphics[width=0.49\textwidth]{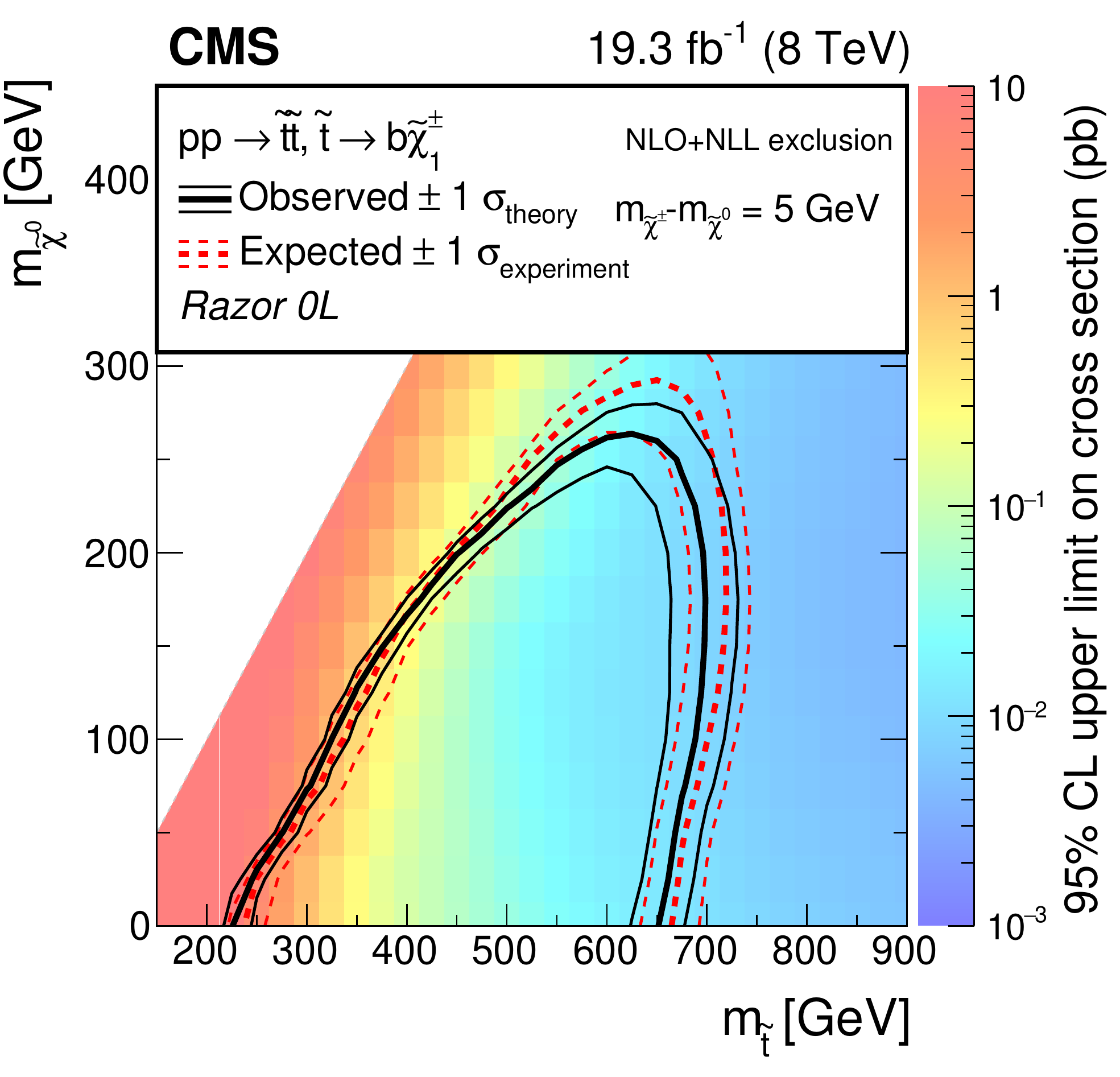}
\includegraphics[width=0.49\textwidth]{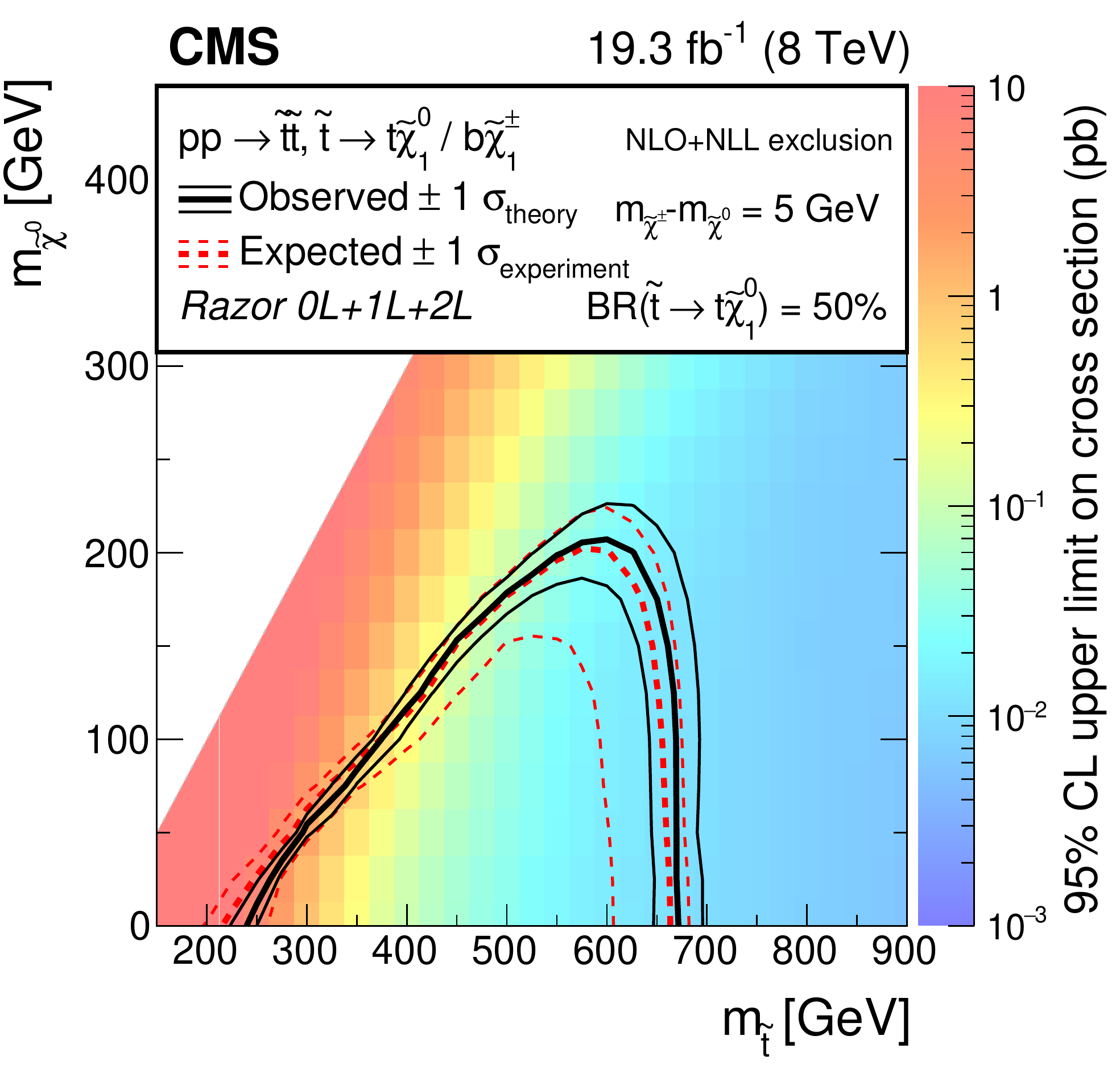}
\includegraphics[width=0.49\textwidth]{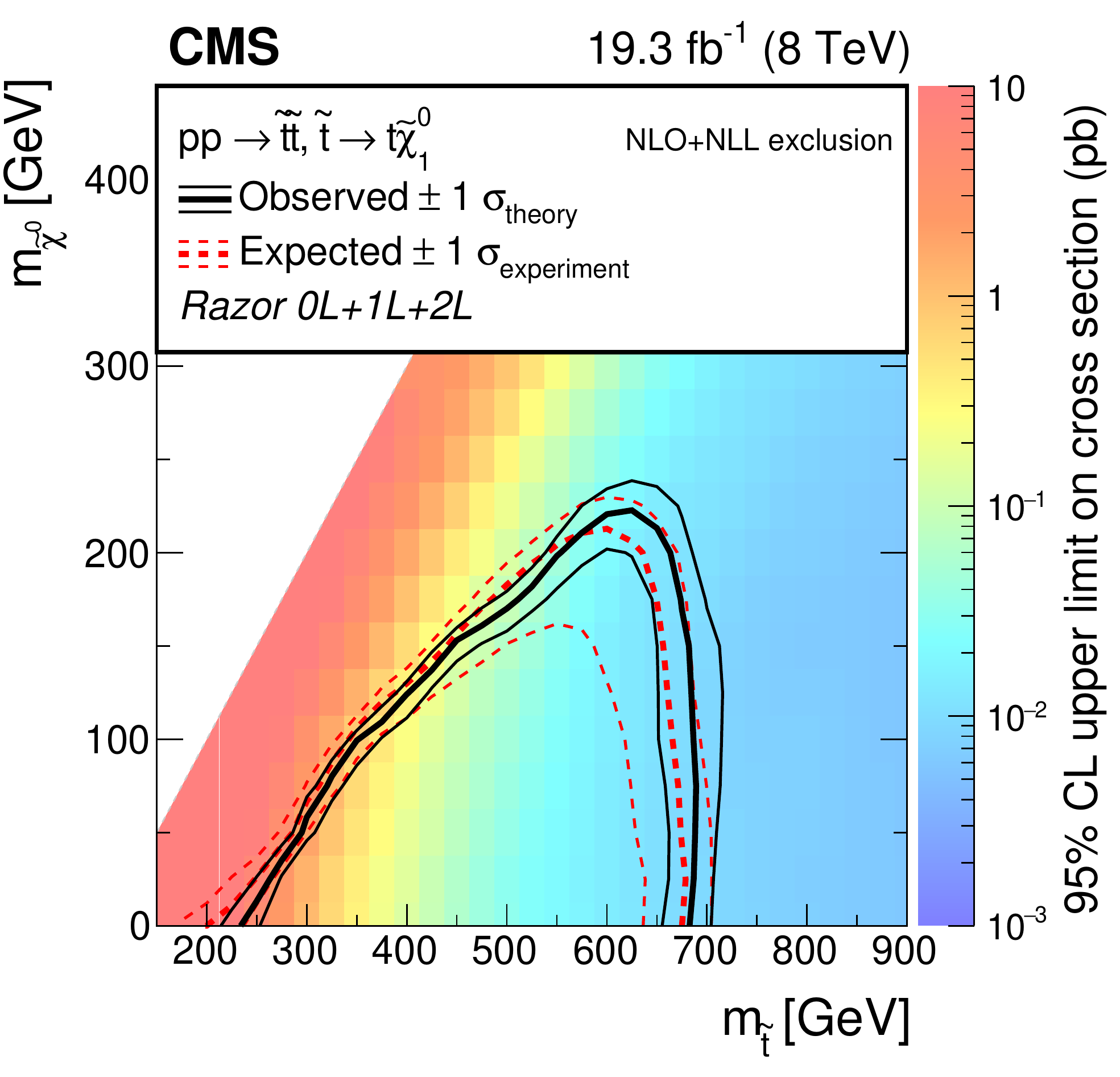}
\caption{Interpretation of the inclusive search with razor variables
  in the context of top-squark pair production models: (\cmsUpperLeft)
  T2bW$^{\ast}$, (\cmsUpperRight) T2tb, and (bottom) T2tt.  The limit for T2bW$^{\ast}$
  is derived using only the hadronic boxes, while the limits for the
  remaining models are derived using all nine boxes. The meaning of
  the color coding and the displayed contours is explained in the
  caption of Fig.~\ref{fig:gluino2D}.\label{fig:squark2D}}
\end{figure*}

\begin{figure}[tb!]
\centering
 \includegraphics[width=\cmsFigWidth]{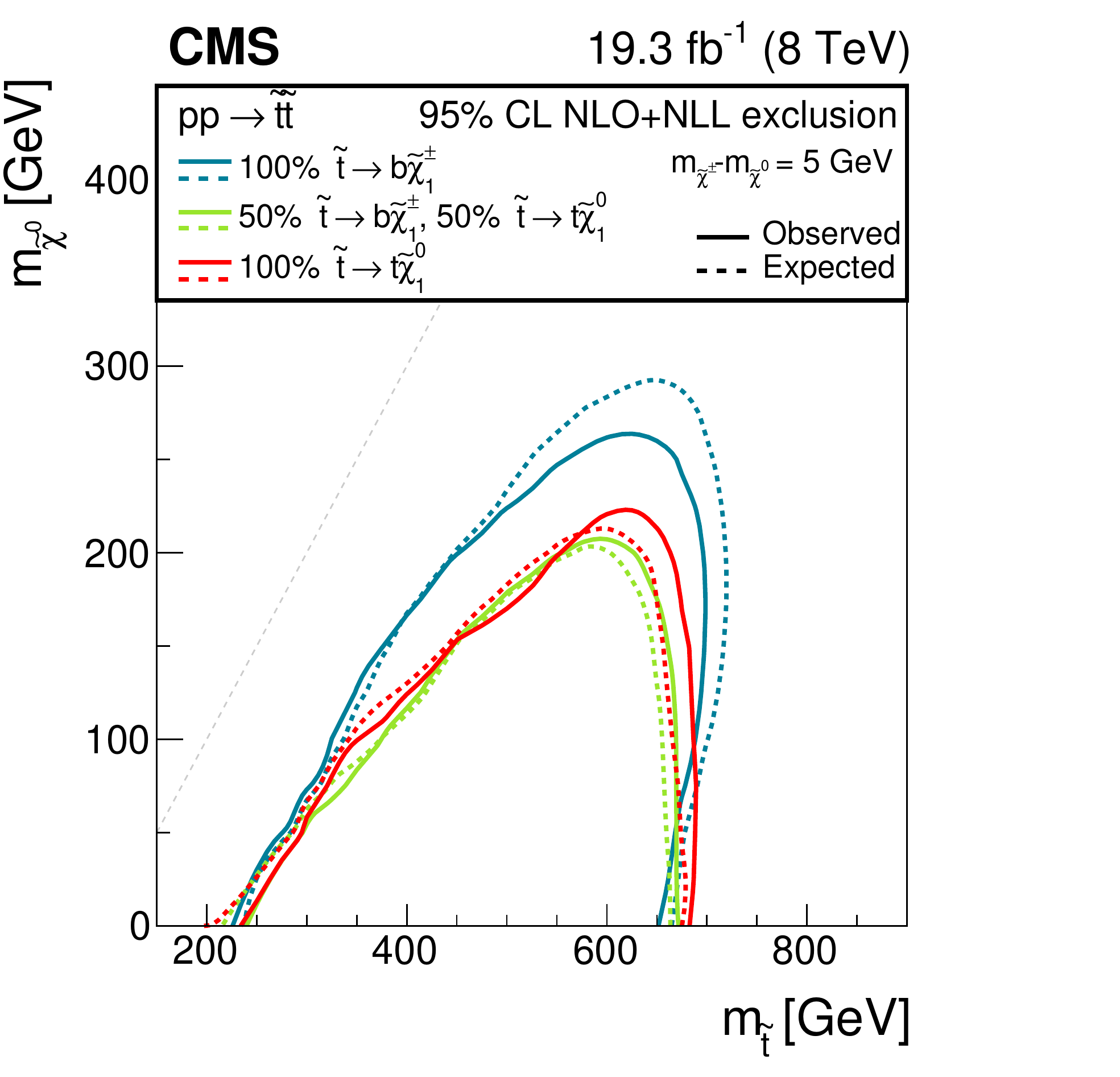}
 \caption{Top-squark mass limit at a 95\% \CL, obtained for different
   squark pair production models with the inclusive razor analysis in the
   context of the natural SUSY spectrum of
   Fig.~\ref{fig:spectrum}.\label{fig:stopSUMMARY}}

\end{figure}

Derived limits on squark pair production from the razor variables in
the T2bW$^{\ast}$, T2tb, and T2tt scenarios are presented in
Fig.~\ref{fig:squark2D} and compared in
Fig.~\ref{fig:stopSUMMARY}. As in the
case of the gluino interpretation, the expected limit from the razor
search improves as the number of top quarks in the decay topology
decreases.
For an LSP mass of 100\GeV, top-squark mass values larger than
400\GeV and smaller than 650\GeV are excluded in all three
top-squark branching fraction scenarios.

Within the considered scenarios, a top-squark decay
to a chargino (neutralino) is topologically similar to a bottom-squark
decay to a neutralino (chargino). In the limit of degenerate charginos
and neutralinos, the decay products of the chargino are
generically too soft to be detected and this correspondence is
exact. However, for large mass differences between the squarks and the
chargino, the chargino decay products may be boosted enough to become
observable, breaking the correspondence. For the models
with the intermediate decay to charginos, there is a migration
of reconstructed events from the low-background 2b-Jet box to the
high-background MultiJet box and a consequently weaker
limit with respect to the simplified model without decays to charginos.

\begin{figure}[tb!]
\centering
\includegraphics[width=0.49\textwidth]{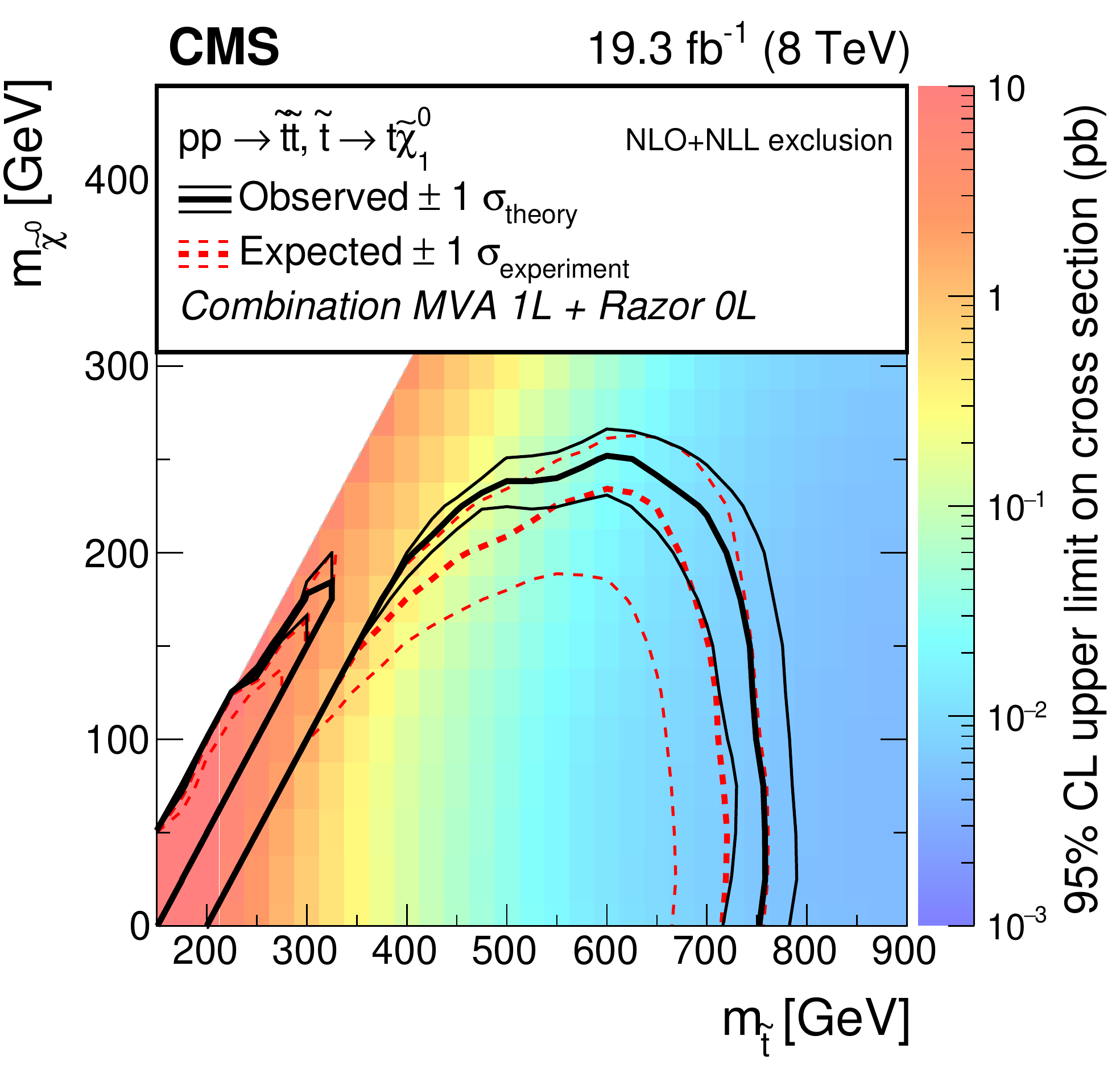}
\includegraphics[width=0.49\textwidth]{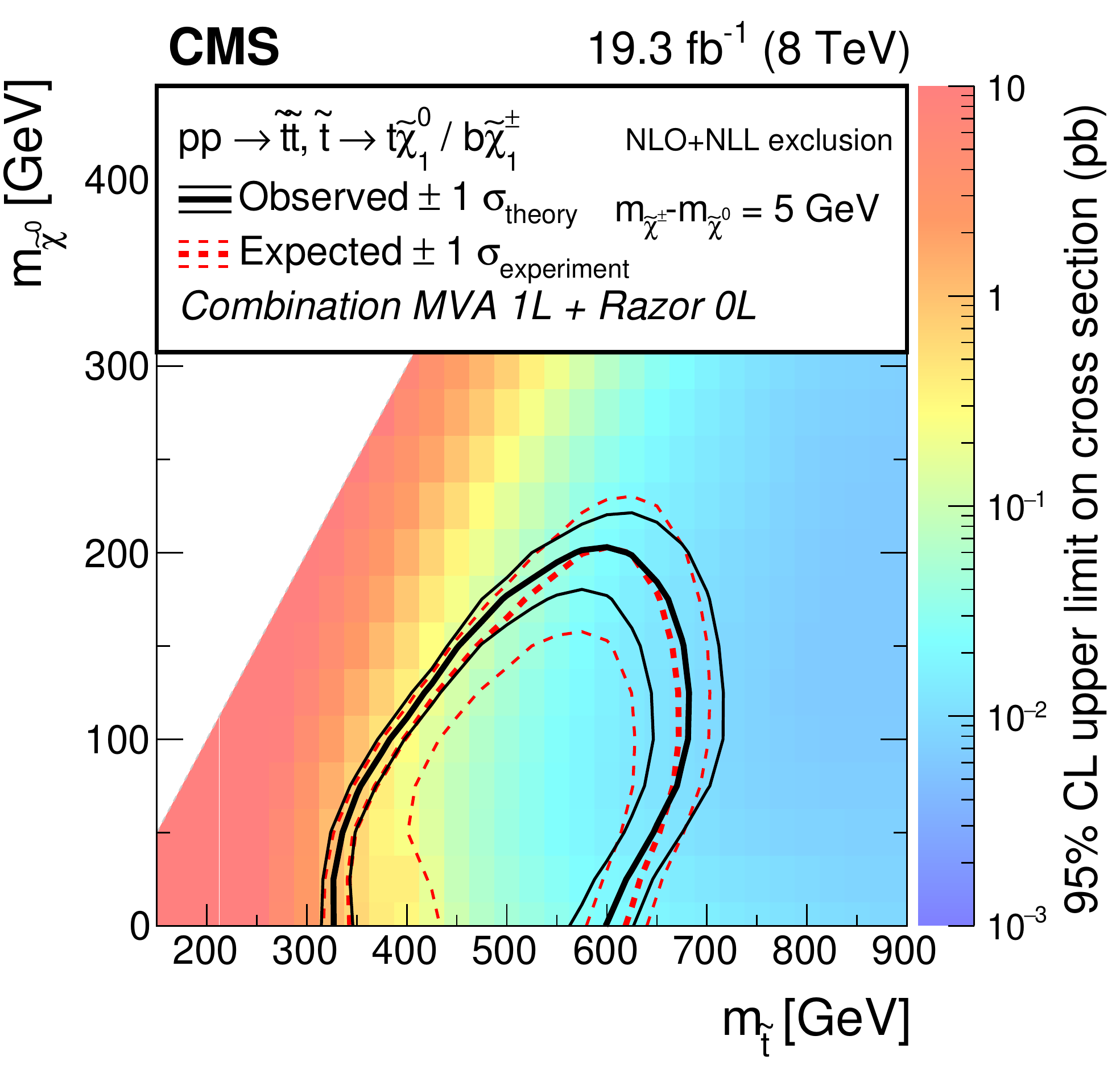}
\caption{Top-squark mass limit at a 95\% \CL, obtained combining the
  result of the hadronic razor boxes with the result of
  Ref.~\cite{1LepMVA} for (\cmsLeft) T2tt and (\cmsRight) independent of
  the branching fraction choice. The meaning of
  the color coding and the displayed contours is explained in the
  caption of Fig.~\ref{fig:gluino2D}.\label{fig:combinedLimit}}
\end{figure}

A stronger limit on top-squark pair production is derived by combining
the hadronic boxes of the razor search with the results of the
exclusive single-lepton analysis~\cite{1LepMVA}. The exclusive
single-lepton search is conservatively assumed to only have
sensitivity when both top squarks decay to a top quark and a
neutralino. Figure~\ref{fig:combinedLimit} (\cmsLeft) presents the combined result obtained for the scenario where the top
squark only decays to a top quark and the lightest neutralino. For an
LSP mass of 100\GeV, the combination improves the constraint on the
top-squark mass from 660 to 730\GeV. This result provides the
most stringent limit on this specific simplified model.

Figure~\ref{fig:combinedLimit} (\cmsRight) presents a more generic limit on the
top-squark mass. We consider two decay modes for the top squark, as
indicated in Fig.~\ref{fig:spectrum}. We scan the relative branching
fractions, assuming that no other decay mode is allowed. The largest
excluded cross section (that is, the worst upper limit) is found for
each choice of the top-squark and neutralino mass. A
branching-fraction-independent limit is derived by comparing the
worst-case exclusion to the corresponding top-squark pair production
cross section. In this manner, top squarks decaying to the two
considered decay modes are excluded at a 95\% confidence level for
mass values $>$400\GeV and $<$645\GeV, assuming a neutralino mass of
100\GeV. Unlike other simplified model interpretations, this
interpretation is not based on a specific choice of branching
fractions. While a residual model dependence is present because only
two decay modes are considered, this result is more general than
previous constraints.

\section{Summary}
\label{sec:conclusion}
We present a search for supersymmetric particles using proton-proton
collision data collected by CMS in 2012 at $\sqrt{s} = 8\TeV$. The data set
size corresponds to an integrated luminosity of 19.3\fbinv. We
consider events with at least two jets, at least one of which is
identified as a \PQb-tagged jet, and study the event distribution in the
razor variables ($\MR$, $\Rtwo$). The data are
classified according to the muon, electron, jet, and \PQb-tagged jet
multiplicities.  No significant excess is observed with respect to the
standard model background expectations, derived from a fit to the data
distribution in low-$\MR$ and low-$\Rtwo$ sidebands.

The inclusive razor search is translated into 95\% confidence level
exclusion limits on the masses of the gluino and the top squark, in the
context of simplified ``natural'' SUSY models.  For a neutralino mass
of 100\GeV and depending on the branching fractions, the pair
production of gluinos and top squarks in multi-bottom, multi-top, and
mixed top-plus-bottom quark topologies is excluded for gluino masses
up to 1310\GeV and top-squark masses up to 660\GeV.  Using the
combined likelihood of the hadronic boxes of the razor search and the
single-lepton channels of the exclusive top-squark
search~\cite{1LepMVA}, the exclusion bound on the top-squark mass is
extended to 730\GeV for a top squark decaying to a top quark and to a
neutralino of mass 100\GeV.  Again assuming the neutralino mass to be 100\GeV, top
squarks decaying to the two considered decay modes are excluded at a
95\% confidence level for mass values between 400 and 645\GeV,
independent of the branching fractions.

\begin{acknowledgments}
\hyphenation{Bundes-ministerium Forschungs-gemeinschaft Forschungs-zentren} We congratulate our colleagues in the CERN accelerator departments for the excellent performance of the LHC and thank the technical and administrative staffs at CERN and at other CMS institutes for their contributions to the success of the CMS effort. In addition, we gratefully acknowledge the computing centers and personnel of the Worldwide LHC Computing Grid for delivering so effectively the computing infrastructure essential to our analyses. Finally, we acknowledge the enduring support for the construction and operation of the LHC and the CMS detector provided by the following funding agencies: the Austrian Federal Ministry of Science, Research and Economy and the Austrian Science Fund; the Belgian Fonds de la Recherche Scientifique, and Fonds voor Wetenschappelijk Onderzoek; the Brazilian Funding Agencies (CNPq, CAPES, FAPERJ, and FAPESP); the Bulgarian Ministry of Education and Science; CERN; the Chinese Academy of Sciences, Ministry of Science and Technology, and National Natural Science Foundation of China; the Colombian Funding Agency (COLCIENCIAS); the Croatian Ministry of Science, Education and Sport, and the Croatian Science Foundation; the Research Promotion Foundation, Cyprus; the Ministry of Education and Research, Estonian Research Council via IUT23-4 and IUT23-6 and European Regional Development Fund, Estonia; the Academy of Finland, Finnish Ministry of Education and Culture, and Helsinki Institute of Physics; the Institut National de Physique Nucl\'eaire et de Physique des Particules~/~CNRS, and Commissariat \`a l'\'Energie Atomique et aux \'Energies Alternatives~/~CEA, France; the Bundesministerium f\"ur Bildung und Forschung, Deutsche Forschungsgemeinschaft, and Helmholtz-Gemeinschaft Deutscher Forschungszentren, Germany; the General Secretariat for Research and Technology, Greece; the National Scientific Research Foundation, and National Innovation Office, Hungary; the Department of Atomic Energy and the Department of Science and Technology, India; the Institute for Studies in Theoretical Physics and Mathematics, Iran; the Science Foundation, Ireland; the Istituto Nazionale di Fisica Nucleare, Italy; the Ministry of Science, ICT and Future Planning, and National Research Foundation (NRF), Republic of Korea; the Lithuanian Academy of Sciences; the Ministry of Education, and University of Malaya (Malaysia); the Mexican Funding Agencies (CINVESTAV, CONACYT, SEP, and UASLP-FAI); the Ministry of Business, Innovation and Employment, New Zealand; the Pakistan Atomic Energy Commission; the Ministry of Science and Higher Education and the National Science Centre, Poland; the Funda\c{c}\~ao para a Ci\^encia e a Tecnologia, Portugal; JINR, Dubna; the Ministry of Education and Science of the Russian Federation, the Federal Agency of Atomic Energy of the Russian Federation, Russian Academy of Sciences, and the Russian Foundation for Basic Research; the Ministry of Education, Science and Technological Development of Serbia; the Secretar\'{\i}a de Estado de Investigaci\'on, Desarrollo e Innovaci\'on and Programa Consolider-Ingenio 2010, Spain; the Swiss Funding Agencies (ETH Board, ETH Zurich, PSI, SNF, UniZH, Canton Zurich, and SER); the Ministry of Science and Technology, Taipei; the Thailand Center of Excellence in Physics, the Institute for the Promotion of Teaching Science and Technology of Thailand, Special Task Force for Activating Research and the National Science and Technology Development Agency of Thailand; the Scientific and Technical Research Council of Turkey, and Turkish Atomic Energy Authority; the National Academy of Sciences of Ukraine, and State Fund for Fundamental Researches, Ukraine; the Science and Technology Facilities Council, UK; the US Department of Energy, and the US National Science Foundation.

Individuals have received support from the Marie-Curie program and the European Research Council and EPLANET (European Union); the Leventis Foundation; the A. P. Sloan Foundation; the Alexander von Humboldt Foundation; the Belgian Federal Science Policy Office; the Fonds pour la Formation \`a la Recherche dans l'Industrie et dans l'Agriculture (FRIA-Belgium); the Agentschap voor Innovatie door Wetenschap en Technologie (IWT-Belgium); the Ministry of Education, Youth and Sports (MEYS) of the Czech Republic; the Council of Science and Industrial Research, India; the HOMING PLUS program of Foundation for Polish Science, cofinanced from European Union, Regional Development Fund; the Compagnia di San Paolo (Torino); the Consorzio per la Fisica (Trieste); MIUR project 20108T4XTM (Italy); the Thalis and Aristeia programs cofinanced by EU-ESF and the Greek NSRF; and the National Priorities Research Program by Qatar National Research Fund.
\end{acknowledgments}
\clearpage

\bibliography{auto_generated} 

\cleardoublepage \appendix\section{The CMS Collaboration \label{app:collab}}\begin{sloppypar}\hyphenpenalty=5000\widowpenalty=500\clubpenalty=5000\textbf{Yerevan Physics Institute,  Yerevan,  Armenia}\\*[0pt]
V.~Khachatryan, A.M.~Sirunyan, A.~Tumasyan
\vskip\cmsinstskip
\textbf{Institut f\"{u}r Hochenergiephysik der OeAW,  Wien,  Austria}\\*[0pt]
W.~Adam, T.~Bergauer, M.~Dragicevic, J.~Er\"{o}, M.~Friedl, R.~Fr\"{u}hwirth\cmsAuthorMark{1}, V.M.~Ghete, C.~Hartl, N.~H\"{o}rmann, J.~Hrubec, M.~Jeitler\cmsAuthorMark{1}, W.~Kiesenhofer, V.~Kn\"{u}nz, M.~Krammer\cmsAuthorMark{1}, I.~Kr\"{a}tschmer, D.~Liko, I.~Mikulec, D.~Rabady\cmsAuthorMark{2}, B.~Rahbaran, H.~Rohringer, R.~Sch\"{o}fbeck, J.~Strauss, W.~Treberer-Treberspurg, W.~Waltenberger, C.-E.~Wulz\cmsAuthorMark{1}
% ...
\vskip\cmsinstskip
% ...
1:~~Also at Vienna University of Technology, Vienna, Austria\\
2:~~Also at CERN, European Organization for Nuclear Research, Geneva, Switzerland\\
% ...

\end{sloppypar}
\end{document}